\documentclass[a4paper,11pt]{article}
\pdfoutput=1 

\usepackage{jheppub} 
\newcommand{\be}{\begin{equation}}
\newcommand{\ee}{\end{equation}}
\def \esan{E^{(3)}_{2}}

\newcommand{\hG}{h^\vee_G}
\newcommand{\ep}{\epsilon}
\def\sign{\frac{Q_m}{\sqrt{q_1q_2}}}
\def\hij{h_{ij}}
\usepackage[T1]{fontenc} 
\usepackage{makecell}
\usepackage{wasysym,amsthm}
\usepackage{dynkin-diagrams}
\usepackage{tikz}

\newtheorem*{conj}{Conjecture}
\newtheorem*{thm}{Criterion}
\newtheorem{mydef}{Definition}

\title{ \bf Twisted 6d (2$,$\,0) SCFTs on a Circle}

\author[a]{Zhihao Duan,}

\author[a]{Kimyeong Lee,}

\author[a]{June Nahmgoong,}

\author[b]{Xin Wang}

\affiliation[a]{School of Physics, Korea Institute for Advanced Study, Seoul 02455, Korea}
\affiliation[b]{Quantum Universe Center, Korea Institute for Advanced Study, Seoul 02455, Korea}

\emailAdd{xduanz@gmail.com}
\emailAdd{klee@kias.re.kr}
\emailAdd{junenahmgoong@gmail.com}
\emailAdd{wxin@kias.re.kr}

\abstract{We study twisted circle compactification of 6d $(2,0)$ SCFTs to 5d $\mathcal{N} = 2$ supersymmetric gauge theories  with non-simply-laced gauge groups. We provide two complementary approaches towards the BPS partition functions, reflecting the 5d and 6d point of view respectively. The first is based on the blowup equations for the instanton partition function, from which in particular we determine explicitly the one-instanton contribution for all simple Lie groups. The second is based on the modular bootstrap program, and we propose a novel modular ansatz for the twisted elliptic genera that transform under the congruence subgroups $\Gamma_0(N)$ of $\text{SL}(2,\mathbb{Z})$. We conjecture a vanishing bound for the refined Gopakumar-Vafa invariants of the genus one fibered Calabi-Yau threefolds, upon which one can determine the twisted elliptic genera recursively. We use our results to obtain the 6d Cardy formulas and find universal behaviour for all simple Lie groups. In addition, the Cardy formulas remain  invariant under the twist once the normalization of the compact circle is taken into account.}

\preprint{KIAS-P21003}

\begin{document} 
\maketitle
\flushbottom

\section{Introduction}

Interacting conformal field theories in dimensions larger than four is one of the most remarkable discoveries in string theory. Many of them do not have a Lagrangian description, and to study their dynamics we have to make full use of our knowledge on string theory accumulated in the past few decades. Six is known to be the maximal dimension allowing for their presence \cite{Nahm:1977tg}, so one naturally starts from theories in six dimensions. In particular, 6d $\mathcal{N} = (2,0)$ superconformal field theories (SCFTs) can arise from type IIB string theory on the ALE spaces $\mathbb{C}^2/\Gamma_{ADE}$ \cite{Witten:1995zh, Strominger:1995ac}. Theories of $A$ and $D$ types can also be engineered on parallel M5 branes without or with OM5 plane. These theories in the tensor branch have 1/2 BPS self-dual strings and 1/2 BPS massless tensor multiplets \cite{Haghighat:2013gba}. Furthermore, these 6d $(2,0)$ SCFTs  can be compactified on a circle $S^1$ and the resulting low energy 5d theories are  5d $\mathcal{N}=2$ supersymmetric  gauge theories with simply-laced $ADE$ type gauge groups. The instantons of these 5d theories play the role of Kaluza-Klein (KK) modes of the compactification \cite{Kim:2011mv}. The massless tensor multiplet becomes either 1/2 BPS instantons without electric charge or massless vector multiplet in the Coulomb branch. The winding self-dual strings become  1/2 BPS W-bosons and the unwound self-dual strings become magnetic monopole strings in the Coulomb phase.  

A further compactification of the 5d theory of $ADE$ type on a circle $\tilde{S}^1$   with outer-automorphism twist leads to the 4d KK theories with twisted affine algebra $G^{(n)}$ \cite{Tachikawa:2011ch}. The exchange of ordering of two compactifications is the 4d S-duality and provides the 5d $\mathcal{N}=2$ supersymmetric gauge theory with non-simply-laced group. This 4d S-dual, or  \emph{Langlands dual}, of the twisted affine Dynkin diagram is the untwisted affine Dynkin diagram of non-simple-laced group. Thus  5d $\mathcal{N}=2$ gauge theory can be obtained by the twisted compactification of the 6d $(2,0)$ SCFTs of $A, D$ and $E_6$ types. This will be explained in more detail in section \ref{sec:twist}. The low energy dynamics of these twisted theories are the 5d $\mathcal{N}=2$ theories with the non-simply-laced group $B_r=SO(2r+1), C_r=USp(2r)_0, (C_r)_\pi=USp(2r)_\pi, G_2$ and $F_4$ \cite{Tachikawa:2011ch}. There are two kinds of 5d $ USp(2r) $ gauge theories with adjoint flavor, depending on the discrete theta parameter being $0$ or $\pi$, which is sensitive to the homotopy group \(\pi_4(G)\).   The exact correspondence between the 6d $(2,0)$ theories and the 5d $\mathcal{N}=2$ theories of non-simply-laced groups   and the corresponding insertion of orientifold planes are given in table~\ref{tab:6d5d}.

In this work, we also want to explore the physics of these 5d theories directly from twisting the 6d theory. A very partial list of this approach can be found in \cite{Jefferson:2018irk,Bhardwaj:2018yhy,Bhardwaj:2018vuu,Bhardwaj:2019fzv,Apruzzi:2019vpe,Apruzzi:2019kgb,Apruzzi:2019opn,Apruzzi:2019enx,Bhardwaj:2020gyu,Braun:2021lzt} and references therein. For us, one of the main interests is the elliptic genus of the BPS strings in the 6d $(2,0)$ theories with twisted circle compactification. The 6d partition function for the counting of  BPS states of BPS strings with KK momenta would be equivalent to the 5d partition functions for dyonic instantons. From the 4d S-dual perspective,  magnetic monopole strings in the twisted compactification of the 5d $ADE$ theory with KK momentum but without electric charge and instanton number  would correspond to these self-dual strings.

 \begin{table}[h]
     \centering
           \arraycolsep=5pt \def\arraystretch{1.5}
     \begin{tabular}{c|c|c|c|c}
      6d  theory  & twist & 5d theory & \(\Gamma_0(n_G)\) & orientifold \\ \hline
    \(A_{2r}\)   &\(\mathbb{Z}_2\)   & \(USp(2r)_{\pi}\) &  \(\Gamma_0(4)\) &\(\widetilde{\text O4}{}^+\)\\
    \(A_{2r-1}\) &  \(\mathbb{Z}_2\)   & \(SO(2r+1)\) &  \(\Gamma_0(2)\) &\(\widetilde{\text O4}{}^-\) \\
    \(D_{r+1}\) & \(\mathbb{Z}_2\)   & \(USp(2r)_0\) &  \(\Gamma_0(2)\) & \({\text O4}^+\) \\
    \(D_4\) & \(\mathbb{Z}_3\)   & \(G_2\) &  \(\Gamma_0(3)\) & \\
    \(E_6\) & \(\mathbb{Z}_2\)   & \(F_4\)  &  \(\Gamma_0(2)\) &
     \end{tabular}
     \caption{Twisted compactification of 6d $(2,0)$ theories}
     \label{tab:6d5d}
 \end{table}

First of all, we analyze the BPS objects of these twisted theories from the 5d point of view by studying the blowup equations of the 5d instanton partition function. The original blowup equations were derived by Nakajima and Yoshioka, in order to prove that the
Nekrasov's instanton partition function gives a deformation of the Seiberg-Witten
prepotential for four dimensional $\mathcal{N}=2$ supersymmetric gauge
theory on $\mathbb{C}^2$.  Later on, they generalised them to five dimensional gauge theories and obtained K-theoretic blowup equations \cite{Nakajima:2005fg}, which count K-theoretic Donaldson invariants as discussed in a paper with G\"ottsche~\cite{Gottsche:2006bm}. Based on \cite{Sun:2016obh,Grassi:2016nnt}, in \cite{Huang:2017mis}, a geometric description was proposed, which makes it possible to define blowup equations for non-Lagrangian theories like local $\mathbb{P}^2$. In \cite{Kim:2020hhh}, the geometric description was generalised to all KK theories of 6d SCFTs, and was used to bootstrap the BPS invariants of the KK theories. The blowup equations are functional equations, which turn  out to be very powerful tool to bootstrap the BPS invariants, instanton partition functions/elliptic genera for various 5d/6d SCFTs \cite{Keller:2012da,Kim:2019uqw,Huang:2017mis,Gu:2018gmy,Gu:2019dan,Gu:2019pqj,Gu:2020fem,Kim:2020hhh}. In particular, the elliptic blowup equations for 6d $(2,0)$ $ADE$ M-strings were proposed in \cite{Gu:2019pqj}. However, the number of the equations is not enough to solve the elliptic genera recursively. Thanks to the 5d KK theory description of the 6d theory, we develop novel methods to bootstrap the instanton partition functions and the BPS invariants, which helps to recover the information of the elliptic genera. Utilizing our method, we manage to compute the exact one-instanton partition functions for all exceptional Lie groups with adjoint matters, when turning off gauge fugacities. We also study the (elliptic) blowup equations for the 6d $(2,0)$ KK theory with twisted circle compactifications. 

Secondly, as mentioned before, from the 6d point of view we should study the elliptic genus of effective strings. The elliptic genus is known to transform as a Jacobi modular form of weight zero, which imposes very strong constraints on its structure. To make full use of it, we come up with a physically motivated ansatz compatible with the modular property, and we are left with a finite number of undetermined coefficients. If one can find sufficiently many extra constraints either from gauge theory or geometry, one is able to fix all the unknowns and hence the elliptic genus itself. This approach is often dubbed modular bootstrap \cite{DelZotto:2016pvm,Gu:2017ccq,DelZotto:2017mee, Kim:2018gak,DelZotto:2018tcj,Duan:2018sqe, Duan:2020cta,Duan:2020imo}, and was successfully applied to compute the elliptic genera of various $(1,0)$ and $(2,0)$ SCFTs in 6d. This method is also applicable to compact elliptically fibered Calabi-Yau (CY) threefolds \cite{Huang:2015sta,Huang:2020dbh}, leading to interesting results on Heterotic/Type II string duality \cite{Cota:2019cjx,Cota:2020zse}, black hole physics \cite{Haghighat:2015ega} and swampland conjectures \cite{Lee:2018urn,Lee:2018spm,Lee:2020gvu,Cota:2020zse}, to name a few.

In this paper, we find a lot of evidence that after twisted compactification of $(2,0)$ SCFTs to 5d, there exists a twisted version of the elliptic genera that are covariant under congruence subgroups $\Gamma_0(N)$ of $\text{SL}(2,\mathbb{Z})$, which is shown in  table~\ref{tab:6d5d}. Geometrically, this should correspond to the topological string partition function on genus one fibered CY threefolds upon which M-theory compactifies \cite{Bhardwaj:2019fzv}. We develop novel modular ansatz for  6d twisted \( (2,0)\) theories  and determine the elliptic genera at low base degrees using conjectural vanishing conditions for the corresponding genus one fibered CY threefolds, extending the previous work \cite{Duan:2020cta}. We emphasize that although the twisted elliptic genera are not related to the 6d elliptic genera in an apparent way, the choice of subgroups indeed coincides nicely with one's expectation from twisting the theory, which is not at all obvious from the perspective of 5d instanton dynamics.

This paper is organized as follows. In section \ref{sec:twist}, we analyze in detail the procedure of twisted circle compactification of $(2,0)$ SCFTs to 5d $\mathcal{N} = 2$ gauge theories. We also discuss the effect of twisting on the self-dual strings. Then we move on to determining the partition functions. In section \ref{sec:Blowup}, we tackle this problem from the perspective of instanton counting. In particular, we show how to bootstrap the instanton partition function using the blowup equations with adjoint matters. In section \ref{sec:Modular_bootstrap}, our starting point becomes twisted self-dual strings instead. We propose a novel ansatz for the twisted elliptic genera, which, combined with conjectural vanishing conditions of refined BPS invariants, can also bootstrap the twisted elliptic genera recursively. In section \ref{sec:Cardy}, we consider the Cardy limit of the twisted elliptic genera and find universal asymptotic behavior of the 6d free energy for all simple Lie algebras. In section \ref{sec:conclusion}, we end this paper with concluding remarks and possible future directions.



\section{Twisted circle compactification of 6d $(2,0)$ SCFTs}\label{sec:twist}
In this section, we discuss some basic results about the twisted circle compactification of $(2,0)$ theories. Section \ref{sec:2.1} is devoted to a field theoretic derivation of the twisted compatification. It not only reviews previous works in the literature, but also contains some new discussions useful for later sections. In section \ref{subsec:EG}, we analyze the effect of the twisting on the $ADE$ M-strings. Finally in section \ref{sec:2.3}, we re-examine the results from the geometric engineering perspective.

\subsection{Outer-automorphism twist}\label{sec:2.1}
In this subsection, we discuss the compactification of 6d $(2,0)$ theories on a circle with the outer-automorphism twist. From the classification 6d $(2,0)$ SCFTs only allow the simply-laced chain of M-strings, engineered from IIB string theory on $ADE$ orbifold \(\mathbb{C}^2/\Gamma_{ADE}\). After the circle compactification, the 6d $(2,0)$ $ADE$ SCFTs become 5d \(\mathcal{N}=2\) gauge with $ADE$ gauge group, whose instantons carry KK modes on the circle \cite{Kim:2011mv}. However, one can still consider a 5d $\mathcal{N} = 2$ SYM with a non-simply-laced gauge group. For those theories, their UV completions are given by the 6d $(2,0)$ $ADE$ theories on a circle with outer-automorphism twist \cite{Vafa:1997mh, Tachikawa:2011ch}. 

\par

Since the 6d theory is non-Lagrangian and also $G$ is not its gauge group, it is hard to pin down the twisted circle compactification directly. To circumvent this, we revisit and refine the argument given by Tachikawa~\cite{Tachikawa:2011ch}. We compactify the 6d theory on a torus made of two distinct circles \(S^1\) and \(\tilde{S}^1\).    Now the order of compactification matters, so we shall consider the following two compactification chains,
\begin{align}
\begin{array}{lcl}
  &  \text{6d $(2,0)$ }G\text{-SCFT}     &     \\ [1mm]
  \hspace{1.5cm} {\color{blue}\swarrow}\    \text{on}\  S^1\ \text{of}\ R_6   &   & {\color{red}\searrow }\  \text{on}\ \tilde S^1 \ \text{of} \ \tilde R_6   
  \\ [1.5mm]
 \text{5d } G - \text{MSYM} &        &    \  \text{5d } {G}^{\vee} -  \text{MSYM} \\ [1mm]
  \hspace{1cm}{\color{red}\downarrow}\  \ \text{on} \ \tilde S^1\ \text{of}\ R_5 &  &   \hspace{1.5cm}     {\color{blue}\downarrow} \ \ \text{on} \ S^1 \ \text{of} \ \tilde R_5 \\ [1.5mm]
  \text{4d\ KK}\ G^{(n)} - \text{MSYM}  &   \xleftrightarrow[]{\hspace{5mm}\text{S-dual}\hspace{5mm}} &     \text{4d\ KK}\ {G}^{\vee(1)} - \text{MSYM} 
 \end{array}  
\label{eq127_chain}
\end{align}
where the blue arrow denotes the untwisted compactification on \(S^1\), and the red arrow denotes the twisted compactification on \(\tilde{S}^1\). The change of the compactification, which is the origin of the 4d S-duality, provides the clues to the twisted compactification of the 6d (2,0) theories. The direct compactification along $S^1$ without twist, there is no change of the radius, $R_6=\tilde R_5$. We will see that in the convention where the long roots of the Lie algebra of the gauge groups take the square length two, the circle radius of the twisted circle $\tilde S^1$ changes such that $n_G  R_5=\tilde R_6$ with $n_G=1,2,3,4$, depending on the starting theory and also its twisting $G^{(n)}$.

Let us first consider the left chain. To begin with, we compactify these 6d theories on a circle of radius \(R_6\) with periodic boundary condition \(x_6\sim  x_6+2\pi R_6\), to get 5d \(\mathcal{N}=2\) gauge theories of \(ADE\) type.  As instantons represent the KK modes of mass $1/R_6$, the 5d gauge coupling constant $g_5$ is fixed to satisfy \( 8\pi^2/g_5^2=1/R_6\). 
We further compactify this 5d theory  on   a circle of radius \(R_5\) so that \(x_5\sim  x_5+2\pi R_5\) with twisted boundary condition 
\begin{align}
      \phi(x_5+2\pi R_5) = \sigma(\phi(x_5))\,,
      \label{twist01}
\end{align}   
where \(\sigma\) is an automorphism of order \(n_G\). The twisted 5d theory has the effective 4d coupling constant \( 4\pi/g_4^2= 8\pi^2 R_5/g_5^2 = R_5/R_6 \) at the low energy. In the presence of twisting, the generators of the corresponding Lie algebra get split  to  the eigenstates under the corresponding automorphism \(   \sigma \). Each mode can have KK momentum of \((\mathbb{Z}+s/n_G)/R_5\) with \(n_G\) being the order of the corresponding automorphism and \(s=0,1,\cdots n_G-1\). The adjoint  field of the original 5d theory gets split into various fractional KK modes.  The case for \(A_{2r}\) needs an additional consideration with automorphism including the \(\mathbb{Z}_2\)   outer-automorphism~\cite{kac1990infinite,Tachikawa:2011ch}. With a matrix \(R=(1,i\sigma_2\otimes {\bf 1}_r)\), the automorphism for \(A_{2r}=SU(2r+1)\) is the map \( \sigma(\phi)= -R\phi^T R^{-1}\). As \(R^2=\text{diag}(1,-{\bf 1}_{2r})\), this twisting is actually of order four. See the table \ref{tab:6d5d} for the list of  \(n_G\).
 
 After carefully analyzing the decomposition of the adjoint representation into the representations of its invariant subalgebra as well as the fractional momentum dependence \(e^{(k+s/n_G)x_5/R_5}\),  we obtain the well-known result below: 
\begin{align}
    A^{(2)}_{2r} &: \ {\bf adj} \ {\rm of} \ A_{2r}   &  &\longrightarrow & & {\bf adj}_0\oplus {\bf fund }_\frac14 \oplus {\boldsymbol \Lambda}^2_\frac12 \oplus {\bf 1}_\frac12 \oplus{\bf fund}_\frac34 \ \ {\rm of} \ C_r\,, \nonumber\\
   A^{(2)}_{2r-1} &:\  {\bf adj} \ \  {\rm of} \ A_{2r-1} & & \longrightarrow & & {\bf adj}_0\oplus       {\boldsymbol \Lambda}^2_\frac12  \ \  {\rm of} \ C_r\,, \nonumber\\
   D^{(2)}_{r+1} &:\ {\bf adj} \ {\rm of} \ D_{r+1}   & &  \longrightarrow & &  {\bf adj}_0\oplus     {\bf fund }_\frac12  \ \ {\rm of} \ B_r\,, \nonumber\\
   E^{(2)}_6 &:\ {\bf adj} \ {\rm of} \ E_6    & & \longrightarrow   & &   {\bf adj}_0\oplus    {\bf fund }_\frac12  \ {\rm of} \ F_4\,, \nonumber\\
   D^{(3)}_4  &:\  {\bf adj} \ {\rm of} \ D_4    & &\longrightarrow  & &    {\bf adj}_0\oplus     {\bf fund }_\frac13 \oplus {\bf fund }_\frac23   \ \   \text{of} \ G_2\,,
   \label{momentumtwist}
\end{align}
where \( {\bf adj}_s\), \({\bf fund}_s\), and \({\boldsymbol \Lambda}^2_s\) denote the adjoint, fundamental and rank two anti-symmetric representations. It turns out that the extended weights on the right hand side of figure \ref{momentumtwist}, including also the KK momentum, give rise to the twisted affine algebra \(G^{(n)}\) \cite{kac1990infinite}. The identification of the simple roots of the original Lie algebra \(G\) under the twist is shown on the leftmost of table  \ref{tab166_affine}. Under this identification, the invariant and non-invariant roots under the twist become long and short roots of \(G^{(n)}\), respectively.

The second column in table \ref{tab166_affine} gives the Dynkin diagrams for the twisted affine Lie algebras. They encode the simple roots of the 4d theory plus the affine root with minimally positive momentum.  Ignoring the momentum dependence, the root vectors for the twisted affine algebra are given below:
\begin{align}
 \arraycolsep=5pt \def\arraystretch{1.5}
\begin{array}{|c|c|l|l|l|}
\hline
G^{(n)} &  \text{4d}\     & \qquad  \text{Long} & \qquad   \text{Short} & \ \text{Special} \\ \hline
    A^{(2)}_{2r} &  C'_r &\quad\ \pm \sqrt{2}e_a         &\  \frac{1}{\sqrt{2}} (\pm e_a\pm e_b)  &  \pm  \frac{1}{\sqrt{2}}  e_a    \\
   A^{(2)}_{2r-1} &  C_r &\quad\   \pm\sqrt{2}e_a     &\   \frac{1}{\sqrt{2}} (\pm e_a\pm e_b)    &\quad  $---$  \\ 
   D^{(2)}_{r+1} &  B_r &\quad   \pm e_a\pm e_b  &\qquad \pm  e_a        &   \quad  $---$  \\
   E^{(2)}_6 &  F_4 &\quad  \pm e_a\pm e_b    & \frac12 (\pm e_1\pm ... e_4),\ \pm e_a & \quad  $---$  \\
   D^{(3)}_4  & G_2 &  \scriptstyle \sqrt{2}(\pm 1, 0),  \frac{1}{\sqrt{2}}(\pm 1,\pm \sqrt{3}) &  \scriptstyle \sqrt{\frac23}(0,\pm 1), \ \frac{1}{\sqrt{2}}(\pm 1, \pm \frac{1}{\sqrt{3}}) & \quad  $---$\\
   \hline
\end{array}
\end{align}
In the above table, \(\{ e_1, e_2,\cdots e_r \}\) are \(r\)-dim orthonormal vectors and  \(a\neq b\) . While there is no difference between \(C_r'\) and \(C_r\) at the Lie algebra level, there is a difference in the 4d physics due to the range of the continuous 4d \(\theta\) parameter. We will discuss the detail later on in this subsection. 
The upshot is that we can represent the equation (\ref{momentumtwist}) more succinctly, highlighting the length and the fractional momentum \(e^{i p_5 x_5/R_5}\): 
\begin{align}
    A^{(2)}_{2r} &: \ {\bf adj} \ {\rm of} \ A_{2r}   &  &\longrightarrow & &    \text{long}_k \oplus   \text{short}_{\frac{k}{2}} \oplus \text{special}_{k\pm\frac 14} \oplus 1_{k+\frac12}  \ \ {\rm of} \ C'_r \nonumber\\
   A^{(2)}_{2r-1} &:\  {\bf adj} \ \  {\rm of} \ A_{2r-1} & & \longrightarrow & &  \text{long}_{k} \oplus   \text{short}_{\frac{k}{2}} \  \  {\rm of} \ \ C_r \nonumber\\
   D^{(2)}_{r+1} &:\ {\bf adj} \ {\rm of} \ D_{r+1}   & &  \longrightarrow & &   \text{long}_{k} \oplus   \text{short}_ {\frac{k}{2}}  \    \   {\rm of} \ B_r \nonumber\\
   E^{(2)}_6 &:\ {\bf adj} \ {\rm of} \ E_6    & & \longrightarrow   & &   \text{long}_{k} \oplus   \text{short}_{\frac{k}{2}}   \  \  {\rm of} \ F_4 \nonumber\\
   D^{(3)}_4  &:\  {\bf adj} \ {\rm of} \ D_4    & &\longrightarrow  & &    \text{long}_{k} \oplus   \text{short}_{\frac{k}{3} }  \  \ \text{of} \ G_2\ ,
   \label{eq:momentumtwist}
\end{align}
where  the subscript indicates \(p_5\) with \(k\in\mathbb{Z}\).  For \(A^{(2)}_{2r}\), there are additional special roots with half the length of long roots, which carries non-vanishing KK momenta \(k\pm \frac14, \ k\in \mathbb{Z}\).
There are only \(r\) independent Cartan elements at \(p_5=0\) in the twisted theory  \(G^{(n)}\). The higher momentum modes for the  Cartan elements \(\alpha_{\text{long}}\cdot H\) and  \(   \alpha_{\text{short}} \cdot H\) for long and short simple roots are given exactly by  \(E_{\alpha_{\text{long}}}\) and   \(E_{\alpha_{\text{short}}}\) as shown in the equation (\ref{eq:momentumtwist}).  

The \(A^{(2)}_{2r}\) theory is the intriguing  case. Even for this case, the  \(SU(2)\) generator  elements for the long  simple root  carry integer momentum \(k\in\mathbb{Z}\) while the short simple roots can carry integer or half-integer momenta, namely, \(p_5=\frac{k}{2}, k\in\mathbb{Z}\).
In addition there is an identity element with only half-integer momentum, \(p_5=k+\frac12, k\in\mathbb{Z}\).  Finally, the elements for special roots  can carry only nontrivial momenta \(k\pm \frac14, \ k\in \mathbb{Z}\). The identity element of half-integer momentum is generated, for example, by the commutation relations of special generators with momentum \(k+\frac14\).  

Up to now we have looked at the \(x_5\) dependence of all the modes in the 5d \(\mathcal{N}=2\) gauge multiplet with twisting. If we have not turned on any scalar expectation value \(\langle \phi\rangle\) or gauge holonomy \(\langle A_5\rangle\), then one can quantize each mode in the vanishing coupling limit with gauge fixing, and get the massless and massive modes with KK momentum $p_5/R_5$. The value of the KK momentum for adjoint vector multiplet for each root  is identical to that given in the equation (\ref{eq:momentumtwist}), and so the energy of the mode would be \(E\cong \sqrt{p_1^2+p_2^2+p_3^2+p_5^2/R_5^2}\) for the given 3d spatial momentum \((p_1,p_2,p_3)\).  

\par
 
Let us now consider the  BPS massive spectrum in the Coulomb phase. From the 6d perspective we have turned on the VEV of single scalar field \(\langle\Phi\rangle\) in the 6d (2,0) theory, which has the mass dimension two and decides the tension of self-dual strings. Reduced on $S^1$, the 5d scalar vev \(\langle\phi\rangle=R_6 \langle  \Phi\rangle\) has the mass dimension one and decides the W-boson mass, which is a self-dual string wrapping along the \(x_6\) circle.  Meanwhile the 5d gauge coupling constant is related to the instantons which are KK momentum states via \(8\pi^2/g_5^2=1/R_6\). A further reduction on $\tilde{S}^1$ gives 4d \(\mathcal{N}=4\) KK theory with 4d gauge coupling constant which is related to the 5d one via \(4\pi/g_4^2= 8\pi^2 R_5  /g_5^2 =R_5/R_6\). The scalar vev and the \(A_5\) holonomy get combined into a complex Coulomb moduli \(\mathring{\phi} = \langle \phi + i A_5\rangle\). In this 4d KK theory, there are  1/2 BPS W-bosons, magnetic monopoles, Kaluza-Klein modes for \(x_5\)-compactification and instantons for \(x_6\)-compactification. Both instantons and KK modes get fractionalized in the general Coulomb phase. Let us just consider the W-bosons and magnetic monopoles for the SU(2) of a root in \(\textbf{adj}_0\) of the untwisted or twisted \(G^{(n)}\) 4d KK theory. Their masses are respectively,
\begin{align}
     m_{\text{W}} =   |\alpha\cdot\langle \phi\rangle|    =   R_6 |\alpha\cdot\langle \Phi\rangle |   ,  \ \ 
    m_{\text{M }} =  \frac{4\pi}{g_4^2} |\alpha^\vee \cdot\langle \phi\rangle|   =  \frac{2}{\alpha^2}R_5 |\alpha\cdot\langle \Phi\rangle |\,.   
\label{eq:WbosonMono} 
\end{align}

\par
The 4d magnetic monopoles arise from the 5d magnetic monopole strings wrapping the twisted circle \(\tilde S^1\). Since the 5d monopole strings are the 6d self-dual strings in the 5d theory, the physics of magnetic monopoles with KK momentum along \(\tilde S^1 \) captures the twisted 5d theory on a circle. The magnetic monopoles for roots provide the massive dual W-bosons for the dual gauge theory in the Coulomb branch. The 4d dual magnetic group has the magnetic dual gauge group \(G^\vee\) with the dual roots \(\alpha^\vee=2\alpha/\alpha^2\) made of the  roots \(\alpha\)  of the zero momentum sector of the twisted affine algebra \(G^{(n)}\). 

\par
Now, let us consider the second compactification chain in equation (\ref{eq127_chain}). First, we compactify the original 6d theory on the twisted circle \(\tilde{S}^1\). The resulting 5d theory is also an $\mathcal{N} = 2$ gauge theory with some gauge group \({G}^{\vee}\), which is determined by self-dual strings with twisted compactification.   Then, we further compactify the 5d theory on the untwisted circle \(S^1\) to obtain the 4d \(\mathcal{N} = 4\) KK theory with gauge group ${G}^{\vee(1)}$. Now let us look at the bottom line of equation \eqref{eq127_chain}. Exchanging the order of compactification amounts to the S-duality transformation of the 4d \(\mathcal{N}=4\) KK theory with gauge group $G^{(n)}$, which exchanges the W-bosons and magnetic monopoles and replaces $G^{(n)}$ with the \emph{Langlands dual} gauge group \({G}^{\vee(1)}\), constructed from the affine coroots. Surely, for simply-laced \(ADE\) cases (without actual twisting), their Langlands dual \({G}^{\vee(1)}\) are identical to the original groups \(G^{(n)}\). Then the S-duality in 4d is a self-duality \(4\pi/g_4^2=R_5/R_6 \cong   \tilde{g}_4^2/4\pi=\tilde{R}_6/\tilde{R}_5 \) and we set \( R_5=\tilde{R}_6, R_6=\tilde{R}_5\). 
 
For cases with twisting as in equation (\ref{eq127_chain}), the situation becomes more interesting. Note that we use the convention where the longest roots have the square length two, so we need to scale down the length of all affine coroots by $\sqrt{n_G}$, i.e.,  \(  \alpha_{{G}^{\vee(1)}} = \alpha^\vee /\sqrt{n_G}\).  From \eqref{eq:WbosonMono} for the magnetic monopole mass \( m_\text{M}=n_G R_5 |\langle \alpha_{G^{\vee(1)}}\cdot \Phi/\sqrt{n_G}|\),  we find that the S-duality in these cases exchanges the coupling constant in the following way,
  \begin{align}
  \label{eq:couplingsdual}
     \frac{1}{n_G}\frac{g_4^2}{4\pi} = \frac{R_6}{n_G R_5}\  \cong \ \frac{4\pi}{\tilde g_4^2} =\frac{\tilde R_5}{\tilde R_6}\,,  \
 \end{align}
 which arises from the identification of the compactification radius as follows:
\begin{align} \label{eq:radiusSdual}
    \tilde R_5=R_6, \ \ \tilde R_6=n_G R_5. 
\end{align}
Here there is a scaling of the tensor scalar VEV for W-bosons and magnetic monopoles by \(\sqrt{n_G}\) for the invariance of the mass.  For the earlier works on the S-duality for 4d theories with non-simple-laced group, see, for example, \cite{Dorey:1996hx,Argyres:2006qr,Kim:2004xx}.

The Dynkin diagram of Langlands dual algebra \(G^{(1)}_r\) is shown in the third column of the table \ref{tab166_affine}. Except for the \(A^{(2)}_{2r}\) case, the dual roots are simply the  Dynkin diagram of the dual roots \(\alpha^\vee/\sqrt{n}_G\) of simple roots \(\alpha\) of \(G^{(n)}\)  algebra in the second column of the table \ref{tab166_affine}. 

For \(A^{(2)}_{2r}\), its dual group turns out to be \((C^{(1)}_r)_\pi\) which arise from the 5d \((C_r)_\pi=USp(2r)_\pi\) theory with discrete theta parameter \(\pi\).  We will provide several evidences for this later in this subsection.   The list of explicit expressions for roots of the Lie algebra  $G^{\vee (1)}_r$ with the zero \(\tilde S^1\) momentum are as follows:
\begin{align}
 \arraycolsep=5pt \def\arraystretch{1.5}
\begin{array}{|c|c|c|c|}
\hline
G^{(n)}   & G^\vee_r &     \text{Long} &    \text{Short}  \\ \hline
A^{(2)}_{2r} &     (C_r)_\pi &   \pm \sqrt{2}e_r         &\  \frac{1}{\sqrt{2}}  (\pm e_a\pm e_{b} )    \\
A^{(2)}_{2r-1} &       B_r &\    \pm e_a\pm e_b      &  \pm e_a       \\ 
D^{(2)}_{2r-1} &       (C_r)_0 &    \pm \sqrt{2}e_a  & \frac{1}{\sqrt{2}} (\pm e_a\pm e_b)         \\
E^{(2)}_6 &      F_4 &   \pm e_a\pm e_b    & \frac12 (\pm e_1\pm ... e_4),\ \pm e_a   \\
 D^{(3)}_4 &     G_2 &  \scriptstyle \sqrt{2}(\pm 1, 0),  \frac{1}{\sqrt{2}}(\pm 1,\pm \sqrt{3}) &  \scriptstyle \sqrt{\frac23}(0,\pm 1), \ \frac{1}{\sqrt{2}}(\pm 1, \pm \frac{1}{\sqrt{3}}) \\
   \hline
\end{array}
\end{align}

For all cases, the Langlands dual gauge group \({G}^{\vee(1)}\) in 4d actually becomes untwisted, meaning that all affine roots have the same KK mode. This can be seen from \eqref{eq:momentumtwist}. Moreover, since descending from 5d to 4d involves no twisting, it is straightforward to find out what is ${G}^{\vee}$. The final result is listed in table \ref{tab166_affine}. Note that \({G}^{\vee}\) is not a subalgebra of \(G\) if \({G}^{\vee}\in\{B_r, \ C_r\}\), which is only possible if 6d (2,0) theory with\(G\) is not a  gauge theory.

 \begin{table}[ht]
\centering
 \arraycolsep=5pt \def\arraystretch{1.5}
\begin{tabular}{|c|c|c|c|}\hline
\(G \ / \ \text{Out}(G)\) & \(G^{(n)}\)\ (4d \ \(G'\) ) & \( {G}^{\vee(1)}\) & 5d \ \( {G}^\vee\) \\ \hline \hline
\(A_{2r}\ /\ \mathbb{Z}_2\) & \(A_{2r}^{(2)}\)\ (4d \(C_r'\) ) &  \((C_r^{(1)})_\pi\)   & \( (C_r)_\pi \) \\
\dynkin[%
edge length=.75cm, involution/.style={blue,stealth-stealth},
labels*={\alpha_1,\alpha_{r-1},\alpha_r,\alpha_{r+1},\alpha_{r+2},\alpha_{2r} },
involutions={16;25;34}]{A}{*.****.*}
 &  \dynkin A[2]{even}  &   \dynkin C[1]{}  & 
 \dynkin[%
edge length=.75cm, 
labels*={\alpha_1,\alpha_2,\alpha_{r-1},\alpha_r}]{C}{**.**}
\\
& \(\widetilde{\text{O3}}{}^-\)  \hfill \(\widetilde{\text{O3}}{}^+\) & \(\text{O3}^+\)  \hfill \( \widetilde{\text{O3}}{}^+\)
& \( \widetilde{\text{O4}}{}^+\)
 \\
\hline
\(A_{2r-1}\ /\ \mathbb{Z}_2\) & \(A_{2r-1}^{(2)}\)\ (4d \(C_r\) ) & \(B_r^{(1)}\) & \(B_r\)  \\
 \dynkin[%
edge length=.75cm, involution/.style={blue,stealth-stealth},
labels*={\alpha_1,\alpha_{r-1},\alpha_r,\alpha_{r+1},\alpha_{2r-1} },
involutions={15;24}]{A}{*.***.*}
&\dynkin A[2]{odd}
&\dynkin B[1]{}
&\dynkin[%
edge length=.75cm, 
labels*={\alpha_1,\alpha_2,\alpha_{r-1},\alpha_r}]{B}{**.**}
\\
& \(\text{O3}^-\)  \hfill \(\text{O3}^+\) & \(\text{O3}^-\)  \hfill \(\widetilde{\text{O3}}{}^-\)
& \(\widetilde{\text{O4}}{}^-\)
\\ \hline
\(D_{r+1}\ /\ \mathbb{Z}_2\) & \(D_{r+1}^{(2)}\)\ (4d \(B_r\) ) & \((C_r^{(1)})_0\) & \( (C_r)_0\) \\
\dynkin[%
edge length=.75cm, involution/.style={blue,stealth-stealth},
labels={\alpha_1,\alpha_2,\alpha_{r-1},\alpha_r,\alpha_{r+1} },
involutions={54}]{D}{**.***}
&\dynkin D[2]{}
&\dynkin C[1]{}
&\dynkin[%
edge length=.75cm, 
labels*={\alpha_1,\alpha_2,\alpha_{r-1},\alpha_r}]{C}{**.**}
\\
& \(\widetilde{\text{O3}}{}^-\)  \hfill \(\widetilde{\text{O3}}{}^-\) & \(\text{O3}^+\)  \hfill \( \text{O3}^+\)
& \( \text{O4}^+\)
\\ \hline
\(D_4\ /\ \mathbb{Z}_3\) & \(D_4^{(3)}\)\ (4d   \(G_2\) )  & \(G_2^{(1)}\) & \(G_2\)  \\ 
\dynkin[%
edge length = .75cm, involution/.style={blue,stealth-stealth},
labels={\alpha_1,\alpha_2,\alpha_3,\alpha_4},
involutions = {31;43;14}]{D}{4}
&\dynkin D[3]{4}
&\dynkin G[1]{2}
&\dynkin[%
labels*={\alpha_1,\alpha_2},
edge length=.75cm]{G}{2}
\\ \hline
\(E_6\ /\ \mathbb{Z}_2\) & \(E_6^{(2)}\)\ (4d \(F_4\) )  & \(F_4^{(1)}\) & \(F_4\) \\
\dynkin[%
edge length = .75cm, involution/.style={blue,stealth-stealth},
labels*={\alpha_1,\alpha_6,\alpha_2,\alpha_3,\alpha_4,\alpha_5},
involutions = {16;35}]{E}{6}
&\dynkin E[2]{6}
&\dynkin F[1]{4}
&\dynkin[%
edge length=.75cm,
labels*={\alpha_1,\alpha_2,\alpha_3,\alpha_4}]{F}{4} 
\\ 
\hline
\end{tabular}
\caption{Dynkin diagrams for (affine) Lie algebras along the compactification chain (\ref{eq127_chain}). Brane interpretation is given when possible. Note that \(G^{(n)}\) and \( {G}^{\vee(1)}\) are related by the S-duality between 4d theories. The \(A^{(2)}_{2r}\) case is subtler and its S-duality is explained in the text.   } 
\label{tab166_affine}
\end{table}

It is also productive to understand the above results from the brane systems. Recall that the 5d \(\mathcal{N}=2\) theories with \(B_r, (C_r)_0,  (C_r)_\pi\) and \( D_r \) gauge groups  arise from multiple D4-branes on \(\widetilde{\text{O4}}{}^-,\)  \(\text{O4}^+,\)  \(\widetilde{\text{O4}}{}^+\), and  \( \text{O4}^-\) planes, respectively \cite{Hori:1998iv,Tachikawa:2011ch}.  Note that after a circle compactification and T-duality, we get the interpretation in terms of O3-planes \cite{Gimon:1998be,Hanany:2000fq}. The S-dual of these O3-planes for the 5d theory on a circle without twist is exactly the O3-planes for the twisted   \(G^{(n)}\) shown in table \ref{tab166_affine}.  One important point is that in 4d, there are two kinds of \(C_r=USp(2r)\) theories depending on the range of the 4d continuous \(\theta\)-parameter. The standard 4d \(C_r=USp(2r)\) theory on D3-branes with  \( \text{O3}^+\)is S-dual to 4d \(B_r=SO(2r+1)\) theory on D3-branes with \(\widetilde{\text{O3}}{}^-\).

Another 4d
\(C_r'=USp(2r)\) theory on D3-branes with  \(\widetilde{\text{O3}}{}^+\) is self-dual. This   \(C_r'\) theory has the continuous theta \(\theta=2\pi\), such that the magnetic monopole for the long simple root has nontrivial electric charge due to additional Witten effect~\cite{Witten:1979ey}. On the other hand, the  dyonic  bound state of two such monopoles,  a.k.a. Sen's state~\cite{Sen:1994yi} with unit electric charge, has no net electric charge when one includes also the Witten effect. Thus there exists a 1/2 BPS massive vector multiplet of zero electric charge  and twice magnetic charge for the long simple root, making it self-dual under the S-transformations.  1/2 BPS single magnetic monopole states  for short simple roots have zero electric charge sector when one considers both the dyonic BPS states and the  Witten effect together. Thus this theory is S-duality invariant. Note that \(\text{O3}^+\) and \(\widetilde{\text{O3}}{}^+\) are related by the T-transformation of the $\text{SL}(2,\mathbb{Z})$. 
Some studies of the twisted theories and magnetic monopole spectrum were done before \cite{Kim:2004xx,Tachikawa:2011ch}. 

The S-duality of \(A^{(2)}_{2r}\) is  a bit complicated as its affine Dynkin diagram is self-dual under the \(\alpha\leftrightarrow \alpha^\vee=2\alpha/\alpha^2\). Its S-dual theory is the 4d KK theory with affine \((C_r^{(1)})_\pi\). From the brane perspective, the S-dual of \(A^{(2)}_{2r}\) theory  arising from D3 branes between \(\widetilde{\rm O3}{}^-\) and \(\widetilde{\rm O3}{}^+\) is the \((C^{(1)}_r)_\pi\) theory  arising from D3 branes between \( \text{O3}{}^+\) and \(\widetilde{\rm O3}^+\).  
From it we learn that the  \(A_{2r}^{(2)}\) 4d KK theory cannot be self-dual. 
The Montonen-Olive duality is a quantum symmetry and so is sensitive to full quantum spectrum. 
  
There is an additional subtlety in the S-duality. In the left-side line of the compactification  \eqref{eq127_chain}  to 4d, we have 1/2 BPS W-bosons, instantons, magnetic monopoles and \(x_5\) KK momentum. We are interested in the magnetic monopoles with \(x_5\) KK momentum, which would correspond to the 5d W-bosons and instantons of the dual theory with non-simple-laced group, obtained by the right-side line of the compactification \eqref{eq127_chain}.  These 5d W-bosons and instantons are the wrapped self-dual strings with KK momentum of the twisted compactification of 6d (2,0) theories, which are of our primary interest.  

In 4d \(\mathcal{N}=4\) theories, in the Coulomb branch with only a single scalar field vev turned on \(\langle \phi\rangle\), the simple roots of the Lie algebra appear naturally.  While  W-bosons of mass \(\alpha\cdot\langle\phi\rangle\) exist as 1/2 BPS elementary particle for all positive roots, classically only for each simple root there exists  an \emph{elementary} 1/2 BPS  magnetic monopole with four zero modes, three for the 3d position and one for the internal phase. Once we quantize the moduli space of these elementary monopoles, the threshold bound states without electric charge appear for each positive roots, forming the W-boson of the dual gauge theory\cite{Lee:1996if,Gauntlett:1996cw}.  
Along the left-side of the compactification to 4d, we want to consider these elementary BPS magnetic monopole  for each simple roots with zero \(x^5\) momentum and mass in  \eqref{eq:WbosonMono}. We want to add the \(x^5\) momentum by simply combining the KK momentum carried by the Cartan for each simple roots as given in \eqref{eq:momentumtwist}. Namely, one is imagining chargeless KK momentum combined with magnetic monopoles for each simple root \(\alpha\) of \(G^\vee\). This would generate a simple prescription for elementary monopoles with arbitrary \(x_5\) KK momentum and without any electric and 
instanton charge. The full dynamics of these objects is at least as complicated as self-dual strings with KK momentum on the twisted circle.

In the right-side of the compactification chain, they correspond to 5d W-bosons with instantons for the twisted compactification, with the radius \(\tilde R_6=n_G R_5\).  This makes the \( x^6\) momentum \(n_G p_5\) integer-valued. On the other hand, a single  
\(SU(2)\) instanton embedded  in each simple root \(\alpha\)    carries the instanton number \(2/\alpha^2\). Both the 4d monopole and 5d instanton description give the same result for the allowed instanton number of a single W-boson or the KK momentum of an elementary string, corresponding to each simple root in 5d \( G^\vee\) as follows: 
\begin{align}
 \arraycolsep=5pt \def\arraystretch{1.5}
\begin{array}{|c|c|c|c|c|c|} 
\hline
\text{Twist} & \text{5d}\ G^\vee & \text{Long} & \text{Short} & \text{Special}  \\ [1mm]
\hline
A^{(2)}_{2r} & (C_r)_\pi   & [\alpha_r]_{2k,4k} &  [\alpha_a]_{2k}   &   [\frac{1}{2}\alpha_r]_{4k\pm 1}   \\ [1mm]
   A^{(2)}_{2r-1} &  B_r & [\alpha_a]_k    & [\alpha_r]_{2k}  & $---$   \\ [1mm]
   D^{(2)}_{r+1} &(C_r)_0 &    [\alpha_r]_k & [\alpha_a]_{2k}    & $---$  \\      [1mm]
   E^{(2)}_6 &  F_4   & [\alpha_1]_k, [\alpha_2]_{k}  & [\alpha_3]_{2k}, [\alpha_4]_{2k}   & $---$    \\ [1mm]
   D^{(3)}_4  & G_2  &   [\alpha_1]_{k} & [\alpha_2]_{3k}  & $---$  \\ [1mm]
   \hline
\end{array}
   \label{eq:6dtwist}
\end{align} 
where $a=1,2,\cdots r-1$.  
Before we move on to $A^{(2)}_{2r}$ case, one would ask if any role is played by monopoles related to the affine simple root of \(G^{(n)}\). One could imagine an \(SU(2)\) generator \( (\alpha\cdot H, E_\alpha e^{ip_5/R_5}, E_{-\alpha}e^{-ip_5/R_5}) \) and building up a magnetic monopole solution. It is static and has \(x_5\) dependence, so it naturally has nontrivial \(F_{i5}\) component with an instanton number along the left-side compactification in \eqref{eq127_chain}. Thus they are not of our current interest.  
 
Let us finally consider the 5d \((C_r)_\pi\) case arising from twisting   \(A^{(2)}_{2r} \).  After the twisted compactification from 5d to 4d,  the equation \eqref{eq:momentumtwist} shows that the possible KK momentum states are \(k,k/2, k\pm 1/4, k+1/2\) with \(k\in \mathbb{Z}\),  and the possible magnetic charges  are  long, short and special. When one combines them together and takes the S-dual picture with momentum factor \(n_G=4\), The possible BPS states one gets in 5d turn out to be given in the equation   \eqref{eq:6dtwist} . Without KK momentum, there are states for each root of \(C_r\). Here we take the view from the 6d and so they appear as a composite of  elementary   strings for simple roots of \(C_r\). These elementary strings for simple roots can carry KK momentum \(2k, k\in\mathbb{Z}\). In addition, there exist strings of half electric charge \(\frac12\alpha_r\) of the long simple root. This string has to carry nontrivial KK momentum \(4k\pm 1,  k\in\mathbb{Z}\) so that it does not appear in the gauge multiplet spectrum of 5d theory. This is consistent with the known instanton dynamics. Single instanton for \((C_r)_\pi\) case carrying nontrivial electric charge belongs to fundamental representation of \(USp(2r)=C_r\). For the Cartan elements, the \(x_6\) KK momentum is just 4 times of what is given in \eqref{eq:momentumtwist}: the \(\alpha_r\cdot H\) mode  carries the KK momentum  \(4k\),   the  \(\alpha_a\cdot H, a=1,...r-1\) modes carry the KK momentum \(2k\), and finally the identity element carries the KK momentum \(4k+2\) with integer \(k\). 

\subsection{Elliptic genera of twisted $ADE$ M-strings}\label{subsec:EG}
Now, let us consider the partition function of the twisted 6d $(2,0)$ theories. More precisely, we will study the partition function of 6d $(2,0)$ theory on \(\mathbb{R}^4\times T^2\) where a spatial circle in \(T^2\) gives the outer-automorphism twist. As we have discussed so far, it is equivalent to the partition function of 5d $\mathcal{N} = 2$ gauge theory on \(\mathbb{R}^4\times S^1\) with a non-simply-laced gauge group. The full \(\mathbb{R}^4\times T^2\) partition function is defined schematically as follows,
\begin{align}
Z_{\sigma}(\tau,\epsilon_{1,2},m,\mathbf{v})=\text{Tr}\Bigr[(-1)^F e^{2\pi i \tau P} e^{-\epsilon_1 J_1-\epsilon_2  J_2}e^{-(\epsilon_+ +m)R_1-(\epsilon_+ -m)R_2} e^{-\mathfrak{n}\cdot \mathbf{v}}
\Bigr].
\label{eq163_BPSR}
\end{align}
Here, $\sigma$ resembles the twisting operation, \(P\) is the KK momentum along the spatial circle in \(T^2\), \(J_{1,2}\) are two angular momenta on \(\mathbb{R}^4\), and \(R_{1,2}\) are two charges of \(SO(5)\) R-symmetry. $m$ can be regarded as a mass deformation, known as the M-string mass. Lastly, \(\mathfrak{n}\) is the charge of the twisted M-strings in 6d, which becomes the electric charge of W-bosons in 5d. Henceforth, for simplicity we will omit $\sigma$ in the subscript of $Z$, with the correct twisting understood from the context.

\par 

 The 6d BPS partition function (\ref{eq163_BPSR}) admits two different descriptions: the instanton partition function and the elliptic genus. First, one can study a 6d $(2,0)$ theory from the 5d $\mathcal{N} = 2$ gauge theory with KK instantons. Then, the 6d partition function admits the following expansion,
 \begin{align}\label{eq:instanton_expansion}
 Z(\tau,\epsilon_{1,2},m,\mathbf{v})=Z_\text{pert}(\epsilon_{1,2},m,\mathbf{v}) \Bigr(1+\sum_{k=1}^\infty Z_k^{\text{inst}}(\epsilon_{1,2},m,\mathbf{v}) q^k\Bigr), \quad (q=e^{2\pi i \tau}).
 \end{align}
Here, \(Z_\text{pert}\) is the perturbative partition function which captures the perturbative W-bosons in 5d, and its form is given as follows,
\begin{align}
Z_\text{pert}=\text{PE}\Bigr[{\sinh{m\pm\epsilon_+\over 2}\over \sinh{\epsilon_{1,2}\over 2}} \sum_{\alpha\in\Delta^+(G)}e^{-\alpha\cdot \mathbf{v}} \Bigr],
\end{align}
where \(\Delta^+(G)\) is the positive root system of 5d gauge algebra \(\mathfrak{g}\). Also, \(Z_k\) is the \(k\)-instanton partition function, which captures the non-perturbative degrees in 5d with \(k\)-unit of KK momentum. For the classical gauge groups, the instanton partition functions are well-studied with ADHM construction \cite{Nekrasov:2004vw, Hwang:2014uwa, Hwang:2016gfw}. On the other hand, the exceptional gauge groups do not have ADHM construction, and their instanton partition functions are much less understood.

\par 

Instead of the 5d description, one can study the 6d partition function from the BPS spectra on the twisted M-strings on \(T^2\). Then, the 6d partition function admits the following expansion,
\begin{align}
Z=Z_0\times\Bigr(1+\sum_{\mathfrak{n}} \mathbb{E}_{\mathfrak{n}}(\tau,\epsilon_{1,2},m) e^{-\mathfrak{n}\cdot \mathbf{v}}\Bigr).
\label{eq202_egexp}
\end{align}
In (\ref{eq202_egexp}), the expansion parameter is the string fugacity \(e^{-\mathfrak{n}\cdot \mathbf{v}}\), and the coefficient \(Z_{\mathfrak{n}}\) is called the elliptic genus with charge \(\mathfrak{n}=(n_1,n_2,...,n_r)\). Here, \(Z_0\) is call the Abelian contribution which is the neutral part of the partition function independent of the tensor VEV \(\mathbf{v}\). The Abelian contribution of general gauge group takes the following form,
\begin{align}\label{eq:abl1}
U(1): \quad Z_0&=\text{PE}\Bigr[{\sinh{m\pm\epsilon_-\over 2}\over \sinh{\epsilon_{1,2}\over 2}}{q\over 1-q} \Bigr],
\nonumber \\
A_r, \ D_r, \ E_r: \quad Z_0&=\text{PE}\Bigr[{\sinh{m\pm\epsilon_-\over 2}\over \sinh{\epsilon_{1,2}\over 2}}{rq\over 1-q} \Bigr],
\nonumber \\
B_r: \quad Z_0&=\text{PE}\Bigr[{\sinh{m\pm\epsilon_-\over 2}\over \sinh{\epsilon_{1,2}\over 2}}\Bigr((r-1){q\over 1-q}+{q^2\over 1-q^2} \Bigr) \Bigr],
\nonumber \\
(C_r)_0: \quad Z_0&=\text{PE}\Bigr[{\sinh{m\pm\epsilon_-\over 2}\over \sinh{\epsilon_{1,2}\over 2}}\Bigr((r-1){ q^2\over 1-q^2}+{q\over 1-q} \Bigr) \Bigr],
\nonumber \\
(C_r)_\pi: \quad Z_0&=\text{PE}\Bigr[{\sinh{m\pm\epsilon_-\over 2}\over \sinh{\epsilon_{1,2}\over 2}}\Bigr((r-1){ q^2\over 1-q^2}
+{q^4\over 1-q^4}+{q^2\over 1-q^4}\Bigr)\Bigr],
\nonumber \\
G_2: \quad Z_0&=\text{PE}\Bigr[{\sinh{m\pm\epsilon_-\over 2}\over \sinh{\epsilon_{1,2}\over 2}}\Bigr({q\over 1-q}+{q^3\over 1-q^3} \Bigr) \Bigr],
\nonumber \\
F_4: \quad Z_0&=\text{PE}\Bigr[{\sinh{m\pm\epsilon_-\over 2}\over \sinh{\epsilon_{1,2}\over 2}}\Bigr({2q\over 1-q}+{2q^2\over 1-q^2} \Bigr) \Bigr].
\end{align}
Note that \(Z_0\) of non-simply-laced gauge group is made of \(U(1)\) partition functions with different momentum fugacities. In general, it can be read off from the decomposition rule (\ref{eq:momentumtwist}). In this paper, we will mostly focus on the non-Abelian part \(Z/Z_0\) and the elliptic genera \(\mathbb{E}_{\mathfrak{n}}\).

\par 

For 6d $(2,0)$ $A$-type theories, the elliptic genus can be computed from 2d quiver gauge theories on M-strings \cite{Haghighat:2013gba}. For other types of theories, such 2d gauge theory description has been yet unknown. Instead, \(\text{SL}(2,\mathbb{Z})\) modular property of the elliptic genera can be used to determine them. Such procedure is called the `modular bootstrap', and it has been studied in various 6d theories \cite{DelZotto:2016pvm,Gu:2017ccq,DelZotto:2017mee, Kim:2018gak,DelZotto:2018tcj,Duan:2018sqe, Duan:2020cta,Duan:2020imo}, including $(2,0)$ $D$, $E$-type theories \cite{Gu:2017ccq, Duan:2020cta}. In this paper, we will focus on the twisted circle compactification of $(2,0)$ theories whose elliptic genera are unknown. Specifically, we will extend the modular bootstrap program to be applicable for non-simply-laced theories also.

\par 

As will be explained later, for the twisted M-strings of type \({G}^{\vee}\) given by the \(\mathbb{Z}_{n_G}\) twist of \(G\), the elliptic genus is not the Jacobi form of \(\text{SL}(2,\mathbb{Z})\), but of the particular congruence subgroup of \(\text{SL}(2,\mathbb{Z})\). Such subgroup \(\Gamma_0(n_G)\footnote{We use both letters $n_G$ and $N$ for congruence subgroups in this paper, hoping no confusion will occur.}\subset \text{SL}(2,\mathbb{Z})\) is given as follows,
\begin{align}
\Gamma_0(n_G)=\Bigr\{
\begin{pmatrix}
a & b \\ c & d
\end{pmatrix} \in \text{SL}(2,\mathbb{Z}): c \equiv 0\ (\text{mod}\ n_G)
\Bigr\}\,.
\end{align}
Under the \(\Gamma_0(n_G)\) action, as a Jacobi modular form the elliptic genus transforms as follows,
\begin{align}
\mathbb{E}_{\mathfrak{n}}\Bigr({a\tau+b\over c\tau+d}\Bigr|{z\over c\tau+d}\Bigr)
=\exp\Bigr[-{\pi i c\over c\tau+d}\mathfrak{i}_{\mathfrak{n}}(z) \Bigr]\cdot \mathbb{E}_{\mathfrak{n}}(\tau|z),\quad
\begin{pmatrix}
a & b \\ c & d 
\end{pmatrix}\in \Gamma_0(n_G),
\end{align}
where \(z\) collectively denotes the elliptic parameters \(\epsilon_\pm\) and \(m\). The elliptic genus is a modular form with index \(\mathfrak{i}_{\mathfrak{n}}\) and weight 0. For the untwisted case with \(n_G=1\), the index can be obtained from the 6d anomaly polynomial, through the anomaly inflow from 6d to 2d \cite{Kim:2016foj}. Then, the index of the untwisted M-strings are given as follows, 
\begin{align}
\mathfrak{i}_{\mathfrak{n}}(z)={\epsilon_1 \epsilon_2\over 2}\mathfrak{n}^T  \Omega \mathfrak{n}+(m^2-\epsilon_+^2)\sum_{a=1}^r n_a, \quad(n_G=1),
\label{eq237_untwisted}
\end{align}
where \(\Omega\) is the Cartan matrix of the simply-laced Lie algebra \(\mathfrak{g}\).

\par 

For the elliptic genus of twisted M-strings, its index can be read off from the index of the untwisted M-strings. In the last subsection, we explained how the roots transform under the twisted compactification. From table \ref{tab166_affine}, one can obtain the following mapping between the twisted and the untwisted string,
\begin{align}
&B_r: \ \hat{\mathfrak{n}}=(n_1,n_2,...,n_r) & &\leftarrow & &A_{2r-1}: \ \mathfrak{n}=(n_1,n_2,...,n_r,n_{r-1},...,n_1)
\nonumber \\
&(C_r)_{0}: \ \hat{\mathfrak{n}}=(n_1,n_2,...,n_r) & &\leftarrow & &D_{r+1}: \ \mathfrak{n}=(n_1,n_2,...,n_{r-1},n_r,n_r)
\nonumber \\
&G_2: \ \hat{\mathfrak{n}}=(n_1,n_2) & &\leftarrow & &D_{4}: \ \mathfrak{n}=(n_2,n_1,n_2,n_2)
\nonumber \\
&F_4: \ \hat{\mathfrak{n}}=(n_1,n_2,n_3,n_4) & &\leftarrow & &E_6: \ \mathfrak{n}=(n_1,n_2,n_3,n_2,n_1,n_4)
\nonumber \\
&(C_r)_{\pi}: \ \hat{\mathfrak{n}}=(n_1,n_2,...,n_r) & &\leftarrow & &A_{2r}: \ \mathfrak{n}=(n_1,...,n_r,n_r,...,n_1).
\label{eq385_folding}
\end{align}
See table \ref{tab166_affine} for our convention of the node enumeration. We claim that the index of the twisted M-strings of type \({G}^{\vee}\) and charge \(\hat{\mathfrak{n}}\) can be obtained from the index of the untwisted M-strings of type \(G\) and charge \(\mathfrak{n}\) as follows,
\begin{align}
\mathfrak{i}_{\hat{\mathfrak{n}}}(z) \text{ of }{G}^{\vee}
=\frac{1}{n_G} \Bigr(\mathfrak{i}_{\mathfrak{n}}(z) \text{ of }G\Bigr)\,.
\label{eq264_twisted}
\end{align}
\par 

Using (\ref{eq237_untwisted}) and (\ref{eq264_twisted}), one can universally write down the index of the M-strings in any Lie algebra \(\mathfrak{g}\) in the following expression,
\begin{align}\label{eq: twistedMstringindex}
\mathfrak{i}_{\mathfrak{n}}
&={\epsilon_1 \epsilon_2\over 2}\mathfrak{n} (\Omega  D)\mathfrak{n}^T
+({m^2-\epsilon_+^2})\sum_{a=1}^r D_{aa} \mathfrak{n}_a
,&
\mathfrak{n}&=
\begin{cases}
(n_1,n_2,...,{n_r\over 2})\text{ if }G= (C_r)_\pi
\\
(n_1,n_2,...,{n_r})\text{ otherwise}
\end{cases}.
\end{align}
Here, except for \( (C_r)_\pi \) theories, \(\Omega\) is the Cartan matrix of \(\mathfrak{g}\) and the matrix \(D\) is defined as follows,
\begin{align}
D_{ij}=2{\delta_{ij}\over (\alpha_i,\alpha_i)}.
\end{align}
The above index will play an important role only in the section \ref{sec:Modular_bootstrap}, and readers can see table \ref{tab832_symcartan} for the explicit values of \(\Omega D\). Lastly, the vector \(\mathfrak{n}\) is a usual string charge vector except for \( (C_r)_\pi \) case where the string charge at the last node is halved.

\subsection{Geometric engineering}\label{sec:2.3}
In this section, we want to understand the physics from the point of view of geometry. To start with, the 6d $(2,0)$ SCFTs arise from F-theory on a elliptic-fibered CY threefold, where the base is the ALE space, i.e, the resolution of $\mathbb{C}^2/\Gamma$ with $\Gamma$ a finite subgroup of $SU(2)$. Since the base itself is a CY twofold, the resulting 6d theory has $(2,0)$ supersymmetry.\footnote{However, in order to turn on the M-string mass, one needs to modify this picture, rendering the fibration non-trivial.} The resolution of the singularity in the base gives rise to a class of compact divisors $S_i$ and a class of compact curves $\Sigma_j$, where the negative of the intersection matrix
\be
\Omega_{ij}=-(S_i\cdot \Sigma_j),
\ee
is exactly the Cartan matrix of $ADE$ type.

In the geometric picture, BPS strings arise naturally from D3-branes wrapping on two-cycles in the base, which also inherits the self-duality condition of the D3-branes in ten dimensions. The partition function under the 6d $\Omega$-background localizes to the BPS strings wrapped on the $T^2$ in the spacetime, giving us an expansion in terms of elliptic genera \eqref{eq202_egexp}. Through the F-theory/M-theory duality, the elliptic genus captures the essential part of the refined topological string partition function, i.e., $Z_{\text{GV}}$ in \eqref{eq:refinedGV}, for the elliptically-fibered CY threefold.\footnote{For general $\mathcal{N} = (1,0)$ SCFTs they are the same only up to overall factors \cite{Gu:2017ccq,DelZotto:2017mee}.}

This brings us to another crucial ingredient of the story, i.e., topological strings. Given a CY threefold $X$, if one denotes the K\"{a}hler parameter of two cycles in $ H_2(X, \mathbb{Z})$ as $\mathbf{t}$ with its exponential $\mathbf{Q} = \exp(- \mathbf{t})$, we can associate the free energy as a formal power series in $g_s$,
\be
\mathcal{F}_{\text{top}}(g_s) = \sum\limits_{g \geq 0}\, F_{g}\, g_s^{2g-2}\,,
\ee
where the genus $g$ free energy $\mathcal{F}_g$ can be expanded in terms of $\mathbf{Q}$ with coefficients the celebrated Gromov-Witten invariants. Furthermore, through lifting type IIA string theory to M-theory \cite{Gopakumar:1998ii, Gopakumar:1998jq,Iqbal:2007ii,Aganagic:2011mi}, for $X$ non-compact, we are able to rewrite and refine topological string free energy in the following way, 
\begin{equation}\label{eq:refinedGV}
    \begin{aligned}
    \mathcal{F}_{\text{ref}} &=F_{\text{poly}} +  \log Z_{\text{GV}}(\epsilon_1,\epsilon_2,\mathbf{t}) \\
    & =F_{\text{poly}}+ \sum\limits_{2j_{\pm} \in \mathbb{N}} \sum\limits_{d \geq 1} \sum\limits_{\alpha \in H_2(X, \mathbb{Z}) } N^{\alpha}_{j_-j_+} {(-1)^{2 (j_-+j_+)} \, \chi_{j_-}(u^d) \chi_{j_+}(v^d) \over v^d + v^{-d} - u^d - u^{-d}} \frac{\mathbf{Q}^{d \alpha}}{d}\,,
    \end{aligned}
\end{equation}
where $\epsilon_1$ and $\epsilon_2$ are two equivariant parameters related to the Cartan subalgebra of $SO(4)$, originated from instanton counting under the $\Omega$-background in five dimensional gauge theory with $\mathcal{N} = 1$ supersymmetry \cite{Nekrasov:2002qd,Nekrasov:2003rj}. $F_{\text{poly}}$ takes the structure, 
\be\label{eq:F_poly}
F_{\text{poly}}=\frac{a_{ijk}t_it_jt_k}{6\ep_1\ep_2}+b_{i}^{(0,1)}t_i+\frac{(\ep_1+\ep_2)^2}{\ep_1\ep_2}b_i^{(1,0)}t_i\,.
\ee
Moreover, for the second part in \eqref{eq:refinedGV}, the $\chi_j$ is the character of an irreducible $SU(2)$ highest-weight representation with spin $j$,
\begin{equation}\label{eq:spinj}
    \chi_{j}(x) = x^{-2j} + x^{-2j+2}+ \cdots + x^{2j}\,,
\end{equation}
and $v=\exp(-\frac{\epsilon_1 + \epsilon_2}{2}), u = \exp(-\frac{\epsilon_1 - \epsilon_2}{2})$. Last but not least, $N^{\alpha}_{j_- j_+}$ is the number of BPS states in 5d with spin $j_-$ and $j_+$, arising from M2-branes wrapped on curves in $X$. They are integers known as the refined Gopakumar-Vafa (GV) or BPS invariants.

For the type of CY threefold considered in this paper, we can divide the K\"{a}hler parameters $\mathbf{t}$ into three different categories:
\be
\mathbf{t} = (-2\pi i\tau; \mathbf{v} = \{v_1, v_2,\cdots v_r\} ; m  )\,,
\ee
where they correspond to the volume of the elliptic fiber (instanton counting parameter in 5d), the compact curves in the base (Coulomb parameters associated to the simple roots of the gauge groups in 5d) and the M-string mass (mass of the adjoint matter in 5d) respectively. This also provides a unifying view on two types of expansion of $Z(\tau,\epsilon_{1,2},m,\mathbf{v})$ in \eqref{eq163_BPSR}: If one sums up $\tau$ and $m$ while expands in terms of $\mathbf{v}$, this gives \eqref{eq202_egexp} which is an elliptic genus expansion and naturally fits the 6d perspective; If one sums up $\mathbf{v}$ and $m$ while expands in terms of $\tau$, this gives instead \eqref{eq:instanton_expansion} which is an instanton expansion and arises from the 5d perspective. This is plausible since they are simply the UV/IR description of the same theory, and are supposed to share the same protected quantities.

To put it in another way, the upshot of the above discussion is that the elliptic genera or the instanton partition function enjoys a more constraining expansion \eqref{eq:refinedGV}, henceforth referred as the BPS or GV expansion.

Finally, we sketch a heuristic picture to understand the twisted compactification. We also consider the two compactification chains in figure \ref{eq127_chain}, and let us first concentrate on the left one. 6d $\mathcal{N} = (2,0)$ SCFTs compactified on $S^1$ is the same as M-theory compactification on the same Calabi-Yau due to F-theory/M-theory duality. Furthermore, When compactifying on a twisting circle, M-theory descends to type IIA theory, but the geometry gets modified according to the action $\sigma$ depicted in section \ref{sec:2.1}. The first step is to identify the curve $\Sigma_{i}$ and $\Sigma_{\sigma(i)}$ in the same orbit. Denote $\alpha,\beta$ the orbit of the action $\sigma$, we select one node $i$ in the orbit $\alpha$ and the intersection matrix naturally becomes \cite{Bhardwaj:2019fzv}
\be
\Omega_{\alpha \beta} =-\sum_{j\in\beta}(S_i\cdot (\Sigma_{j}))\,.
\ee
Remember that the elliptically fibered CY threefolds engineering 5d or 6d theories of $ADE$ type have only one section. When we identify the curves, the section over them corresponds to different points in the fiber in general, hence the one section should be merged into an $N$-section in the terminology of \cite{Cota:2019cjx}. As a result, the threefold is transformed to a genus one fibered CY manifold. This procedure is also explained, for example in \cite{Bhardwaj:2019fzv}.

Except for $A_{2l}$, the reduced intersection matrix becomes the Cartan matrix of the invariant subalgebra of the twisting. Interestingly, this gives a different result for $A_{2l}$ group, where the negative of the intersection matrix becomes,\footnote{Notice that the last diagonal element is not two, so it cannot possibly be identified as a Cartan matrix. In the terminology of \cite{Bhardwaj:2019fzv}, there is a self-edge attached to the last node in the graph.}
\be\label{eq:A_2l}
\Omega=\left(
\begin{array}{cccccc}
 2 & -1 & 0 & 0 & 0 & 0 \\
 -1 & 2 & -1 & 0 & 0 & 0 \\
 0 & \ddots & \ddots & \ddots & 0 & 0 \\
 0 & 0 & -1 & 2 & -1 & 0 \\
 0 & 0 & 0 & -1 & 2 & -1 \\
 0 & 0 & 0 & 0 & -1 & 1 \\
\end{array}
\right).
\ee

Remember that this does not give the 5d theory after twisted compactification. Instead, as discussed in the section \ref{sec:2.1}, they are S-dual to the real 5d theory with adjoint matters, which lives on the right compactification chain. Recall that S-duality includes exchanging the W-bosons and magnetic monopoles. Therefore, in the geometric language, it is equivalent to exchanging the role of divisors and curves, since D2-branes wrapping on compact curves give W-bosons while D4-branes on compact divisors give magnetic monopoles. With this in mind, after taking the S-dual the new intersection matrix becomes its transpose, giving the same matrix $\Omega$ in \eqref{eq: twistedMstringindex}. Here we are embedding the intersection matrix into a larger intersection matrix of a putative genus one fibered CY threefold, such that the column corresponds to rational curves while the row represents toric divisors. For all cases except $(C_r)_\pi$, this is nothing but a rewriting of the procedure in the section \ref{sec:2.1}. Furthermore, for the $(C_r)_\pi$ theory it gives the precise matrix $\Omega$ that is identified quite indirectly in the sections \ref{sec:2.1} and \ref{subsec:EG}. 
\section{Blowup equations}\label{sec:Blowup}
The twisted circle compactification of a 6d $(2,0)$ SCFT gives rise to a 5d $\mathcal{N}=2$ theory, as explained in detail in \ref{sec:2.1}. In section \ref{sec:3.1}, we first review the general setups for the $5d$ blowup equations. In section \ref{sec:3.2} we write down the blowup equations for 5d $\mathcal{N}=2$ theories of all $ABCDEFG$ types, and discuss how to solve them in section \ref{sec:3.3}. Then in section \ref{sec:3.4}, we re-write them in the elliptic version, which also helps to fix/verify the index of elliptic genera for twisted theories.  

\subsection{Review of the blowup equations}\label{sec:3.1}
The idea of blowup equation for instanton counting is to consider the equivariant localization on $\hat{\mathbb{C}}^2$, which is constructed from $\mathbb{C}^2$ by blowing up the origin to create a compact 2-cycle $\mathbb{P}^1$. One can regard the geometry as a total space of the tautological line bundle $\mathcal{O}(-1)$ over the $\mathbb{P}^1$, which can be parametrized by the homogeneous coordinates $(z_0,z_1,z_2)$. We are interested in the $U(1)^2$ equivariant partition function, with the equivariant action on the homogeneous coordinates as
\be
(z_0,z_1,z_2)\mapsto (z_0,e^{\epsilon_1}z_1,e^{\epsilon_2}z_2).
\ee
In the blowup geometry, the instantons are located at two fix points, the north and south poles of $\mathbb{P}^1$, whose coordinates are $(z_0,z_1,z_2)=(0,1,0)$ and $(0,0,1)$. Around these fix points, they can be locally described by the $\mathbb{C}^2$ coordinates $(z_0 z_1,z_2/z_1)$ and $(z_0z_2,z_1/z_2)$ respectively, leading to the equivalent action 
\begin{equation}\begin{split}
(z_0z_1,z_2/z_1)&\mapsto (e^{\epsilon_1}z_0z_1,e^{\epsilon_2-\epsilon_1}z_2/z_1), \,\,\,\,\,\,\quad \text{near the north pole},\\
(z_0z_2,z_1/z_2)&\mapsto (e^{\epsilon_2}z_0z_2,e^{\epsilon_1-\epsilon_2}z_1/z_2), \,\,\,\,\,\,\quad  \text{near the south pole}.
\end{split}\end{equation}
The full partition function $\hat{\mathcal{Z}}_{\hat{\mathbb{C}}^2}$ on $\hat{\mathbb{C}}^2$ can be written as the product of the partition function $\hat{\mathcal{Z}}^{\text{N}/\text{S}}$ around these two fix points,
\be
\hat{\mathcal{Z}}_{\hat{\mathbb{C}}^2}(\vec{v},\epsilon_1,\epsilon_2,q,\vec{m})=\sum_{\vec{n}\in Q^{\vee}}(-1)^{|\vec{n}|}\hat{\mathcal{Z}}^{\text{N}}(\vec{n})\hat{\mathcal{Z}}^{\text{S}}(\vec{n}),
\ee
by summing up all the fluxes $\vec{n}$ on the two-cycle $\mathbb{P}^1$. Here $Q^{\vee}$ is the co-root lattice, $\vec{v}$ are the Coulomb parameters, $\vec{m}$ are masses of matters and $q$ is the instanton counting parameter. Around the North/South poles, they can be locally treated as $\mathbb{C}^2$. In terms of $\mathbb{C}^2$ partition function $\hat{\mathcal{Z}}_{{\mathbb{C}}^2}$, we have
\be\label{eq:ZNS}\begin{split}
\hat{\mathcal{Z}}^{\text{N}}(\vec{n})&=\hat{\mathcal{Z}}_{\mathbb{C}^2}\left(\vec{v}+\vec{n}\epsilon_1+\vec{\lambda}_G\epsilon_1,\epsilon_1,\epsilon_2-\epsilon_1,q e^{r_b\epsilon_1},\vec{m}+\vec{\lambda}_F\epsilon_1\right),\\
\hat{\mathcal{Z}}^{\text{S}}(\vec{n})&=\hat{\mathcal{Z}}_{\mathbb{C}^2}\left(\vec{v}+\vec{n}\epsilon_2+\vec{\lambda}_G\epsilon_2,\epsilon_2-\epsilon_1,\epsilon_2,q e^{r_b\epsilon_2},\vec{m}+\vec{\lambda}_F\epsilon_2\right).\\
\end{split}\ee

Note that the partition function should be invariant under Weyl transformation of its Lie algebra $\mathfrak{g}$, and $\vec{n}+\vec{\lambda}_G$ is actually the co-weight lattice $P^{\vee}$. The number of nonequivalent choices $\vec{\lambda}_G$ depends on the number of elements in the quotient $|P^{\vee}/ Q^{\vee}|=\det \Omega$, where $\Omega$ is the Cartan matrix of $\mathfrak{g}$. For special cases, like $(C_r)_{\pi}$, $\Omega$ should be interpreted as the intersection matrix of compact divisors and curves in the CY geometry.

Another remark here is that the partition function $\hat{\mathcal{Z}}_{{\mathbb{C}}^2}$ is not the usual partition function $\mathcal{Z}_{{\mathbb{C}}^2}$ used in gauge theories. They are the same only up to a shift. As reviewed in section \ref{sec:2.3}, a five dimensional gauge theory can be engineered from the M-theory compactification on a CY manifold, where the total partition function $\mathcal{Z}=\exp \mathcal{F}$ can be written as the product of the GV or BPS expansion $Z_{\text{GV}}$ \eqref{eq:refinedGV} plus singular terms $e^{F_{\text{poly}}}$ \eqref{eq:F_poly} at large volume limit. Then we define  $\hat{\mathcal{Z}}=e^{F_{\text{poly}}}\hat{Z}_{\text{GV}}$, with\footnote{Note that $r_b$ could be a half-integer, so that $(-1)^{2r_b}$ is not necessarily $1$.}
\be\begin{split}
\hat{Z}_{\text{GV}}(\vec{v},\epsilon_1,\epsilon_2,q,\vec{m})&={Z}_{\text{GV}}(\vec{v},\epsilon_1+2\pi i,\epsilon_2,q,\vec{m})\\
&={Z}_{\text{GV}}(\vec{v}+2\pi i \vec{\lambda}_G,\epsilon_1,\epsilon_2,q(-1)^{2r_b} ,\vec{m}+2 \pi i \vec{\lambda}_F),
\end{split}\ee
where the second line comes from the \textit{checker-board} pattern for the spin $(j_L,j_R)$ BPS invariants at degree $(d_G,d_F,d_b)$ in the BPS expansion,\footnote{This is also known as the B-field condition in \cite{Hatsuda:2013oxa,Wang:2015wdy}.}
\be\label{fluxcondition}
(-1)^{2\vec{\lambda}_{G}\cdot d_G+2\vec{\lambda}_{F}\cdot d_F+2r_b\cdot d_b}=(-1)^{2j_L+2j_R+1}.
\ee
As a remark, the checkerboard pattern means that for a given curve class $\{d_G, d_F, d_b\}$, if one enumerates all non-zero BPS invariants with two axes $2j_+$ and $2j_-$, any two occupied blocks are either disconnected or connected through a diagonal. In other words, the combination $\left(2j_-+2j_+\right)$ is always even or odd. Clearly this is necessary for the equation \eqref{fluxcondition} to hold. This pattern was first noticed in \cite{Choi:2012jz} and holds true for all non-compact CY threefolds that the authors know of.

Next, in order to get the blowup equation for ordinary gauge theory partition function, one has to {\it{shift back}} the extra fluxes. This will give rise to an extra phase $(-1)^{|\phi|}$, coming from rearranging the singular part $e^{F_{\text{poly}}}$ in the blowup equation. The explicit expression is a bit complicated and for 6d SCFTs it can be found in \cite{Gu:2020fem}. Here we do not need the explicit form so we omit its expression. Now the blowup equation takes the form,
\be
{\mathcal{Z}}_{\hat{\mathbb{C}}^2}(\vec{v},\epsilon_1,\epsilon_2,q,\vec{m})=\sum_{\vec{n}\in Q^{\vee}}(-1)^{|\vec{n}|+|\phi|}{\mathcal{Z}}^{\text{N}}(\vec{n}){\mathcal{Z}}^{\text{S}}(\vec{n})\,.
\ee

If one consider the limit where the size of blow-up $\mathbb{P}^1$ is small, the resulting geometry is nothing but the original $\mathbb{C}^2$, so one concludes immediately that ${{\mathcal{Z}}}_{\hat{\mathbb{C}}^2}\sim {\mathcal{Z}}_{{\mathbb{C}}^2}$. In most cases, the two partition functions are exactly identical. However, in certain special cases, there could be an extra factor $\Lambda$ coming from the perturbative part. It was further explained in \cite{Huang:2017mis} that $\Lambda$ does not depend on Coulomb VEVs $\vec{v}$, so we have
\be
{\mathcal{Z}}_{\hat{\mathbb{C}}^2}(\vec{v},\epsilon_1,\epsilon_2,q,\vec{m})=\Lambda(\epsilon_1,\epsilon_2,q,\vec{m}){{\mathcal{Z}}_{\mathbb{C}^2}}(\vec{v},\epsilon_1,\epsilon_2,q,\vec{m}).
\ee
If such an equation holds, we could define 
\be\label{lambda0}
\Lambda(\epsilon_1,\epsilon_2,q,\vec{m})= \lim_{Q_i\rightarrow 0}\frac{{{\mathcal{Z}}}_{\hat{\mathbb{C}}^2}(\vec{v},\epsilon_1,\epsilon_2,q,\vec{m})}{{{\mathcal{Z}}}_{{\mathbb{C}}^2}(\vec{v},\epsilon_1,\epsilon_2,q,\vec{m})},
\ee
where $Q_i$ is the exponential of the Coulomb VEVs $Q_i=e^{-v_i}$. For a 5d/6d SCFT that can be engineered from geometry, the instanton partition function admits a BPS expansion, counting the BPS states of M2-brane wrapping curves inside the CY such that the degree of $Q_i$'s is always non-negative. This indicates that most of the instanton contributions in \eqref{lambda0} vanish, leaving only the neutral part $Z^{\text{inst}}_{\text{neutral}}=\left.Z^{\text{inst}}\right|_{Q_i\rightarrow 0}$, times the contribution from $e^{F_{\text{poly}}}$. 

It seems that one can use arbitrary $\vec{\lambda}_{G/F},r_b$ in \eqref{eq:ZNS} to define an equation. Nevertheless, notice that the perturbative part is divergent in this limit, so one has to choose $\vec{\lambda}_{G/F},r_b$ properly to cancel these divergences. The cancellation condition gives a strong constraint on $\vec{\lambda}_{G/F},r_b$, which turns out to give all the possible choices. Among them, we choose those that satisfy the aforementioned condition \eqref{fluxcondition} for all BPS states.
\subsection{Blowup equations for $G+1{\text{\bf{Adj}}}$}\label{sec:3.2}
Now let us focus on 5d theories with an adjoint matter $G+1{\text{\bf{Adj}}}$. Recall that the adjoint representation is the same as the root system of its Lie algebra $\mathfrak{g}$. We denote $m$ the mass of the adjoint matter, $v_\alpha$ the Coulomb VEVs $\alpha \cdot \vec{v}  $, $R^+=\{(v_\alpha,\pm m)|\alpha\in \Delta^+\}$ the positive weights where $\Delta^+$ is the set of positive roots. Following the gauge theory language, we separate the partition function into the classical, 1-loop and instanton part. We modify slightly the notations in \cite{Gu:2020fem} to adapt to our present convention, 
\be
Z^{\text{class}}=\exp\left[-\frac{2\pi i\tau}{2\epsilon_1\epsilon_2}\sum_{ij}v_i v_j h_{ij}\right],
\ee
\be\begin{split}
Z^{\text{1-loop}}=&\exp\left[\frac{m^2}{2\epsilon_1\epsilon_2}\sum_{{{\alpha}} \in \Delta^+}v_{\alpha}+\frac{(\epsilon_1+\epsilon_2)^2}{8\epsilon_1\epsilon_2}\sum_{{{\alpha}} \in \Delta^+}v_{\alpha}\right]\\
&\times \mathrm{PE}\left(\frac{\sum_{\alpha \in \Delta^+}\left(Q_{\alpha}(Q_m+Q_m^{-1})-\left(v + v^{-1}\right)Q_{\alpha}\right)}{\left(q_1^{1/2}-q_1^{1/2}\right)\left(q_2^{1/2}-q_2^{1/2}\right)} \right),
\end{split}\ee
\begin{equation}
  Z^{\text{inst}} = \text{PE}{\Bigg[
  \sum_{j_-,j_+=0}^\infty\sum_{\beta}(-1)^{2(j_-+j_+)}
  {N^{\beta}_{j_-,j_+}} f_{(j_-,j_+)}(q_1,q_2)Q_{\beta}
  \Bigg]},
\end{equation}
where PE stands for the plethystic exponential and $h_{ij}$ is the Killing form of $\mathfrak{g}$, 
\be\label{eqdeffjlfr}
  f_{(j_-,j_+)}(q_1,q_2) = \frac{\chi_{j_+}(v)\chi_{j_-}(u)}
  {(q_1^{1/2}-q_1^{-1/2})(q_2^{1/2}-q_2^{-1/2})}\,.
\ee
One can easily recognize that $Z^{\text{inst}}$ is nothing but the $Z_{\text{GV}}$ defined in \eqref{eq:refinedGV}.

As discussed in the last section, the number of choices of $\vec{\lambda}_G$ is determined by the determinant of the Cartan matrix or the negative of the intersection matrix. Therefore we can immediately get the number of flux $\vec{\lambda}_G$, summarized in table \ref{tab:numrG}.
\begin{table}[h!]
\begin{center}
 \def\arraystretch{1.3}
	\begin{tabular}{|c|c|c|c|c|c|c|c|c|c|c|c|c|c|c|c|c|c|c|c|c|c|c|c|c|c|c|}
		\hline
	        $G$
                &$A_r$&$B_r$&$(C_r)_{0}$&$(C_r)_{\pi}$&$D_r$&$E_6$&$E_7$&$E_8$&$F_4$&$G_2$\\ \hline
		\#$\vec{\lambda}_G$ &$r+1$ & 2 & 2&1& 4  &  3  & 2  &  1         & 1   &1   \\
		\hline
	 \end{tabular}
\end{center}
\caption{The number of possible fluxes $\vec{\lambda}_G$ for simple Lie algebras.}
\label{tab:numrG}
\end{table}

Among all the cases except for $(C_r)_{\pi}$, there is a common one: $\vec{\lambda}_G=0,\lambda_F=\pm\frac{1}{2},r_b=0$. For simplicity, we only consider $\lambda_F=-\frac{1}{2}$. Together with the neutral part in \eqref{eq:abl1}, a short computation shows that the expression for $\Lambda$ is 
\be\label{lambda5d1}
\Lambda(m,\epsilon_1,\epsilon_2,\tau)=\mathfrak{f}(\tau) \sum_{{\vec{n}\in \mathbb{Z}^r}}{e}^{i\pi \tau {\hij}n_in_j  +(m-\epsilon_+) |\vec{n}|},
\ee
where ${\hij}=(\Omega D^{-1})_{ij}$ and $D$ is the matrix $D_{ij} = 2\frac{\delta_{i,j}}{\langle\alpha_i,\alpha_i\rangle}$. One can easily verify from the expression that $\phi=2\pi i \lambda_F |\vec{n}|$ cancels with the phase $(-1)^{|\vec{n}|}$. The prefactor $\mathfrak{f}(\tau)$ comes from the neutral part of instanton partition function described in section \ref{subsec:EG}, for classical Lie groups,
\be\label{ffunction}
\mathfrak{f}(\tau)=\begin{cases}
\prod_{k=1}^{\infty}(1-q^k)^{-r},& \text{for } A_r, D_r \ \text{and}\ E_r,\\
\prod_{k=1}^{\infty}(1-q^{k})^{-r+1}(1-q^{2k})^{-1},              & \text{for } B_r,\\
\prod_{k=1}^{\infty}(1-q^{2k})^{-r+1}(1-q^{k})^{-1},              & \text{for } (C_r)_{0},\\
\prod_{k=1}^{\infty}(1-q^{2k})^{-r},              & \text{for } (C_r)_{\pi},\\
\prod_{k=1}^{\infty}(1-q^{k})^{-1}(1-q^{3k})^{-1},              & \text{for } G_2,\\
\prod_{k=1}^{\infty}(1-q^{k})^{-2}(1-q^{2k})^{-2},              & \text{for } F_4.
\end{cases}
\ee
Note that $\mathfrak{f}(\tau)$ is not important in our current cases, so one can set it to be one if one does not care about the neutral part.

Finally, for the $(C_r)_{\pi}$ theory, the negative of the intersection matrix is not the Cartan matrix of $C_r$ and $\vec{\lambda}_G=0$ breaks the flux quantization condition \eqref{fluxcondition}. It turns out that there is only one blowup equation in this case. The correct choice is to simply replace the co-root lattice in \eqref{lambda5d1} with another co-weight lattice by the shift,
\be\label{rep_pi}
n_i\rightarrow n_i+\frac{i}{2},\quad i=1,\cdots,r,
\ee
which gives the blowup equation for both $(C_r)_{0,\pi}$. As a summary, the blowup equation for 5d $G+1{\text{\bf{Adj}}}$ theory is
\be\label{blowup5d3}\begin{split}
\Lambda&(m,\epsilon_1,\epsilon_2,\tau)Z_{}^{\mathrm{inst}}\left(\ep_1,\ep_2,v_i,\tau,m \right)=\sum_{\vec{n}\in \mathbb{Z}^r}{e}^{i\pi \tau {\hij}n_in_j +( m-\epsilon_+) |\vec{n}|}\mathcal{A}^0(\ep_1,\ep_2,v_i,m,n_i)\\
&\times Z_{}^{\mathrm{inst}}\left(\ep_1-\ep_2,\ep_2,v_i+\Omega_{ij}n^j \ep_2,\tau,m- \frac{\ep_2}{2}\right)Z_{}^{\mathrm{inst}}\left(\ep_1,\ep_2-\ep_1,v_i+\Omega_{ij}n^j \ep_1,\tau,m- \frac{\ep_1}{2}\right).\\
\end{split}\ee
For a given root $\alpha$, define $Q_\alpha =e^{-v_{\alpha}}=\prod_i Q_i^{\alpha_i}$ and $ R_{\alpha}=-\alpha_i \Omega_{ij}n_j$,
\begin{small}
\be
\mathcal{A}^0=
\begin{cases}
{\displaystyle \prod_{\alpha \in \Delta^+}\prod_{\displaystyle \substack{m,n\geq 0\\m+n\leq |R_{\alpha}|-1}}}\frac{\left(1-Q_{\alpha} Q_m^{-1}q_1^{m+1/2}q_2^{n+1/2}\right)^{}}{\left(1-Q_{\alpha} q_1^mq_2^n\right)^{}}{\displaystyle\prod_{\displaystyle \substack{m,n\geq 0\\m+n\leq |R_{\alpha}|-2}}}\frac{\left(1-Q_{\alpha} Q_m^{}q_1^{m+1/2}q_2^{n+1/2}\right)^{}}{\left(1-Q_{\alpha} q_1^{m+1}q_2^{n+1}\right)^{}},& R_{\alpha}\geq 0\\
{\displaystyle    \prod_{\alpha \in \Delta^+}\prod_{\displaystyle \substack{m,n\geq 0\\m+n\leq |R_{\alpha}|-1}}}\frac{\left(1-Q_{\alpha} Q_m^{}q_1^{-m-1/2}q_2^{-n-1/2}\right)^{}}{\left(1-Q_{\alpha} q_1^{-m}q_2^{-n}\right)^{}}{\displaystyle\prod_{\displaystyle \substack{m,n\geq 0\\m+n\leq |R_{\alpha}|-2}}}\frac{\left(1-Q_{\alpha} Q_m^{-1}q_1^{-m-1/2}q_2^{-n-1/2}\right)^{}}{\left(1-Q_{\alpha} q_1^{-m-1}q_2^{-n-1}\right)^{}},              & R_{\alpha}< 0.
\end{cases}
\ee
\end{small}
For the $(C_r)_{\pi}$ theory, we should make the replacement as in \eqref{rep_pi}. We verify \eqref{blowup5d3} for $A_1+1\text{\bf{Adj}}$ up to 4-instantons with instanton partition function from ADHM description, and for $(C_r)_{0,\pi}$, $r\leq 4$ at the one-instanton level.

\subsection{Solving blowup equations}\label{sec:3.3}
Here we give a description of our recipe to solve the blowup equation.
\subsubsection*{Solving BPS invariants}
To solve the BPS invariants, one can expand the blowup equation with instanton counting parameter $q$ and the Coulomb parameter $\mathbf{v}$. With the ansatz on the BPS expansion with unknown BPS invariants, we can bootstrap all the BPS invariants, except for the spin $(0,0)$ and $(0,1/2)$ invariants. See \cite{Huang:2017mis,Gu:2019pqj} for the proof and \cite{Kim:2020hhh} for a recent discussion on this approach.

Following the definition in \eqref{eqdeffjlfr}, define 
\begin{equation}\label{Bldef}
  Bl_{(j_-,j_+,R)}(q_1,q_2) = f_{(j_-,j_+)}(q_1,q_2/q_1)q_1^R +
  f_{(j_-,j_+)}(q_1/q_2,q_2)q_2^R - f_{(j_-,j_+)}(q_1,q_2),
\end{equation}
which is a Laurent polynomial of $q_1,q_2$. It is proved in \cite{Huang:2017mis,Gu:2019pqj} that for a fixed spin $R$, $Bl_{(j_-,j_+,R)}$ are mutually independent for all possible $(j_-,j_+)$, except for $Bl_{(0,0,\pm 1/2)}(q_1,q_2)=Bl_{(0,1/2,0)}(q_1,q_2)=0$. Then finding the BPS invariants becomes a problem of decomposing a Laurent polynomial into the basis $Bl_{(j_-,j_+,R)}(q_1,q_2)$. In our case, we set $\vec{\lambda}_G=0,\vec{\lambda}_F=-\frac{1}{2},r_b=0$, so that the only unfixed BPS invariants are $N_{(0,0)}^{d_m=\pm 1}$ and $ N_{(0,1/2)}^{d_m=0}$. In principle, one can go to higher degrees to fix these invariants, but here we use another condition to fix them more conveniently.

It is known that if we set the mass of the adjoint matter to $\ep_-$, the partition function is trivially one. In this limit,
\be
f_{(0,0)}Q_m^{\pm 1}\rightarrow \frac{u^{\pm 1}}{{(q_1^{1/2}-q_1^{-1/2})(q_2^{1/2}-q_2^{-1/2})}}, \ f_{(0,1/2)} Q_m^0 \rightarrow \frac{v + v^{-1}}{{(q_1^{1/2}-q_1^{-1/2})(q_2^{1/2}-q_2^{-1/2})}}\,,
\ee
which are linearly independent. This means that we can use the triviality condition to fix the remaining BPS invariants. In practice, we extract the BPS invariants up to 10-instantons, at lower degree expansion of the Coulomb parameters, which are compatible with the results using the modular bootstrap in section \ref{sec:Modular_bootstrap}. Part of the BPS invariants we solved can be found in appendix \ref{App:EG}.

\subsubsection*{Solving instanton partition functions}
As reviewed in section \ref{subsec:EG}, from the 5d point of view, the partition function can be expanded in terms of instanton sectors. To begin with, we expand the blowup equation at the $k$-instanton order and the resulting equation is
\be\label{blowup5d2}\begin{split}
&\sum_{ \substack{\mathbf{n}\in \mathbb{Z}^r\\ k_1+k_2+\frac{1}{2}{\hij}n_in_j =k}} \left(\sign\right)^{| \vec{n} |}\mathcal{A}^0(\ep_1,\ep_2,v_i,m,n_i)\\
&\times Z_{k_1}^{\mathrm{inst}}\left(\ep_1-\ep_2,\ep_2,v_i+\Omega_{ij}n_j \ep_2,m- \frac{\ep_2}{2}\right)Z_{k_2}^{\mathrm{inst}}\left(\ep_1,\ep_2-\ep_1,v_i+\Omega_{ij}n_j \ep_1,m- \frac{\ep_1}{2}\right)\\
=&\sum_{\substack{\mathbf{n}\in \mathbb{Z}^r\\ l+k'+\frac{1}{2}\Omega_{ij}n_in_j =k}}\left(\sign\right)^{| \vec{n} |}
\mathfrak{f}_l Z_{k'}^{\mathrm{inst}}\left(\ep_1,\ep_2,v_i,m \right)\,.
\end{split}\ee

Here $\mathfrak{f}_l$ is the $l$-th Fourier coefficient of $\mathfrak{f}(\tau)$, which can be defined from the coefficients of the $-r$ power of Dedekind eta function $p_{-r}(l)$
\be
\eta(\tau)^{-r}:=q^{-\frac{r}{24}}\prod_{n=1}^{\infty}(1-q^n)^{-r}=q^{-\frac{r}{24}}\sum_{l=0}^{\infty}p_{-r}(l)q^l\,.
\ee
When $r>0$, $p_{-r}(l)$ are always positive integers for $l \geq 0$. For the first few terms, we have 
\be
p_{-r}(0)=1,\,p_{-r}(1)=r,\,p_{-r}(2)=\frac{1}{2} r(r+3),\,p_{-r}(3)=\frac{1}{6} r (r^2+9r+8),\cdots.
\ee

(\ref{blowup5d2}) can be alternatively written as
\be\label{blowup5d}
I_{k}=Z_{k}^{\mathrm{inst}}\left(\ep_1,\ep_2,v_i,m \right)-Z_{k}^{\mathrm{inst}}\left(\ep_1-\ep_2,\ep_2,v_i,m- \frac{\ep_2}{2}\right)-Z_{k}^{\mathrm{inst}}\left(\ep_1,\ep_2-\ep_1,v_i,m- \frac{\ep_1}{2}\right),
\ee
where $I_{k}$ is the contribution from lower instanton numbers
\be\begin{split}
I_{k}=&-\sum_{\substack{ l+k'+\frac{1}{2}{\hij}n_in_j =k\\ \mathbf{n}'\in \mathbb{Z}^r,k'<k}}\left(\sign\right)^{| \vec{n} |}
p_{-r}(l)Z_{k'}^{\mathrm{inst}}\left(\ep_1,\ep_2,v_i,m \right)\\
&+\sum_{ \substack{ k_1+k_2+\frac{1}{2}{\hij}n_in_j =k\\ \mathbf{n}\in \mathbb{Z}^r,k_1,k_2 <k}} \left(\sign\right)^{| \vec{n} |}\mathcal{A}^0(\ep_1,\ep_2,v_i,m,n_i)\\
&\times Z_{k_1}^{\mathrm{inst}}\left(\ep_1-\ep_2,\ep_2,v_i+\Omega_{ij}n_j \ep_2,m- \frac{\ep_2}{2}\right)Z_{k_2}^{\mathrm{inst}}\left(\ep_1,\ep_2-\ep_1,v_i+\Omega_{ij}n_j \ep_1,m- \frac{\ep_1}{2}\right).\\
\end{split}\ee
In order to solve the equation, we can first make the ansatz
\be\label{Zkansatz}
Z_k^{\text{inst}}=\frac{\mathcal{N}_k(Q_i,Q_m,q_1,q_2)}{\mathcal{D}_k(Q_i,q_1,q_2)}.
\ee
where $\mathcal{N}_k(Q_i,Q_m,q_1,q_2)$ is a Laurent polynomial of $Q_i,Q_m,q_1,q_2$ with unknown coefficients, and $\mathcal{D}_k(Q_i,q_1,q_2)$ can be determined from the pole structure of $I_k$. Then substituting the ansatz \eqref{Zkansatz} into the $k$-instanton blowup equation \eqref{blowup5d}, we can solve the coefficients in the $\mathcal{N}_k(Q_i,Q_m,q_1,q_2)$ with extra information. At one-instanton level, the problem becomes much simpler and easier to solve. 

When $k=1$, (\ref{blowup5d}) becomes
\be\label{blowup1}\begin{split}
&-\mathfrak{f}_1+\sum_{ \substack{\mathbf{n}\in \mathbb{Z}^r\\ \frac{1}{2}{\hij}n_in_j =1}} \left(\sign\right)^{| \vec{n} |}\left(\mathcal{A}^0(\ep_1,\ep_2,v_i,m,n_i)-1\right)\\
=&Z_{1}^{\mathrm{inst}}\left(\ep_1,\ep_2,v_i,m \right)-Z_{1}^{\mathrm{inst}}\left(\ep_1-\ep_2,\ep_2,v_i,m- \frac{\ep_2}{2}\right)-Z_{1}^{\mathrm{inst}}\left(\ep_1,\ep_2-\ep_1,v_i,m- \frac{\ep_1}{2}\right)\, ,
\end{split}\ee
where $\mathfrak{f}_1$ coincides with the number of long simple roots according to \eqref{ffunction}.
Here the sum in the first line can be understood as a sum over length-2 (co-)root. In our convention they correspond precisely to long roots, hence there are only poles at the long roots in the first line of \eqref{blowup1}. This analysis is enough to conjecture the denominator of $Z_{1}^{\mathrm{inst}}\left(\ep_1,\ep_2,v_i,m \right)$ for a simple Lie algebra $\mathfrak{g}$ as
\be
\mathcal{D}_1(Q_i,q_1,q_2)=(1-q_1)(1-q_2)\prod_{\alpha \in \Delta_l}(1-Q_{\alpha}q_1 q_2),
\ee
where $\Delta_l$ is the set of long roots. Based on the zeros and properties of one-instanton partition function, one can further make an ansatz for the numerator
\be
\mathcal{N}_1(Q_i,Q_m,q_1,q_2)=(\sqrt{q_2}-\sqrt{q_1}Q_m)(\sqrt{q_2}-\sqrt{q_1} Q_m^{-1})\sum_{n,i,k,c_i} a_{n,i,k,c_i}(Q_m^n+Q_m^{-n})Q_i^{c_i}(q_1q_2)^{\frac{k}{2}}.
\ee
Define $Z_1^{\text{inst}}=\frac{\left(\sqrt{q_1} Q_m-\sqrt{q_2}\right) \left(\sqrt{q_1}-\sqrt{q_2}
   Q_m\right)}{\left(1-q_1\right) \left(1-q_2\right) Q_m} \tilde{Z}_1$ and $v=(q_1 q_2)^{-\frac{1}{2}}$, we solve $\tilde{Z}_1$ for gauge groups $A_1,A_2$ and $A_3$ model with gauge fugacities turned off by setting $Q_i=1$, 
\begin{align}
\tilde{Z}_{1,A_1}&=&\frac{v^2}{\left(1-v^2\right)^2 }&\left(-4\chi_{(0)}+\chi_{(1)}(v+v^{-1})\right),\nonumber \\
\tilde{Z}_{1,A_2}&=&\frac{v^4}{\left(1-v^2\right)^4 }&\left(\chi_{(0)}(- v^2+32 - v^{-2})+\chi_{(1)}(v^3-13 v-13 v^{-1}+v^{-3})\right. \nonumber \\
&&&\left.+\chi_{(2)}(v^2+4+v^{-2})\right),\nonumber \\
\tilde{Z}_{1,A_3}&=&\frac{v^6}{\left(1-v^2\right)^6 }&\left(\chi_{(0)}(- v^4-14v^2-250 -14v^{-2}- v^{-4}) \right. \nonumber \\
&&&\left.+\chi_{(1)}(v^5-7 v^3+146 v+146 v^{-1}-7 v^{-3}+v^{-5})\right.\nonumber \\
&&&\left.+\chi_{(2)}(v^4-22 v^2-78-22 v^{-2}+v^{-4})+\chi_{(3)}(v^3+9 v+9 v+v^{-3})\right),\nonumber
\end{align}
and for $(C_1)_\pi$,
\begin{align*}
&\begin{aligned}\tilde{Z}_{1,(C_1)_\pi}&=\frac{v^{2}}{\left(1-v^2\right)^{2}}
\Big(2\chi_{(0)}({v}+ v^{-1})-2\chi_{(1)}\Big). \end{aligned}
\end{align*}
For all cases, they have the structure
\be\label{eq:oneinstanton}
\tilde{Z}_{1,G}=\frac{v^{2h_G^{\vee}-2}}{(1-v^2)^{2h_G^{\vee}-2}}P(v,\chi_{(n)}),\,
\ee
with $P(v,\chi_{(n)})$ always a Laurent polynomial in $v$, which is symmetric under $v \leftrightarrow v^{-1}$.
Here $\chi_{(n)}:=\chi_{(n)}(m)$ is the character of the spin $\frac{n}{2}$ representation of the $SU(2)$ flavor group. Furthermore,
the structure \eqref{eq:oneinstanton} still holds for other gauge groups, and in particular for $ABCD$ type, we verified that the results agree with those computed from ADHM descriptions. For theories without an ADHM description, we list $\tilde{Z}_{1,G}$ in appendix \ref{app:Z1}.

Notice that for most cases, we turn off the gauge fugacities $Q_i=1$, but our method works also with all gauge fugacities turned on. For $G_2$, we first compute the exact result and then expand it in terms of $v$,

\be\begin{split}
\tilde{Z}_{1,G_2}=&\chi_{(3)}^{SU(2)}\left(\sum_{n=0}^{\infty}\chi_{(0,n)}v^{3+2n}\right)-\chi_{(2)}^{SU(2)}\left(\sum_{n=0}^{\infty}\left(\chi_{(0,n)}+\chi_{(0,n+1)}+\chi_{(3,n)}v^2\right)v^{4+2n}\right)\\
&+\chi_{(1)}^{SU(2)}\left(-\chi_{(1,0)}v^3+\sum_{n=0}^{\infty}\left(\chi_{(0,n+1)}+\chi_{(3,n)}(1+v^2)+\chi_{(4,n)}v^2\right)v^{5+2n}\right)\\
&+\chi_{(0)}^{SU(2)}\left(\chi_{(1,0)}(v^2+v^4)-\sum_{n=0}^{\infty}\left(\chi_{(4,n)}(1+v^2)+\chi_{(3,n)}\right)v^{6+2n}\right),\\
\end{split}\ee
where $\chi_{(n_1,n_2)}$ is the character of $G_2$ with highest weight $(n_1,n_2)$. As a remark, all the constant coefficients are $\pm 1$ in the expansion, which agrees with the conjecture in \cite{Gu:2020fem}. Similarly, for $F_4$ we have
\be\begin{split}
\tilde{Z}_{1,F_4}=&-\chi_{(8)}^{SU(2)}\left(\sum_{n=0}^{\infty}\chi_{(n,0,0,0)}v^{8+2n}\right)\\
&+\chi_{(7)}^{SU(2)}\left(\sum_{n=0}^{\infty}\left(\chi_{(n,1,0,0)}v^2+\chi_{(n+1,0,0,0)}+\chi_{(n,0,0,0)}\right)v^{9+2n}\right)\\
&+\cdots.\\
\end{split}\ee
For $E$ type gauge groups, the computation becomes too lengthy to reproduce here. 

\subsection*{Elliptic blowup equations}\label{sec:3.4}
In our 5d KK theory description, there is no cubic terms for the Coulomb branch parameter $a_i$ in the prepotential. According to a theorem in  \cite{MR1228584,MR1314743}, geometrically, they are described by the genus one fibered CY three-folds on twisted ALE singularities. This already indicates that this is actually a 6d theory and the Coulomb branch becomes the tensor branch in 6d. One can then re-write the blowup equation to its elliptic version  

\begin{equation}
\begin{split}
&\sum_{\mathfrak{n}'+\mathfrak{n}{''}=\mathfrak{n}}\sum_{\vec{n}\in \mathbb{Z}^r}{e}^{i\pi \tau {\hij}n_in_j +( m-\epsilon_+) |\vec{n}|+\mathfrak{n}_i'\Omega_{ij}n_j \ep_1+\mathfrak{n}_i{''}\Omega_{ij}n_j \ep_2}\\
&\times \mathbb{E}_{\mathfrak{n}'}\left(\ep_1-\ep_2,\ep_2,m- \frac{\ep_2}{2}\right)\mathbb{E}_{\mathfrak{n}{''}}\left(\ep_1,\ep_2-\ep_1,m- \frac{\ep_1}{2}\right)\\
=&\sum_{{\vec{n}\in \mathbb{Z}^r}}{e}^{i\pi \tau {\hij}n_in_j  +(m-\epsilon_+) |\vec{n}|} 
\mathbb{E}_{\mathfrak{n}}\left(\ep_1,\ep_2,m \right).
\label{eq677_EBW}
\end{split}\end{equation}
By comparing with the elliptic blowup equation in \cite{Gu:2019pqj}, one observes that for general simple groups the formula is the same if we do the replacement $\Omega_{ij}\rightarrow {\hij}=(\Omega D^{-1})_{ij}, \  k\rightarrow k\cdot D$. We conclude that the index $\mathfrak{i}_{\mathfrak{n}}$ for $Z_{\mathfrak{n}}$ is
\be
\mathfrak{i}_{\mathfrak{n}}={\epsilon_1 \epsilon_2\over 2}(\mathfrak{n}\cdot D) (\Omega  D^{-1})(\mathfrak{n}\cdot D)^T
+({m^2-\epsilon_+^2})\sum_{a=1}^r D_{aa} \mathfrak{n}_a
,
\ee
agree with the results in section \ref{subsec:EG}.
For $(C_r)_{\pi}$ theory, the blowup equation is the same as the second blowup equation of $(C_r)_{0}$, so they should have the same index. The only difference between them is that there are half-integer string charges, such that we need to normalize $\mathfrak{n}\rightarrow \frac{\mathfrak{n}}{2}$ in the definition. Again, we recover the index for $(C_r)_{\pi}$ theories as described in section \ref{subsec:EG}.

If we focus on a single node in the base, i.e., $\mathfrak{n}_I>0\ \text{and}\ \mathfrak{n}_{J}=0,J\neq I$, \eqref{eq677_EBW} becomes
\begin{equation}
\begin{split}
&\sum_{\mathfrak{n}_I'+\mathfrak{n}_I^{''}=\mathfrak{n}}\sum_{{n}\in \mathbb{Z}}{e}^{i\pi D_{II}\tau n^2 +( m-\epsilon_+) n-2\mathfrak{n}_I'n \ep_1-2\mathfrak{n}_I^{''}n \ep_2}\\
&\times \mathbb{E}_{\mathfrak{n}'_I}\left(\ep_1-\ep_2,\ep_2,m- \frac{\ep_2}{2},\tau\right)\mathbb{E}_{\mathfrak{n}_I^{''}}\left(\ep_1,\ep_2-\ep_1,m- \frac{\ep_1}{2},\tau\right)\\
=&\sum_{{\vec{n}\in \mathbb{Z}^r}}{e}^{i\pi D_{II}\tau n^2  +(m-\epsilon_+) n} 
\mathbb{E}_{\mathfrak{n}_I}\left(\ep_1,\ep_2,m,\tau \right).
\label{eq677_EBW2}
\end{split}\end{equation}
One can observe that equation \eqref{eq677_EBW2} is the same as $A_1$ M-string blowup equation with the replacement $\tau\rightarrow D_{II}\tau$, such that the solution of \eqref{eq677_EBW2} is
\be\label{eq336_EBW3}
\mathbb{E}_{\mathfrak{n}_I}\left(\ep_1,\ep_2,m,\tau \right)=\mathbb{E}^{A_1}_{\mathfrak{n}_I}\left(\ep_1,\ep_2,m,D_{II}\tau \right)
\ee
\section{Modular bootstrap}\label{sec:Modular_bootstrap}
In this section, we will talk about the modular bootstrap approach towards the twisted elliptic genera. In section \ref{subsec:4.1}, we give a quick overview of the modular group of twisted elliptic genus. In section \ref{subsec:bootstrap}, we discuss in detail how to bootstrap all non-simply-laced cases except $(C_r)_{\pi}$. Section \ref{sec:thetapi} is devoted to the exceptional case $(C_r)_{\pi}$.

\subsection{Modular group of twisted elliptic genus}\label{subsec:4.1}

A 6d SCFT can be described by F-theory compactification on an elliptically fibered Calabi-Yau threefold. After twisted circle compactification, since the $\mathbb{Z}_N$ automorphism should act on the fiber as well, the geometry becomes a geometry with $N$-sections. The geometry is expected to be a genus one fibered Calabi-Yau threefold \cite{Bhardwaj:2019fzv}. The genus one fibration is a fibration with $N$-sections, which is defined by $N$ points that can be identified with each other and are transformed to each other by monodromies in the base. As proved in \cite{Schimannek:2019ijf,Cota:2019cjx}, the monodromy is an element in $\Gamma_1(N)$,
\begin{align}
\Gamma_1(N)=\Bigr\{
\left(\begin{array}{cc}a & b \\ c & d\end{array} \right)\in \mathrm{SL}(2,\mathbb{Z}): a,d \equiv 1 (\text{mod}\ N),\quad c \equiv 0\ (\text{mod}\ N)
\Bigr\}\,,
\end{align}
which is a subgroup of $\Gamma_0(N)$. Moreover, it is known that
\begin{equation}
    \Gamma_0(2) = \Gamma_1(2),\quad  \Gamma_0(N) = \langle\ \Gamma_1(N), \begin{pmatrix} -1&0\\0&-1\\ \end{pmatrix}\ \rangle, \ \text{for}\ 3 \leq N \leq 4\,.
\end{equation}
Although the minus of the identity matrix acts trivially on $\tau$, its presence eliminates non-trivial modular forms of odd weights. In other words, if one considers modular forms invariant under $\Gamma_1(N)$ for $3 \leq N \leq 4$, in additional to those invariant under $\Gamma_0(N)$ one in principle also has to include odd-weight modular forms. But since those are absent in the elliptic genera, we think it better to choose the modular group to be $\Gamma_0(N)$ rather than $\Gamma_1(N)$. This should be understood as applying only to particular CY threefolds considered in this paper.\footnote{We thank Thorsten Schimannek for answering a relevant question concerning this point.}

In short, we expect the elliptic genus to be a Jacobi form in $\Gamma_0(N)$, where \(N=2\) for \(B_r\), \((C_r)_{0}\), \(F_4\) algebras, \(N=3\) for \(G_2\) algebra and $N= 4$ for $(C_r)_{\pi}$. 

In order to bootstrap the elliptic genera, one should use the ring of $\Gamma_0(N)$, which is reviewed in appendix \ref{appendix_modular_forms}. However, the full ring of $\Gamma_0(N)$ modular forms contains object $E_2^{(N)}$ with smaller weight, which makes the number of generators growing too fast as we increase the string numbers. In order to simplify our computation, we treat the geometry after twisted circle compactification as a gluing of single nodes in the base, such that along each node the modular group is either $\Gamma_0(N)$ or $\text{SL}(2,\mathbb{Z})$. Another key observation from the end of last section is that, restricting to a single base node, the elliptic genera always resemble the elliptic genera of $A_1$ M-strings. If the node is a long simple root of the Lie algebra, then they are the same, while for a short simple root we simply make the replacement $\tau\rightarrow N\tau$. Such an observation makes us conclude that along a single short node, the $E_2^{(N)}$ actually disappear and we could just use the modular forms arising from $\text{SL}(2,\mathbb{Z})$ Jacobi forms instead. For general string numbers not necessarily restricted to a single node, except for $(C_r)_{\pi}$,  
we further conjecture that the numerator of the elliptic genera can still be written in terms of the following set of elliptic modular forms and Jacobi forms, 
\begin{align}
&\Bigr\{E_4(\tau), \ E_6(\tau), \ A(\tau|z), \ B(\tau|z), \ 
E_4(N\tau), \ E_6(N\tau), \ A(N\tau|z), \ B(N\tau|z) \Bigr\},&
&z\in \{m,\epsilon_+,\epsilon_-\}.
\label{eq108_rings}
\end{align}
Recall that \(E_4(\tau)\) and \(E_6(\tau)\) are Eisenstein series with weight 4 and 6 under \(\Gamma_0(1) = \text{SL}(2,\mathbb{Z})\). Second, \(A(\tau|z)=\varphi_{0,1}(\tau,z)\) and \(B(\tau|z)=\varphi_{-2,1}(\tau,z)\) are weak Jacobi forms with weight 0, -2 and index 1. Those functions generate the ring of elliptic modular forms and weak Jacobi forms for \(\Gamma_0(1)\) respectively. For more details, one can refer appendix \ref{appendix_modular_forms}. In order to keep the expression simple, we will use the following notation
\begin{align}
&\widehat{E}_4 \equiv E_4(N\tau), &
&\widehat{E}_6 \equiv E_6(N\tau), &
&\widehat{A}(z) \equiv A(N\tau|z), &
&\widehat{B}(z) \equiv B(N\tau|z)
\end{align}
from now on. Nevertheless, for the $(C_r)_{\pi}$ case, we propose a different modular ansatz, which will be discussed in section \ref{sec:thetapi}.

\par 
As a remark, notice that the generators in \eqref{eq108_rings} are not independent. 
For all base degrees computed in this paper, we find that the unknown part of the elliptic genera indeed lies in the subring \eqref{eq108_rings}, and is completely fixed after imposing all the identities among them. For definiteness, we list those identities explicitly in appendix \ref{appendix_modular_forms} for $N = 2$ and $3$ which are used in this paper. We also indicate how to prove them for the case $N = 2$. Again the $N = 4$ case is very different and will be discussed separately in section \ref{sec:thetapi}.

Finally, we explain how the elliptic genus generated from (\ref{eq108_rings}) transforms under \(\Gamma_0(N)\). In general, one can express the elliptic genus in the following form, 
\begin{align}
\mathbb{E}_{\mathfrak{n}}(\tau,z)=\sum_k f^{(k)}(\tau,z)\cdot \widehat{f}^{(k)}(N\tau,z).
\label{eq79_ansatz}
\end{align}
Here, we consider the monomial expansion of the elliptic genus such that the \(k\)'th monomial is given by \(f^{(k)}(\tau,z)\cdot \widehat{f}^{(k)}(N\tau,z)\). The functions \(f^{(k)}(\tau,z)\) and \(\widehat{f}^{(k)}(\tau,z)\) are modular functions with weight \(w\), \(\hat{w}\) and index \(\mathfrak{i}\), \(\hat{\mathfrak{i}}\) respectively. They transforms as follows under \(\Bigr(\begin{smallmatrix}
a & b \\ c & d
\end{smallmatrix}\Bigr) \in \text{SL}(2,\mathbb{Z})\),
\begin{align}
f^{(k)}({a\tau+b\over c\tau+d})&=
(c\tau+d)^{w} \cdot \exp\Bigr[-{\pi i c\over c\tau+d} \mathfrak{i}(z) \Bigr]
\cdot f^{(k)}(\tau,z),
\nonumber \\
\hat{f}^{(k)}({a\tau+b\over c\tau+d})&=
(c\tau+d)^{\hat{w}} \cdot \exp\Bigr[-{\pi i c\over c\tau+d} \hat{\mathfrak{i}}(z) \Bigr]
\cdot \widehat{f}^{(k)}(\tau,z).
\end{align}
Then, under \(\Bigr(\begin{smallmatrix}
a & b \\ c & d
\end{smallmatrix}\Bigr) \in \Gamma_0(N)\), the elliptic genus (\ref{eq79_ansatz}) transforms as follows, 
\begin{align}
\mathbb{E}_{\mathfrak{n}}({a\tau+b\over c\tau+d},{z\over c\tau+d})
&=\sum_k f^{(k)}({a\tau+b\over c\tau+d},{z\over c\tau+d}) \cdot 
 \widehat{f}^{(k)}({a(N\tau)+Nb\over {c\over N}(N\tau)+d},{z\over {c\over N}(N\tau)+d})
\nonumber \\
&=(c\tau+d)^{w+\hat{w}}\cdot
\exp\Bigr[{-\pi i c\over c\tau+d}\Bigr(\mathfrak{i}(z)+{\hat{\mathfrak{i}}(z)\over N}\Bigr) \Bigr]\cdot
\mathbb{E}_{\mathfrak{n}}(\tau,z).
\label{eq723_gamma}
\end{align}
Note that \(c\equiv 0\) (mod \(N\)) plays a critical role in the transformation above. As a result, the full elliptic genus has a weight \(w+\hat{w}\) and index \(\mathfrak{i}+{\hat{\mathfrak{i}}\over N}\) under \(\Gamma_0(N)\). This also shows that the index of the hatted Jacobi forms should be divided by $N$.

\par 

\subsection{Bootstrap from the vanishing bound}\label{subsec:bootstrap}
Recall that from the end of section \ref{subsec:EG}, the weight and index of the elliptic genus are
\begin{align}
w_{\mathfrak{n}}&=0, \quad 
\mathfrak{i}_{\mathfrak{n}}={\epsilon_1\epsilon_2\over 2}\mathfrak{n}^T(\Omega D)\mathfrak{n}
+(m^2-\epsilon_+^2)\sum_{a=1}^r D_{aa} \mathfrak{n}_a\,,
\end{align}
where $\Omega$ is the Cartan matrix for the Lie algebra $ \mathfrak{g}$ and $D$ is the matrix $D_{ij} = 2\frac{\delta_{i,j}}{\langle\alpha_i,\alpha_i\rangle}$. For $\mathfrak{g}$ simply-laced, $D$ is the identity matrix and we are back to the case considered in \cite{Duan:2020cta}, while for $\mathfrak{g}$ non-simply-laced which is the focus of this paper, it is non-trivial such that the combined matrix $\Omega^{\text{s}} = \Omega D$ becomes
\begin{equation}
    \Omega^{\text{s}}_{ij} = 4\frac{\langle\alpha_i,\alpha_j\rangle}{\langle\alpha_i,\alpha_i\rangle \langle\alpha_j,\alpha_j\rangle}\,,
\end{equation}
which is manifestly symmetric. For reader's convenience, we give the explicit form of $\Omega^{\text{s}}$ for $B_n$, $(C_r)_{0}$, $F_4$ and $G_2$ in table \ref{tab:Omegasym}. For the $(C_r)_{\pi}$ theory, $\Omega$ is the negative of the intersection matrix \eqref{eq:A_2l}, $D_{ij}=\frac{1}{2}\delta_{i,j}$.

\begin{table}[h]\label{tab:Omegasym}
    \centering
    \def\arraystretch{1.3}
    \begin{tabular}{c|c|c|c}
    $ B_n$ & $ (C_r)_{0}$ & $ F_4$ & $ G_2$\\ \hline
    $\scriptstyle \begin{pmatrix}
    2 & -1 & 0&\cdots&0 &0 \\
    -1  & 2 &-1 & \cdots & 0 &0\\
    0  & -1 &2& \cdots & 0 &0\\
    \vdots & \vdots & \vdots& \vdots &\vdots&\vdots \\
    0 & \cdots & \cdots &0 & 2 &-1  \\
    0 & \cdots & \cdots &0 & -1 & 1\\
\end{pmatrix}$ & $\scriptstyle \begin{pmatrix}
    1 & -\frac{1}{2} & 0&\cdots&0 &0 \\
    -\frac{1}{2}  & 1 &-\frac{1}{2} & \cdots & 0 &0\\
    0  & -\frac{1}{2} &1& \cdots & 0 &0\\
    \vdots & \vdots & \vdots& \vdots &\vdots&\vdots \\
    0 & \cdots & \cdots &0 & 1 &-1  \\
    0 & \cdots & \cdots &0 & -1 & 2\\
\end{pmatrix}$  & $\scriptstyle \begin{pmatrix}
    2 & -1 &0 &0\\
    -1  & 2 &-1&0\\
    0&-1&1&-\frac{1}{2}\\
    0&0&-\frac{1}{2}&1\\
\end{pmatrix}$ & $\scriptstyle \begin{pmatrix}
    \frac{2}{3} & -1 \\
    -1  & 2\\
\end{pmatrix}$
    \end{tabular}\caption{$\Omega^{\text{s}}$ for non-simply-laced $\mathfrak{g}$.}
    \label{tab832_symcartan}
    \end{table}
Before doing actual computation, we emphasize that our modular ansatz is motivated by the picture of twisting at the level of the elliptically fibered CY threefold, briefly explained in section \ref{sec:2.3}. Note that the number $N$ should coincide with the order of twisting mentioned in section \ref{sec:2.1}.

Although we do not have a detailed understanding of how the geometry transforms, and for the cases we computed we actually do not need the full ring of $\Gamma_0(N)$ modular forms, we believe that this geometric picture is supposed to support our modular ansatz for the twisted elliptic genera.

Without further ado, we claim that the modular ansatz takes the following form,
\begin{align}\label{sec_MB:ellipticgenus}
\mathbb{E}_{\mathfrak{n}}&={\mathcal{N}_{\mathfrak{n}}\over \mathcal{D}_{\mathfrak{n}}}, \quad
\mathcal{D}_{\mathfrak{n}}=\prod_{a=1}^r \prod_{k=1}^{n_a} 
{\theta_1(2\tau/\Omega^{\text{s}}_{aa},k\epsilon_1)\over \eta^3(2\tau/\Omega^{\text{s}}_{aa}) } \cdot 
{\theta_1(2\tau/\Omega^{\text{s}}_{aa},k\epsilon_2)\over \eta^3(2\tau/\Omega^{\text{s}}_{aa}) }\,,
\end{align}
and the numerator $\mathcal{N}_n$ is an element in the ring of $\Gamma_0(N)$ Jacobi modular forms,
\begin{equation}\label{eq115_numerator} 
    \mathcal{N}_{\mathfrak{n}} = \sum_{\mathfrak{p},\widehat{\mathfrak{p}} }C_{\mathfrak{p},\widehat{\mathfrak{p}}} \left(E_4^{p_4} E_6^{p_6} \prod_{i=1}^3 A(z_i)^{p_{-2,i}} B(z_i)^{p_{0,i}}\right) \times \left(\widehat{E}_4^{\widehat{p}_4} \widehat{E}_6^{\widehat{p}_6} \prod_{i=1}^3 \widehat{A}(z_i)^{\widehat{p}_{-2,i}} \widehat{B}(z_i)^{\widehat{p}_{0,i}}\right)\,,
\end{equation}
where as before the overall hat means $\widehat{f}(\tau,...) = f(N\tau,...)$ and $z_i$ represent the chemical potentials as $(z_1,z_2,z_3)=(\epsilon_+,\epsilon_-,m)$.

As promised above, further constraints can be imposed to reduce the number of terms in \eqref{eq115_numerator}. The basic idea behind is that since we have roots of different length, the BPS string wrapping the corresponding curves should to some extent see the difference. For extreme cases, we learn from the last part of the section \ref{sec:Blowup} that the $\mathcal{N}_n$ with non-zero $n$ only for roots dual to unfolded roots (corresponding to those with $\Omega^{\text{s}}_{aa} \neq 2$) only depends on $\text{SL}(2,\mathbb{Z})$ forms, while with only non-vanishing wrapping degrees for folded roots it only depends on hatted forms.

Although in general cases both forms are needed, we can still try to separate their contributions. Let us look again at \eqref{sec_MB:ellipticgenus}. Note that the numerator $\mathcal{N}_{\mathfrak{n}}$ has weight and index
\begin{equation}\label{eq:weight_index}
\begin{aligned}
        w_{\mathcal{N}_{\mathfrak{n}}} &= -2\sum_{a=1}^r \sum_{k = 1}^{n_a} k\,,\\
        i_{\mathcal{N}_{\mathfrak{n}}} &= \frac{\epsilon_+^2 - \epsilon_-^2}{2}\mathfrak{n}^T\Omega ^{\text{s}}\mathfrak{n} +(m^2-\epsilon_+^2)D\cdot\mathfrak{n} +( \sum_{a=1}^r D_{aa}\sum_{k = 1}^{n_a} k^2) (\epsilon_+^2 + \epsilon_-^2)\,.
\end{aligned}
\end{equation}
The weight $w_{\mathcal{N}_{\mathfrak{n}}}$ is naturally factorized, hence we can demand that the total weight for two types of roots should match the weight of the two products in \eqref{eq115_numerator} separately. The index $i_{\mathcal{N}_{\mathfrak{n}}}$ is a bit more complicated. Since $D$ is a diagonal matrix, the index for $m$ is factorized, and we can ask for separation as above. However, because of the non-diagonal form of $\Omega^{\text{s}}$, the index for $\epsilon_{\pm}$ mixes $n_a$ for both types. The best we can demand is to separate the sum of indices $i_{\epsilon_+} + i_{\epsilon_-}$ into roots different types. For all the cases computed in appendix \ref{App:EG}, this set of constraints is satisfied.

To give a concrete example, let us consider the case $\mathfrak{g} = F_4$ ($N = 2$). From table \ref{tab:Omegasym} we learn that nodes 1 and 2 are long roots dual to the folded roots of $E_6$, while nodes 3 and 4 correspond to short roots dual to the unfolded roots of $E_6$. A potentially confused reader can look up the table \ref{tab166_affine} for clarification. We rewrite the weight and index polynomial in the following way, 
\begin{equation}
\begin{aligned}
    w_{\mathcal{N}_{\mathfrak{n}}} &= -2\left(\frac{n_1(n_1 + 1)}{2} + \frac{n_2(n_2 + 1)}{2}\right) -2\left(\frac{n_3(n_3 + 1)}{2} + \frac{n_4(n_4 + 1)}{2}\right)\,,\\
    i_{\mathcal{N}_{\mathfrak{n}}} &= (m^2 - \epsilon_+^2)(n_1 + n_2) + \frac{1}{2}(m^2 - \epsilon_+^2)(n_3 + n_4)\\
    &+ \frac{\epsilon_+^2 - \epsilon_-^2}{2}\mathfrak{n}^T\Omega ^{\text{s}}\mathfrak{n} +( \sum_{k = 1}^{n_1} k^2 + \sum_{k = 1}^{n_2} k^2) (\epsilon_+^2 + \epsilon_-^2) + \frac{1}{2}( \sum_{k = 1}^{n_3} k^2 + \sum_{k = 1}^{n_4} k^2) (\epsilon_+^2 + \epsilon_-^2)\,.
\end{aligned}
\end{equation}
Then we require the weight and index $i_{m}$ involving $n_1$ and $n_2$ to match the first parenthesis in \eqref{eq115_numerator}, and those of $n_3$ and $n_4$ to match the second. Moreover, the sum of indices $i_{\epsilon_+} + i_{\epsilon_-}$ containing $n_1$ and $n_2$ should match the sum in the first parenthesis in \eqref{eq115_numerator} while those of $n_3$ and $n_4$ should match the second one.

Given the modified ansatz \eqref{eq115_numerator} and the weight and index polynomial \eqref{eq:weight_index}, one can impose constraints to determine the coefficients which are finite. In earlier works \cite{Gu:2017ccq,DelZotto:2017mee,Duan:2020cta}, the authors successfully employed the vanishing conditions to fix them in many 6d SCFTs. To be more precise, the elliptic genus of BPS strings in 6d $\mathcal{N} = (2,0)$ SCFTs can be identified with the topological string partition function on the elliptically fibered CY manifold, thanks to the M-theory/F-theory duality. The latter is known to have a GV expansion, encapsulating integral enumerative invariants known as the GV invariants. Those invariants turn out to satisfy vanishing conditions, which immediately give us infinitely many constraints on the coefficients. After twisted circle compactification, we expect the elliptically-fibered CY manifold to change to an $N$-section geometry, and the twisted elliptic genus should still be identified with the topological string partition function. Therefore, it's still legitimate to apply the vanishing condition to determine the elliptic genera. As a final remark, we stress that from the point of view of topological strings, the existence of the structure of Jacobi forms \eqref{sec_MB:ellipticgenus} is non-trivial and signals special $Sp(n,\mathbb{Z})$ monodromies in the extended K\"{a}hler moduli space. The latter is demonstrated beautifully in \cite{Cota:2019cjx} based on homological mirror symmetry.

Here we present a self-contained summary of the vanishing bound for GV invariants. For more details, the reader can refer \cite{Gu:2017ccq,DelZotto:2017mee,Duan:2020cta}. First, recall that the refined free energy enjoys the GV expansion \eqref{eq:refinedGV},
\begin{equation}\label{TS:refinedGV}
    \begin{aligned}
    \mathcal{F}_{\text{GV}} &= \log Z_{\text{GV}}(\epsilon_1,\epsilon_2,\mathbf{t}) \\
    & = \sum\limits_{2j_{\pm} \in \mathbb{N}} \sum\limits_{d \geq 1} \sum\limits_{\alpha \in H_2(X, \mathbb{Z})} N^{\alpha}_{j_-j_+} {(-1)^{2 (j_-+j_+)} \, \chi_{j_-}(u^d) \chi_{j_+}(v^d) \over v^d + v^{-d} - u^d - u^{-d}} \frac{\mathbf{Q}^{d \alpha}}{d}\,,
    \end{aligned}
\end{equation}
with $N^{\alpha}_{j_-j_+}$ the refined GV invariants. They count the number of BPS states with spin $j_{\pm}$ in five dimensions, so they are always non-negative integers. Mathematically they can be defined in terms of the \textit{motivic} enumerative invariants \cite{MR2927365,Choi:2012jz}, but they are quite complicated to understand. To argue for the vanishing property, we adopt the route taken in \cite{Duan:2020cta} which instead starts from the unrefined GV invariants $I^{\alpha}_g$. The unrefined or self-dual limit sets $v = 1$, and we have 
\begin{equation}\label{eq:refinedUnrefined}
 \sum_{j_{\pm} \in \mathbb{N}/2} (-1)^{2 j_+} (2j_+ + 1) \chi_{j_-}(u)\,N^{\alpha}_{j_- j_+}  = \sum_{g=0}^{\infty} I^{\alpha}_{g} (u^{\frac{1}{2}} + u^{-\frac{1}{2}})^{2 g}\,,
\end{equation}
where the symbol $\chi_{j}(x)$ denotes the character of an irreducible $SU(2)$ highest-weight representation with spin $j \in \mathbb{N}/2$ \eqref{eq:spinj}.

In mathematics, $I^{\alpha}_g$ is closely related to enumerative invariants known as stable pair or Pandharipande-Thomas (PT) invariants $P_{n,\alpha}$ \cite{MR2545686}. To define the latter, we start from a stable pair consisting of torsion sheaves $F$ having dimension one support and a non-trivial holomorphic section $s$, which altogether captures the D6-D2-D0 bound states in physics. The moduli space of all possible stable pairs with a fixed holomorphic Euler characteristic $n$ and a fixed class of support $\alpha$ carries a perfection obstruction theory \cite{MR2545686}, and the virtual fundamental cycle happens to be zero dimensional for a CY threefold. Hence the PT invariants $P_{n,\alpha}$ just count the number of points inside the virtual fundamental cycle. For a fixed $\alpha$, there exists an $n$ small enough such that $P_{m,\alpha}$ vanishes for any $m\leq n$. This is simply because a curve with given homology class can not have arbitrarily small Euler characteristic, hence the moduli space is empty.

On the other hand, $P_{n,\alpha}$ can be expressed in terms of $I_{g}^{\alpha}$ \cite{MR2545686}. For simplicity, if we consider an irreducible class $\beta \in H_2(X,\mathbb{Z})$, they are related in the following way,
\begin{equation}
    \sum_{n} P_{n,\beta}\, q^n = \sum_{g \geq 0} I_{g}^{\beta} q^{1-g} (q + 1)^{2g-2}\,.
\end{equation}

In order to satisfy the vanishing property of $P_{n,\beta}$, $I_{g}^{\beta}$ must vanish when $g$ is sufficiently large. Inductively, we can extend the vanishing property for $I_{g}^{\beta}$ to all curve classes in $H_2(X,\mathbb{Z})$.

Furthermore, we assume that the refined GV invariants satisfy the so-called checkerboard pattern \cite{Choi:2012jz}: for a given curve class $\alpha$, the combination $\left(2j_-+2j_+\right)$ is always even or odd. This pattern also plays an important role in the blowup equations, as explained in the section \ref{sec:3.1}.

Let us look back at the equation \eqref{eq:refinedUnrefined}. Decompose $\chi_{j_-}(u)$ into a polynomial with variable $(u^{\frac{1}{2}} + u^{-\frac{1}{2}})^2$, based on the above assumption it is not hard to derive a \textit{generic} vanishing bound for the refined GV invariants,
\begin{equation}
    N^{\alpha}_{j_- j_+} = 0\quad \text{for}\ j_-\ \text{or}\ j_+ \gg 0\,.
\end{equation}

However, the precise vanishing bound for $j_{\pm}$ depends on the detailed knowledge of the geometry. At this stage, it is not clear to the authors how to construct a useful presentation of the genus-one fibered CY manifolds needed in this paper, let alone deriving the precise bound from the first principle. Therefore, we adopt the same strategy as in \cite{Duan:2020cta}. Namely, we make some guess on the bound by looking at the BPS data available from section \ref{sec:Blowup}. Moreover, since the CY threefolds consider here are obtained from certain twist of the geometries that engineer 6d $\mathcal{N} = (2,0)$ SCFTs, we expect the vanishing conditions to be mostly unchanged. In fact, we essentially use the same ones found in \cite{Duan:2020cta}, with only minor simplifications. 

To fix our notation, we specialize the general formula \eqref{TS:refinedGV},
\begin{equation}
    \mathcal{F}_{\text{GV}} = \sum\limits_{\mathfrak{n},k,\mu} \sum\limits_{2j_{\pm} \in \mathbb{N}} \sum\limits_{d \geq 1} \frac{N_{j_-j_+}^{\,\mathfrak{n},k,\mu}}{k} {(-1)^{2 (j_-+j_+)} \, \chi_{j_-}(u^d) \chi_{j_+}(v^d) \over v^d + v^{-d} - u^d - u^{-d}}  Q_m^{\, d\mu}\, q^{dk}\, \mathbf{Q}^{d \mathfrak{n}}\,,
\end{equation}
with $Q_m=e^{-m}$, $q=e^{2\pi i \tau}$ and $\mathbf{Q}=e^{-\mathbf{v}}$ formal exponential of K\"{a}hler parameters associated to two-cycle of the M-string mass, elliptic fiber and exceptional curves inside the base respectively. We choose the parametrization such that $\mu$ can be negative but with a $\mathbb{Z}_2$ symmetry $N_{j_-j_+}^{\,\mathfrak{n},k,\mu} = N_{j_-j_+}^{\,\mathfrak{n},k,-\mu}$. Our conjecture for the vanishing bound of the refined GV invariants goes as follows:
\begin{conj}
For genus one fibered CY threefolds that engineer 5d $\mathcal{N} = 2$ gauge theories arising from twisted circle compactification of 6d $\mathcal{N} = (2,0)$ $ADE$  SCFTs, their refined GV invariants obey uniformly the vanishing bound\footnote{This bound is slightly smaller than that used in \cite{Duan:2020cta}.},
\begin{equation}\label{Vanish:bound}
N_{j_-j_+}^{\,\mathfrak{n},k,\mu \geq 0}=0\ {\rm for}\ \left\{\begin{array}{rl} 2 j_- &> k \cdot \max \{n_i\} - \frac{\mu-k-1}{2} \cdot H( \mu - k -2),\\ & 
		{\rm \ \ \ \ \ \ \ \ \ \ \ \ \ \ \ \ or}  \\   
		2 j_+&> (k + 1) \cdot \max \{n_i\} - (\mu - k) \cdot H( \mu - k -1),\\ &
		{\rm \ \ \ \ \ \ \ \ \ \ \ \ \ \ \ \ or}  \\
		n_i &=0\ \forall\, i \ \text{with}\ \Omega^{\text{s}}_{ii} = 2 \ \text{and}\ N \nmid k.
	\end{array} \right.
	\end{equation}
\end{conj}
The function $H(x)$ is the Heaviside step function,
\begin{equation}
    \begin{aligned}
        H(x) = \begin{cases}0, & \text{if}\ x < 0\, \\
      1, & \text{otherwise}\,,
    \end{cases}
    \end{aligned}
\end{equation}
and the last condition comes simply from the observation that the free energy $\mathcal{F}_{\text{ref}}$ is always a power series in $q^N$ in that case.

Then we explain how to use \eqref{Vanish:bound} to determine the elliptic genus recursively. This starts from the following obvious relation,
\begin{equation}\label{vanishing:ZtoF}
    \ln \mathcal{Z}_{\text{GV}} = \ln\left(1 + \sum_{\mathfrak{n} \neq 0\, \in \mathbb{Z}_{\geq 0}^r} \mathbb{E}_{\mathfrak{n}}\, \mathbf{Q}^{\mathfrak{n}}\right) =  \mathcal{F}_{\text{GV}} = \sum_{\mathfrak{n} \neq 0\, \in \mathbb{Z}_{\geq 0}^r} F_{\mathfrak{n}} \mathbf{Q}^{\mathfrak{n}}\,,
\end{equation}
with both sides regarded as power series in $\mathbf{Q}$. Expand the logarithm in \eqref{vanishing:ZtoF}, we have
\begin{equation}\label{Vanishing:Boot}
  F_{\mathfrak{n}_\alpha} = \mathbb{E}_{\mathfrak{n}_\alpha} + \sum_{j=2}^\infty
  \frac{1}{j!}
  \sum_{\mathfrak{n}_1, \ldots, \mathfrak{n}_j > \boldsymbol{0} \atop\sum \mathfrak{n}_i = \mathfrak{n}_{\alpha}}
  \kappa_{(\mathfrak{n}_\alpha)}^{\mathfrak{n}_1,\ldots,\mathfrak{n}_j} \prod_{i=1}^j
  \mathbb{E}_{\mathfrak{n}_i}\,,
\end{equation}
where $\kappa_{(\mathfrak{n}_\alpha)}^{\mathfrak{n}_1,\ldots,\mathfrak{n}_j}$ are some integers that can be computed order by order. Note that the second term in the right-hand side only involves $Z_\mathfrak{n}$ with base degree strictly smaller than $\mathfrak{n}_{\alpha}$.

It turns out that this equation is very constraining. We first supply a few refined GV invariants at the lowest base degrees as the input data, which e.g., can be obtained from the blowup equations in section \ref{sec:Blowup}. Then we impose simultaneously the vanishing bound in the left-hand side and the modular ansatz in the right-hand side. This already gives us enough constraints to completely fix $F_{\mathfrak{n}_\alpha}$ and hence $\mathbb{E}_{\mathfrak{n}_{\alpha}}$. Increasing the degree one step at each time, we are able to bootstrap the elliptic genus for any value of $\mathfrak{n}_{\alpha}$ in principle.

As always, we need to clarify one issue: is the solution to the above bootstrap equation unique? Surely enough, the uniqueness of the solution is equivalent to determining all the unknowns in the ansatz. For the case of 6d $\mathcal{N} = (2,0)$ SCFTs, we have the following powerful criterion, which is discussed in detail in \cite{Gu:2017ccq}.

\begin{thm}\label{Vanishing:thm}
If either $i_{\epsilon_+}$ or $i_{\epsilon_-}$ is strictly smaller than $-1$, then the solution to \eqref{Vanishing:Boot} must be unique. Otherwise, the ambiguous term can be enumerated explicitly.
\end{thm}

After looking at their proof carefully, it is not difficult to show that the above criterion still holds for the modified ansatz and the difference just lies in the set of ambiguous terms.

In our situation, the index of $\varepsilon_-$ can be written as
\begin{equation}
    i_{\epsilon_-} = -\frac{1}{2} \mathfrak{n}\cdot\Omega^{\text{s}}\cdot\mathfrak{n}^{\text{T}}\,, \quad \Omega^{\text{s}}_{ij} = \langle\alpha_{i}^{\vee},\alpha_{j}^{\vee}\rangle\,,
\end{equation}
where $\alpha^{\vee}_{i}$ correspond to the coroots of $\mathfrak{g}$. Then the symmetric matrix $\Omega^{\text{s}}$ defines a positive-definite lattice and hence the exceptional cases are always finite. For them, the number of ambiguous terms is always very small after we further separate the contribution for different length of roots, explained earlier in this subsection. So we just need to supply a few BPS invariants if necessary.

Finally, it is worth emphasizing that finding the suitable ansatz to make the most of the vanishing bound is nontrivial. For example, if one uses the generators of the $\Gamma_0(N)$ modular forms mentioned in section \ref{subsec:4.1}, at some low base degrees where we cannot invoke the above criterion, there appear to be many coefficients not determined by the vanishing condition \eqref{Vanish:bound}. On the contrary, for cases listed in appendix \ref{App:EG}, we find that the ansatz used here can fix all the unknowns after imposing identities among them, although the number of terms in both ansatze are of roughly the same order.
 
In table \ref{tab_numerator}, we list some useful information about the numerator of the elliptic genera at low base degrees for gauge groups $G_2$, $B_3$, $(C_3)_0$ and $F_4$. The numerator is determined by the modular bootstrap whose explicitly form is shown in appendix \ref{App:EG}. As an independent check, the BPS invariants extracted from the elliptic genera is in perfect agreement with those obtained from the blowup equations.
 
\begin{table}[ht]
\def\arraystretch{1.5}
\centering
\begin{tabular}{|c|c|c|c|c|}\hline
Gauge group & Base degree & Index & Weight & Unknowns\\ \hline 
$G_2$& \dynkin[reverse arrows, labels*={0,1}]G2 & $\epsilon_+^2 + m^2$ & -2 & 2\\ \hline
$G_2$&\dynkin[reverse arrows, labels*={1,0}]G2 & $\frac{1}{3}\epsilon_+^2 +  \frac{1}{3} m^2$ & -2 & 2\\ \hline
$G_2$&\dynkin[reverse arrows, labels*={1,1}]G2 & $\frac{1}{3}\epsilon_+^2 + \epsilon_-^2 + \frac{4}{3} m^2$ & -4 & 4\\ \hline
$G_2$&\dynkin[reverse arrows, labels*={1,2}]G2 & $\frac{16}{3}\epsilon_+^2 + 3\epsilon_-^2 + \frac{7}{3} m^2$ & -6 &132\\ \hline
$G_2$&\dynkin[reverse arrows, labels*={2,1}]G2 & $\frac{4}{3}\epsilon_+^2 + \frac{7}{3}\epsilon_-^2 + \frac{5}{3} m^2 $ &-6 &226\\ \hline

$B_3$&\dynkin[labels*={1,1,1}]B3 & $\frac{1}{2}\epsilon_+^2 + 2\epsilon_-^2 + \frac{5}{2} m^2$ & -6 & 8\\ \hline
$B_3$&\dynkin[labels*={2,1,1}]B3 & $\frac{11}{2}\epsilon_+^2 + 4\epsilon_-^2 + \frac{7}{2} m^2$ & -8 &220\\ \hline
$B_3$&\dynkin[labels*={1,2,1}]B3 & $\frac{9}{2}\epsilon_+^2 + 5\epsilon_-^2 + \frac{7}{2} m^2 $ &-8 &220\\ \hline
$(C_3)_0$&\dynkin[labels*={1,1,1}]C3 & $\frac{9}{2}\epsilon_+^2 + 5\epsilon_-^2 + \frac{7}{2} m^2$ & -6 & 10\\ \hline
$(C_3)_0$&\dynkin[labels*={1,1,2}]C3 & $\frac{11}{2}\epsilon_+^2 + \frac{7}{2}\epsilon_-^2 + 3 m^2$ & -8 &330\\ \hline

$F_4$&\dynkin[labels*={1,1,1,1}]F4 & $\frac{1}{2}\epsilon_+^2 + \frac{5}{2}\epsilon_-^2 + 3 m^2 $ & -8&20\\ \hline
$F_4$&\dynkin[labels*={2,1,1,1}]F4 & $\frac{11}{2}\epsilon_+^2 + \frac{9}{2}\epsilon_-^2 + 4 m^2 $ & -10&550\\ \hline
$F_4$&\dynkin[labels*={1,2,1,1}]F4 &$\frac{9}{2}\epsilon_+^2 + \frac{11}{2}\epsilon_-^2 + 4 m^2 $ & -10&550\\ \hline
\end{tabular}
\caption{The index polynomial, weight and unknown coefficients for the numerator of the elliptic genera except $(C_r)_{\pi}$ case.}
\label{tab_numerator}
\end{table}

\subsection{$(C_r)_{\pi}$ theories}\label{sec:thetapi}
For the last case $(C_r)_{\pi}$, which is called non-geometric in \cite{Bhardwaj:2019fzv}, the situation becomes much subtler, which is the reason why we treat it in a separate subsection.

From \eqref{eq: twistedMstringindex} at the end of section \ref{subsec:EG}, we learn that its weight and index can be written as
\begin{align}\label{eq:4.23}
w_{\mathfrak{n}}&=0, \quad 
\mathfrak{i}_{\mathfrak{n}}={\epsilon_1\epsilon_2\over 2}\mathfrak{n}^T(\Omega^{\text{s}})\mathfrak{n}
+\frac{1}{2}(m^2-\epsilon_+^2)\sum_{a=1}^r \mathfrak{n}_a\,,
\end{align}
where the $\Omega^{\text{s}}$ takes the form,
\begin{equation}
   \Omega^{\text{s}} =  \begin{pmatrix}
    1 & -\frac{1}{2} & 0&\cdots&0 &0 \\
    -\frac{1}{2}  & 1 &-\frac{1}{2} & \cdots & 0 &0\\
    0  & -\frac{1}{2} &1& \cdots & 0 &0\\
    \vdots & \vdots & \vdots& \vdots &\vdots&\vdots \\
    0 & \cdots & \cdots &0 & 1 &-\frac{1}{2}  \\
    0 & \cdots & \cdots &0 & -\frac{1}{2} & \frac{1}{2}\\
\end{pmatrix}\,.
\end{equation}

In this case, all the nodes come from $\mathbb{Z}_2$ outer-automorphism of $A_{2r}$, we would expect that for all possible base degrees the $\tau$ variable in the ansatz will be multiplied by two. As a result, we propose its modular ansatz should be,
\begin{align}
\mathbb{E}_{\mathfrak{n}}&={\mathcal{N}_{\mathfrak{n}}\over \mathcal{D}_{\mathfrak{n}}}, \quad
\mathcal{D}_{\mathfrak{n}}=\prod_{a=1}^r \prod_{k=1}^{n_a} 
{\theta_1(2\tau,k\epsilon_1)\over \eta^3(2\tau) } \cdot 
{\theta_1(2\tau,k\epsilon_2)\over \eta^3(2\tau) }\,.
\end{align}
It's interesting to remark that the $\mathcal{D}_{\mathfrak{n}}$ does not follow the general rule in \eqref{sec_MB:ellipticgenus} which works perfectly for other cases.

The structure of the numerator $\mathcal{N}_{\mathfrak{n}}$ is different, which should enjoy the $\Gamma_0(4)$ modular group rather than the naive $\Gamma_0(2)$ from the Dynkin diagram. The main reason is that, as explained in section \ref{sec:2.1}, the whole automorphism group is actually $\mathbb{Z}_4$, leading to the Jacobi form of $\Gamma_0(4)$.
Following the above logic, we conjecture that the numerator $\mathcal{N}_{\mathfrak{n}}$ is an element 
\begin{align*}
  \mathcal{N}_{\mathfrak{n}} \in &\mathbb{C} [E^{(2)}_2(\tau), E^{(4)}_2(\tau), E_4(\tau), E_6(\tau),  A(4\tau| z_i),B(4\tau| z_i)]\cdot \mathbb{C} [E_4(2\tau), E_6(2\tau),  A(2\tau| z_i), B(2\tau| z_i)],
\end{align*}
such that the total weight and index of $Z_{\mathfrak{n}}$ satisfy \eqref{eq:4.23}. Here $z_i \in (\epsilon_+,\epsilon_-,m)\,.$

Unlike other cases, this ansatz contains too many unfixed terms at low base degrees after imposing the previous vanishing conditions. Also, the number of coefficients in the ansatz is growing so fast that it is already quite hard to determine the case of base degree two. The main reason is due to the full generators we are using for $\Gamma_0(4)$ elliptic modular forms. In practice, we fix all of them consistently by further comparing them with the result obtained from the blowup equations. Below, we will explain this in more detail for the $(C_1)_{\pi}$ theory. Still, it serves as strong evidence for the existence of elliptic genera invariant under the $\Gamma_0(4)$ modular group. In table \ref{tab_numerator_thetapi}, we list some useful information about the $(C_r)_{\pi}$ elliptic genera at low base degrees, whose explicitly form is shown in appendix \ref{App:EG}.
\begin{table}[ht]
\def\arraystretch{1.5}
\centering
\begin{tabular}{|c|c|c|c|}\hline
Base degree & Index & Weight & Unknowns\\ \hline 
\dynkin[labels*={1}]A1 & $\frac{1}{4}\epsilon_+^2 + \frac{1}{4}\epsilon_-^2 + \frac{1}{2}m^2$ & -2 & 30\\ \hline
\dynkin[labels*={2}]A1 & $\frac{5}{2}\epsilon_+^2 + \frac{3}{2}\epsilon_-^2 + m^2$ & -4 & 21707\\ \hline
\dynkin[labels*={1,0}]C2 & $\frac{1}{2}\epsilon_+^2 + \frac{1}{2}m^2$ & -2 & 2\\ \hline
\dynkin[labels*={1,1}]C2 & $\frac{1}{4}\epsilon_+^2 + \frac{3}{4}\epsilon_-^2 +  m^2$ & -4 & 60\\ \hline
\end{tabular}
\caption{The index polynomial, weight and unknown coefficients for the numerator of the $(C_r)_{\pi}$ elliptic genera.}
\label{tab_numerator_thetapi}
\end{table}

As a non-trivial check of our ansatz, we fix the modular expression up to base degree two for the $(C_1)_{\pi}$ theory. In the rank one case, one would expect that the numerator is generated by the following $\Gamma_0(4)$ modular and Jacobi forms
\be
  \mathcal{N}_{\mathfrak{n}} \in \mathbb{C} [E^{(2)}_2(\tau), E^{(4)}_2(\tau), E_4(\tau), E_6(\tau),  A(4\tau| z_i),B(4\tau| z_i)]\,.
\ee
Indeed, as shown in table \ref{tab655_thetapi} of appendix \ref{App:EG}, the one-string elliptic genus\footnote{Here we use the notation $\widehat{A}_{z_i}=\varphi_{-2,1}(4\tau,z_i)$, $\widehat{B}_{z_i}=\varphi_{0,1}(4\tau,z_i)$}
\be
\mathbb{E}_1(\tau,\ep_1,\ep_2,m)= \frac{(E_2^{(2)}-E_2^{(4)})( \widehat{A}_m \widehat{B}_+ - \widehat{A}_+ \widehat{B}_m)
(\widehat{A}_m \widehat{B}_- - \widehat{A}_- \widehat{B}_m)
}{2^8\, 3^2\, \varphi_{-2,1}(2\tau,\ep_1)\varphi_{-2,1}(2\tau,\ep_2)} 
\ee
contains elements $E_2^{(2)}$ and $E_2^{(4)}$ of $\Gamma_0(4)$. We verify that this expression coincides with the instanton partition function from ADHM up to 5-instantons. And we verify further with the one-string elliptic blowup equation
\begin{equation*}\begin{split}
\theta_2(&\tau,m-\ep_+)\mathbb{E}_1(\tau,\ep_1,\ep_2,m)= \\
&\theta_2(\tau,m+\ep_-)\mathbb{E}_1(\tau,\ep_1,\ep_2-\ep_1,m-\ep_1/2)+\theta_2(\tau,m-\ep_-)\mathbb{E}_1(\tau,\ep_1-\ep_2,\ep_2,m-\ep_2/2)
\end{split}\end{equation*}
up to $q^{21}$ order.

For base degree two, from table \ref{tab_numerator_thetapi} we have $21707$ possible generators, making it hard to fix the coefficients. From the gauge theoretic point of view, at the leading order the two strings contain the W-boson contribution, which is the same as a single string for $(C_1)_{0}$ theory. Further noticing that the index of the elliptic genus at two-strings of $(C_1)_\pi$ theory is the same as that of the single $(C_1)_{0}$ string, we may expect that the two-string elliptic genus contains the following contribution\footnote{It is not difficult to show that this expression is invariant under the $\Gamma_0(4)$ modular group.}
\be
\frac{1}{2}\left(\mathbb{E}_1^{(C_1)_{0}}(\tau,\ep_1,\ep_2,m)+\mathbb{E}_1^{(C_1)_{0}}(\tau+1/2,\ep_1,\ep_2,m)\right)\,.
\ee
The statement is indeed true, and we observe that the remaining part can be factorized as
\be\label{E2sp1pi}\begin{split}
\mathbb{E}_2^{(C_1)_{\pi}}(\tau,\ep_1,\ep_2,m)&=\frac{1}{2}\left(\mathbb{E}_1^{(C_1)_{0}}(\tau,\ep_1,\ep_2,m)+\mathbb{E}_1^{(C_1)_{0}}(\tau+1/2,\ep_1,\ep_2,m)\right)\\
&+\frac{( \widehat{A}_m \widehat{B}_+ - \widehat{A}_+ \widehat{B}_m)^2
(\widehat{A}_m \widehat{B}_- - \widehat{A}_- \widehat{B}_m)^2 }{\prod_{k=1}^2 \varphi_{-2,1}(2\tau,k\ep_1)\varphi_{-2,1}(2\tau,k\ep_2)}\mathcal{I}(\tau,\ep_1,\ep_2)\,.
\end{split}\ee
We verify the structure of \eqref{E2sp1pi} by solving the two-string elliptic blowup equation
\begin{equation*}\begin{split}
\theta_2(&\tau,m-\ep_+)\mathbb{E}_2(\tau,\ep_1,\ep_2,m)= \\
&\theta_2(\tau,m+\ep_+)\mathbb{E}_1(\tau,\ep_1,\ep_2-\ep_1,m-\ep_1/2)\mathbb{E}_1(\tau,\ep_1-\ep_2,\ep_2,m-\ep_2/2)\\
&+\theta_2(\tau,m-\ep_++2\ep_1)\mathbb{E}_2(\tau,\ep_1,\ep_2-\ep_1,m-\ep_1/2)+\theta_2(\tau,m-\ep_++2\ep_2)\mathbb{E}_2(\tau,\ep_1-\ep_2,\ep_2,m-\ep_2/2)\,,\\
\end{split}\end{equation*}
and get the Fourier expansion of $\mathcal{I}(\tau,\ep_1,\ep_2)$ up to $q^{20}$ order. Now it is much easier to find the modular expression of $\mathcal{I}(\tau,\ep_1,\ep_2)$, which is expected to be a $\Gamma_0(4)$ Jacobi form with weight $0$ and index $2\epsilon_+^2 + \epsilon_-^2 $, with only $2221$ generators. At the $q^{14}$ order,\footnote{The reason why we have to go to this order is that there exist a combination of the generators which contains a factor $(E_2^{(2)}-E_2^{(4)})^{13}\sim q^{13}$.} we manage to fix all of the unknowns and find
\be
\mathcal{I}(\tau,\ep_1,\ep_2)=(E_2^{(2)}-E_2^{(4)})^2\cdot\mathbb{C}\left[E_2^{(2)},E_2^{(4)}, A(4\tau,\ep_{\pm}), B(4\tau,\ep_{\pm})\right]\,.
\ee
Its expression is listed in appendix \ref{App:EG}. Interestingly, this does not depend on $E_4(\tau)$ and $E_6(\tau)$. We further check it up to $q^{20}$ order with the two-string elliptic blowup equation, and find a perfect agreement.

\section{6d twisted theories in the Cardy limit}\label{sec:Cardy}
In this section, we explore the Cardy limit of 6d $(2,0)$ theories on \(\mathbb{R}^4\times T^2\) with an outer-automorphism twist. The Cardy limit is defined as a limit where momenta are much larger than the central charge of the CFT \cite{Cardy:1986ie, Kim:2019yrz}. The free energy in the Cardy limit is usually called the Cardy formula, and it probes the high energy spectrum of the CFT. Furthermore, Cardy formulas have been used to account the Bekenstein-Hawking entropy of supersymmetric black holes in various dimensions \cite{Strominger:1996sh, Choi:2018hmj}.
In the case of 6d, the Cardy formulas of $(2,0)$ $ADE$ theories on \(\mathbb{R}^4\times T^2\) were studied in \cite{Lee:2020rns, Duan:2020cta}. The authors obtain the Cardy free energy by summing over the elliptic genera using continuum approximation and S-duality. The resulting asymptotic entropy in terms of the charges $(J_1, J_2, P, Q_L, Q_R)$ is
\begin{align}
S\simeq \log Z + \frak{n}\cdot \mathbf{v} + \epsilon_2(J_1 + Q_R) + \epsilon_2(J_2 + Q_R) + (-2\pi i) P + 2m Q_L.
\end{align}
 By solving the saddle point equation from the Cardy formula, the entropy $S$ can be written as a function of charges, and the result for $A_r$-type is\footnote{In \cite{Lee:2020rns}, additional $U(1)^{r+1}$ contribution is considered, here we only consider the non-abelian part.}
\begin{align}
S\simeq 2\pi\sqrt{\sqrt{\frac{2}{3}((r+1)^3-(r+1))P(J_-^2-(J_++Q_R)^2)}-Q_L^2}+2\pi i Q_L,
\label{A_CARDY}
\end{align}
where we define $J_{\pm}=J_1\pm J_2$.
Here, we shall extend their approach to all 6d $(2,0)$ theories, including twisted ones.

\par 
In order to explore the large momenta limit of the BPS partition function (\ref{eq163_BPSR}), we need to set the momentum-conjugate chemical potentials to be small. On \(\mathbb{R}^4\times T^2\), this can be achieved by setting
\begin{align}
|\epsilon_{1,2}|\to 0,\quad |\tau|\to 0\,.
\label{eq69_CARDY}
\end{align}
First, \(|\epsilon_{1,2}|\to 0\) is the well-known Seiberg-Witten prepotential limit \cite{Seiberg:1994rs, Nekrasov:2002qd}. It is also called thermodynamic limit where the volume of \(\mathbb{R}^4\) becomes infinite. In this limit, the leading free energy on \(\mathbb{R}^4\times T^2\) is proportional to \(\text{vol}(\mathbb{R}^4)\sim {1\over \epsilon_1 \epsilon_2}\). Second, \(|\tau|\to 0\) is the limit where KK momentum becomes large. Recall that \(\tau=i{R_\text{time} \over R_\text{KK}}\) is inversely proportional to the KK circle radius. Therefore, the KK circle is effectively decompactified, and the full 6d physics is visible in \(|\tau|\to 0\) limit. In this limit, the leading free energy on \(\mathbb{R}^4\times T^2\) is proportional to the volume of the spatial circle in \(T^2\), i.e. \(\text{vol}(R_\text{KK})\sim {1\over \tau}\) \cite{Cardy:1986ie}. Therefore, the leading free energy in the Cardy limit should be \(\log Z \sim \mathcal{O}({1\over \epsilon_1 \epsilon_2 \tau})\).

\par 

When taking the Cardy limit (\ref{eq69_CARDY}), we should decide the phase of the chemical potentials which can generically have complex values. Here, we shall consider the following regime of the chemical potentials,
\begin{align}
\epsilon_1>0,\quad \epsilon_2<0,\quad 2\pi i \tau<0.
\label{eq691_setting}
\end{align}
For simplicity, we set three fugacities \(Q_1=e^{-\epsilon_1}\), \(Q_2=e^{-\epsilon_2}\), and \(Q_\tau=e^{2\pi i \tau}\) to be purely real. First, setting \(2\pi i \tau < 0\) corresponds to setting \(Q_\tau <1\) which is needed for the convergence of KK expansion. Second, setting different signs for \(\epsilon_{1,2}\) might seem peculiar, but it is nothing but a slight deviation from the unrefined setting \(\epsilon_1=-\epsilon_2\). 

\par 

In order to investigate the Cardy limit, let us take \(\tau\to 0\) limit first. Recall that, according to our ansatz (\ref{eq115_numerator}), the elliptic genus of twisted M-string (except \((C_r)_\pi\) theory) can be written as follows,
\begin{align}
\mathbb{E}_{\mathfrak{n}}(\tau,z)={1\over \mathcal{D}_{\mathfrak{n}}(\tau,z) \cdot \widehat{\mathcal{D}}_{\mathfrak{n}}(n_G\tau,z) }
\times
\sum_{a=1}^{|\mathcal{S}|} \mathcal{N}^{(a)}_{\mathfrak{n}}(\tau,z) \cdot \widehat{\mathcal{N}}^{(a)}_{\mathfrak{n}}(n_G\tau,z).
\label{eq970_form}
\end{align}
Here, unhatted parts have modular parameter \(\tau\), and hatted parts have modular parameter \(n\tau\), where \(n\) is the order of the outer-automorphism. Also, \(a\) is a monomial index which enumerates the monomials in the numerator. The asymptotic behavior of the elliptic genus at \(\tau\to 0\) can be obtained by exploiting its property under the S-transformation. Unlike to the simply-laced case, we have to be careful about the S-transformation. The elliptic genus \(Z_{\mathfrak{n}}\) is a modular form under \(\Gamma_0(n_G)\) which is a congruence subgroup of \(\text{SL}(2,\mathbb{Z})\). If \(n_G\neq 1\), the S-transformation action \((\begin{smallmatrix} 0 & -1 \\ 1 & 0 \end{smallmatrix})\) is not an element of \(\Gamma_0(n_G)\). Therefore, the twisted elliptic genus does not have the S-duality property. 

\par 

However, one can still apply S-transformation to the elliptic genus and observe how it behaves. The elliptic genus (\ref{eq970_form}) has two parts that depend on \(\tau\) and \(n_G\tau\) respectively. Instead of the conventional S-transformation, we shall apply the following `separate S-transformation' to our elliptic genus of the form (\ref{eq970_form}), 
\begin{align}
\mathcal{D}(\tau,z),\ \mathcal{N}(\tau,z)
\quad&\to\quad
\mathcal{D}(-{1\over \tau},{z\over \tau}),\ \mathcal{N}(-{1\over \tau},{z\over \tau})\,,
\nonumber \\
\widehat{\mathcal{D}}(n_G\tau,z),\ \widehat{\mathcal{N}}(n_G\tau,z)
\quad&\to\quad
\widehat{\mathcal{D}}(-{1\over n_G\tau},{z\over n_G\tau}),\ \widehat{\mathcal{N}}(-{1\over n_G\tau},{z\over n_G\tau}).
\end{align}
Recall that \(\tau\)-dependent part and \(n_G\tau\)-dependent part both have zero weights. Let us assume that \(\tau\)-part has index \(\mathfrak{i}\), and \(n_G\tau\)-part has index \(\hat{\mathfrak{i}}\).  By applying the separate S-duality transformation above, the elliptic genus changes as follows,
\begin{align}
\mathbb{E}_\mathfrak{n}(\tau,z)&=
\exp\Bigr[-{1\over 2\pi i \tau}\Bigr( \mathfrak{i}+{\widehat{\mathfrak{i}}\over n_G} \Bigr) \Bigr]
\times 
\sum_{a=1}^{|\mathcal{S}|}
{\mathcal{N}^{(a)}_{\mathfrak{n}}(-{1\over \tau},{z\over \tau})
\widehat{\mathcal{N}}^{(a)}_{\mathfrak{n}}(-{1\over c\tau},{z\over c\tau})
\over 
\mathcal{D}_{\mathfrak{n}}(-{1\over \tau},{z\over \tau})
\widehat{\mathcal{D}}_{\mathfrak{n}}(-{1\over c\tau},{z\over c\tau})}.
\end{align}
As we explained in (\ref{eq723_gamma}), the sum of \(\mathfrak{i}\) and \({\hat{\mathfrak{i}}\over n_G}\) yields the \(\Gamma_0(n_G)\) index of the elliptic genus. Therefore, the above expression can be written as follows,
\begin{align}
\mathbb{E}_\mathfrak{n}(\tau,z)&=
\exp\Bigr[-{1\over 2\pi i \tau}\Bigr( {\epsilon_1 \epsilon_2\over 2} \mathfrak{n}^T (\Omega D)\mathfrak{n}+(m^2-\epsilon_+^2)\vec{D}\cdot \mathfrak{n} \Bigr)\Bigr]
\nonumber \\
&\times 
\sum_{a=1}^{|\mathcal{S}|}
{\mathcal{N}^{(a)}_{\mathfrak{n}}(-{1\over \tau},{z\over \tau})
\widehat{\mathcal{N}}^{(a)}_{\mathfrak{n}}(-{1\over c\tau},{z\over c\tau})
\over 
\mathcal{D}_{\mathfrak{n}}(-{1\over \tau},{z\over \tau})
\widehat{\mathcal{D}}_{\mathfrak{n}}(-{1\over c\tau},{z\over c\tau})}.
\label{eq1011_second}
\end{align}
\par 

Now, the remaining task is to determine the asymptotics of the second line of (\ref{eq1011_second}) in the Cardy limit \(\tau\to i\cdot 0^+\). In the expression, the numerators and the denominators contain the following fugacities,
\begin{align}
&Q_1^D=e^{-{\epsilon_1\over \tau}},\quad
Q_2^D=e^{-{\epsilon_2\over \tau}}, \quad
Q_\tau^D=e^{-{2\pi i \over \tau}}, \quad
Q_m^D=e^{-{m\over \tau}},
\nonumber \\
&\widehat{Q}_1^D=e^{-{\epsilon_1\over n_G\tau}},\quad
\widehat{Q}_2^D=e^{-{\epsilon_2\over n_G\tau}}, \quad
\widehat{Q}_\tau^D=e^{-{2\pi i \over n_G\tau}}, \quad
\widehat{Q}_m^D=e^{-{m\over n_G\tau}}.
\end{align}
In our parameter setting (\ref{eq691_setting}), one can check that \(|Q_{1,2}^D|=|\hat{Q}_{1,2}^D|=1\) and  \(|Q_\tau^D|,|\hat{Q}_\tau^D|\ll 1\). The other fugacities \(Q_m^D\) and \(\hat{Q}_m^D\) can also have extreme value depending on \(\text{Im}[m]\). For simplicity, let us assume that \(m\) is purely imaginary and is in the `canonical chamber' given as,
\begin{align}
0<\text{Im}[m]<2\pi \quad\to\quad Q_\tau^D \ll Q_m^D \ll 1 \ \& \ \widehat{Q}_\tau^D \ll \widehat{Q}_m^D \ll 1.
\label{eq1105_canch}
\end{align}
Then, it is straightforward to obtain the asymptotics of (\ref{eq1011_second}). Our basis of the elliptic genus is given in (\ref{eq108_rings}). Among those functions, we shall only focus on the ones with \(m\). After S-duality, their asymptotics are given as follows,
\begin{align}
A(-{1\over \tau},{m\over \tau})&=(Q_m^D)^{-1}\times
\Bigr(1+\mathcal{O}(Q_m^D,Q_\tau^D/Q_m^D) \Bigr)
\nonumber \\
B(-{1\over \tau},{m\over \tau})&=(Q_m^D)^{-1}\times
\Bigr(1+\mathcal{O}(Q_m^D,Q_\tau^D/Q_m^D) \Bigr)
\nonumber \\
A(-{1\over n_G\tau},{m\over \tau})&=(\widehat{Q}_m^D)^{-1}\times
\Bigr(1+\mathcal{O}(\widehat{Q}_m^D,\widehat{Q}_\tau^D/\widehat{Q}_m^D) \Bigr)
\nonumber \\
B(-{1\over n_G\tau},{m\over \tau})&=(\widehat{Q}_m^D)^{-1}\times
\Bigr(1+\mathcal{O}(\widehat{Q}_m^D,\widehat{Q}_\tau^D/\widehat{Q}_m^D) \Bigr).
\end{align}
For the index \(\mathfrak{i}_{\mathfrak{n}}\) of the \(Z_{\mathfrak{n}}\), a single \(A(\tau,m)\) or \(B(\tau,m)\) contributes \(m^2\), and a single \(A(n_G\tau,m)\) or \(B(n_G\tau,m)\) contributes \({m^2\over n_G}\). Since the \(m\)-dependent part in \(\mathfrak{i}_{\mathfrak{n}}\) is \(m^2 D\cdot \mathfrak{n}\), one can conclude the following asymptotics,
\begin{align}
{\mathcal{N}^{(a)}_{\mathfrak{n}}(-{1\over \tau},{z\over \tau})
\widehat{\mathcal{N}}^{(a)}_{\mathfrak{n}}(-{1\over n_G\tau},{z\over n_G\tau})
\over 
\mathcal{D}_{\mathfrak{n}}(-{1\over \tau},{z\over \tau})
\widehat{\mathcal{D}}_{\mathfrak{n}}(-{1\over n_G\tau},{z\over n_G\tau})}
=(Q_m^D)^{-D\cdot \mathfrak{n}}
\times \Bigr(1+\mathcal{O}(\widehat{Q}_m^D, \widehat{Q}_\tau^D/\widehat{Q}_m^D) \Bigr).
\label{eq1130_mod}
\end{align}
Now, we plug the above result into (\ref{eq1011_second}). Then, one can obtain the following Cardy limit asymptotics of the elliptic genus,
\begin{align}
\mathbb{E}_{\mathfrak{n}}(\tau,z)&=
\exp\Bigr[ -{1\over 2\pi i \tau}\Bigr(
{\epsilon_1 \epsilon_2\over 2}\mathfrak{n}^T(\Omega D)\mathfrak{n}
+[m(m-2\pi i)+\mathcal{O}(\epsilon^2)]\vec{D}\cdot \mathfrak{n}
\Bigr)+\mathcal{O}(\widehat{Q}_m^D, \widehat{Q}_\tau^D/\widehat{Q}_m^D) \Bigr]
\nonumber \\
&\sim 
\exp\Bigr[ -{1\over 2\pi i \tau}\Bigr(
{\epsilon_1 \epsilon_2\over 2}\mathfrak{n}^T(\Omega D)\mathfrak{n}
+[m(m-2\pi i) ]\vec{D}\cdot \mathfrak{n}\Bigr) \Bigr].
\label{eq1082_EGASY}
\end{align}

\par 
Next, let us consider the full 6d index. We shall focus on the non-Abelian part of the index given as follows,
\begin{align}
Z=\sum_{\mathfrak{n}}\mathbb{E}_{\mathfrak{n}}{\mathbf{Q}}^{\mathfrak{n}}.
\end{align}
The electric fugacity \(\mathbf{Q}=e^{-\mathbf{v}}\) probes the tensor branch moduli space of 6d theory. We will focus on the origin of the tensor branch, where conformal symmetry is restored. In the conformal phase, we should set \(\mathbf{v}=0\), i.e. \({\mathbf{Q}}=1\). By using (\ref{eq1082_EGASY}), the 6d index can be written as follows in the Cardy limit,
\begin{align}
Z\sim
\sum_{\mathfrak{n}}
\exp\Bigr[ -{1\over 2\pi i \tau}\Bigr(
{\epsilon_1 \epsilon_2\over 2}\mathfrak{n}^T(\Omega D)\mathfrak{n}
+[m(m-2\pi i) ]\sum_{a=1}^r D_{aa} \mathfrak{n}_a\Bigr) \Bigr].
\label{eq147_sumoa}
\end{align}
The summand has a Gaussian structure with respect to the string charge \(\mathfrak{n}\). Furthermore, the sign of the quadratic term in the exponent is negative, which guarantees the convergence of the summation.

\par 

The summation over \(\mathfrak{n}\) can be evaluated using the continuum approximation.  Let us consider a new variable below,
\begin{align}
\mathfrak{x}\equiv -\epsilon_1 \epsilon_2 D\mathfrak{n},
\quad
(x_a=-\epsilon_1 \epsilon_2 D_{aa} n_a ).
\label{eq1294_x}
\end{align}
Note that the minimal discrete distance \(\Delta\mathfrak{n}=1\) while \(\Delta\mathfrak{x}=-\epsilon_1 \epsilon_2 \ll 1\), and \(\mathfrak{x}\) can be though as an almost continuous variable. Then, the summation (\ref{eq147_sumoa}) can be approximated as the following integral,
\begin{align}
Z\sim 
\int_0^\infty d^r \mathfrak{x}
\exp\Bigr[
{1\over (-2\pi i \tau)\epsilon_1 \epsilon_2 }\Bigr(
{1\over 2}\mathfrak{x}^T(D^{-1}\Omega)\mathfrak{x}
-m(m-2\pi i)(\mathfrak{x}^T\cdot I)
\Bigr)\Bigr].
\label{eq1118_integrand}
\end{align}
Here, we ignored the Jacobian factor which is subleading than \({1\over \tau\epsilon_1 \epsilon_2}\) terms in the exponent. Also, we introduced \(r\)-dimensional identity vector \(I=(1,1,...,1)\). The free energy \(\log Z\) can be computed by using the saddle point approximation of \(\mathfrak{x}\). The saddle point equation and its solution are given by
\begin{align}
(D^{-1}\Omega)\mathfrak{x}-m(m-2\pi i)I=0 \quad\to\quad
\mathfrak{x}=m(m-2\pi i)(\Omega^{-1}D)I.
\label{eq1123_saddle}
\end{align}
By plugging the saddle point value (\ref{eq1123_saddle}) to (\ref{eq1118_integrand}), one can obtain the following 6d free energy,
\begin{align}
\log Z
\simeq -{I^T\cdot \Omega^{-1}D \cdot I\over 2} {m^2(2\pi i -m)^2\over (-2\pi i \tau)\epsilon_1 \epsilon_2 },
\end{align}
which is valid in the Cardy limit. For all Lie algebras including $(C_r)_{\pi}$, one can check that the overall coefficient is equivalent to the following group theoretic constant,
\begin{align}
I^T\cdot \Omega^{-1}D\cdot I={h_G^{\vee}  d_G\over 12},
\end{align}
where \(d_G\) is the dimension and \(h_G^{\vee} \) is the dual Coxeter number of \(G\). The values are listed in table \ref{tab1158_GTH}. As a conclusion, the Cardy free energy of 6d $(2,0)$ theory of type \(G\) is given as follows,
\begin{align}
\log Z
\sim -{h_{G}  d_G  \over 24} {m^2(2\pi i -m)^2\over (-2\pi i \tau)\epsilon_1 \epsilon_2 },
\label{eq1139_6adsfsdfs}
\end{align}
where the 6d theory is twisted if \(G\) is non-simply-laced. Our derivation of \eqref{eq1139_6adsfsdfs} solely depends on the index of the theory, for $C_r$ theory with different theta angles, they have the same index up to a redefinition of the string charge $\mathfrak{n}\rightarrow 
\frac{1}{2}\mathfrak{n}$. In the Cardy limit, the string charge is integrated out, such that at least at the leading order, the Cardy limit is the same for $(C_r)_0$ theory with different theta angles.  Interestingly, one can observe that the 6d Cardy free energy (\ref{eq1139_6adsfsdfs}) (without Abelian contribution) takes a universal form for all Lie algebras, even with the outer-automorphism twist. For the classical algebras \(A_r\), \(B_r\), \(C_r\), and \(D_r\), the free energy scales as \(\log Z \sim r^3\) when \(r\gg 1\). This reiterates the fact that \(N\) M5-branes have \(N^3\) degrees of freedom in the large \(N\) limit \cite{Klebanov:1996un}.

Furthermore, we can compare the Cardy free energy for theories before and after twisting. From the table \ref{tab1158_GTH}, one can see that  for all   the  Langlands dual pairs of Lie algebras \( ( G^{(n)},  G^{\vee(1)}) \) of the twisted compactification in the table \ref{tab:6d5d}, the following identity holds: 
\begin{equation}
    \frac{h_G^{\vee} \, d_G}{n_G} = h_{ G^\vee}^{\vee} \, d_{G^\vee}.
\end{equation}
Plugging it into \eqref{eq1139_6adsfsdfs}, we find that the Cardy free energy remains unchanged after the twist, provided that we take the change of radius into account. This nicely shows that in the large radius limit, the effect of twisting is not visible and the physics should remain the same.

\begin{table}[t!]
 \centering
  \arraycolsep=5pt \def\arraystretch{1.5}
 \begin{tabular}{|c||c|c|c|}\hline
 &  \(d_G\) & \(h_G^{\vee} \)   & \(|\vec\rho|^2\)\\ \hline \hline
\(A_r\)  & \((r+1)^2-1\) & \(r+1\) & \({1\over 12}r^3+{1\over 4}r^2+{1\over 6}r\) \\ \hline 
\(B_r\) &\(2r^2+r\) & \(2r-1\)  & \({1\over 3}r^3-{1\over 12}r\) \\ \hline 
\(C_r\) & \(2r^2+r\) & \(r+1\) & \({1\over 6}r^3+{1\over 4}r^2+{1\over 12}r\) \\ \hline 
\(D_r\)  &\(2r^2-r\) & \(2r-2\) &\({1\over 3}r^3-{1\over 2}r^2+{1\over 6}r\)\\ \hline 
\(G_2\) & 14 & 4 & \({14\over 3}\) \\ \hline 
\(F_4\) & 52 & 9 & 39 \\ \hline 
\(E_6\) & 78 & 12  & 78\\ \hline 
\(E_7\) & 133 & 18 & \({399\over 2}\) \\ \hline 
\(E_8\) & 248 & 30 & 620 \\ \hline 
 \end{tabular}
 \caption{Dimension, dual Coxeter number, square of Weyl vector of Lie algebras }
 \label{tab1158_GTH}
 \end{table}
 
\par 

In the computation above,  the non-zero value of the saddle point (\ref{eq1123_saddle}) has a special importance. Physically, it indicates that the string charge \(\mathfrak{n}\) condensates to a non-zero expectation value in the Cardy limit \cite{Lee:2020rns}. From the saddle point of \(\mathfrak{x}\) in (\ref{eq1123_saddle}), the string charge \(\mathfrak{n}\) in (\ref{eq1294_x}) can be written as follows,
\begin{align}
\langle n_a \rangle= {m(2\pi i -m)\over \epsilon_1 \epsilon_2} \sum_{b=1}^r (D^{-1}\Omega^{-1} D)_{ab}\,,
\end{align}
where \(\langle n_a \rangle\) denotes the value of \(n_a\) at the saddle point. The interesting fact is that the above expression is identical to the Weyl vector of the Lie algebra. To be specific, 
\begin{align}\label{eq:weylvector}
\sum_{b=1}^r (D^{-1} \Omega^{-1} D)_{ab} \vec{\alpha}_b = {1\over 2}\sum_{ \vec{\alpha}\in\Delta^+(G)} \vec{\alpha}=\vec{\rho}\,,
\end{align}
where \(\vec{\alpha}_b\) is the root vector of the \(b\)'th node of the Dynkin diagram of \(G\), and \(\Delta^+(G)\) is the positive root system of \(G\). For the $(C_r)_{\pi}$ theory, the last root vector should be replaced by half of the root vector, one can then verify that \eqref{eq:weylvector} still holds for the $(C_r)_{\pi}$ theory. The half of the sum over the positive root vectors \(\rho\) is called the Weyl vector of \(G\). As a result, for all 6d (2,0) theory with or without twist, the string charge \(\mathfrak{n}\) has the following value at the saddle point,
\begin{align}
\langle \mathfrak{n} \rangle ={m(2\pi i-m)\over \epsilon_1 \epsilon_2}\vec{\rho}\,.
\end{align}
As well as the free energy, the sum of the string charges \(\langle \sum_a n_a \rangle \sim |\vec{\rho}|^2\) also scales as \(r^3\) when \(r\gg 1\). The Weyl vector sum precisely explains the free energy, i.e., \(|\vec{\rho}|^2={h_G^\vee d_G\over 12}\). The values of \(|\vec{\rho}|^2\) are listed in table \ref{tab1158_GTH}.

With all the ingredients we have, we can now perform a similar computation for the asymptotic entropy as was done in \cite{Lee:2020rns}. The result is
\begin{align}
S\simeq 2\pi\sqrt{\sqrt{\frac{2}{3}h_G^{\vee}d_GP(J_-^2-(J_++Q_R)^2)}-Q_L^2}+2\pi i Q_L\,.
\label{G_CARDY}
\end{align}
An interesting remark is that since $h_G^{\vee}d_GP$ is a constant under twist operation, the entropy formula \eqref{G_CARDY} is the same for twisted and untwisted theories.
 \par

Lastly, let us make a brief comment on the periodicity of \(m\). The Cardy formula (\ref{eq1139_6adsfsdfs}) is obtained in the canonical chamber (\ref{eq1105_canch}) where \(0<\text{Im}[m]<2\pi \). One can do a similar computation outside of the canonical chamber, but it needs modification on the asymptotic behavior of (\ref{eq1130_mod}). For $ADE$-type theories, such computation was performed in \cite{Lee:2020rns, Duan:2020cta}, and the result restores the periodicity \(m\sim m+2\pi i\). We expect that the Cardy formulas for twisted $(2,0)$ theories are also periodic under \(m\sim m+2\pi i\).

 \par

 \section{Conclusions}\label{sec:conclusion}
In this paper, we studied 6d $(2,0)$ SCFTs on a circle with outer-automorphism twist and their BPS partition functions. The twisted $(2,0)$ theories are the UV fixed points of 5d $\mathcal{N} = 2$ SYM with non-simply-laced gauge groups: \(B_r\), \(C_r\), \(G_2\), and \(F_4\). We used blowup equations to obtain the BPS spectra and could determine the elliptic genera of twisted M-strings with modular bootstrap. Our results for the elliptic genera of twisted M-strings are completely new since there has been no known method to compute them. Lastly, we obtain universal 6d Cardy formulas for the twisted $(2,0)$ theories, which yields \(\log Z \sim (\text{rank})^3\) when the tensor branch rank is large.

\par 

In this paper, we also determine the modular property of the elliptic genera of twisted M-strings. If we impose outer-automorphism twist \(\mathbb{Z}_n\) on M-strings, its elliptic genus is not a modular form under \(\text{SL}(2,\mathbb{Z})\), but its congruence subgroup. The modular index of the twisted M-string can be obtained by folding the original untwisted M-string, as described in (\ref{eq385_folding}). This modular index was double-checked by comparing with the elliptic blowup equation (\ref{eq677_EBW}). As a result, we can bootstrap the twisted elliptic genus from the ring structure of \(\Gamma_0(N)\). 
 
 \par 
 
 There are several interesting topics that worth pursuing in the near future. The first and most obvious one is to understand the 2d quiver theory living on the twisted BPS string. In particular, for the (2,0) SCFTs of $A$ type one is supposed to perform certain twist or projection on the known 2d quiver \cite{Haghighat:2013gba}. However, at present it is not clear to the authors what are the correct rules, as the elliptic genera on two sides seem not related in a simple way.
 
 Also, one may consider the generalization of our results to 6d little string theories (LSTs) \cite{Aharony:1999ks}. The 6d LSTs are non-local quantum field theories and enjoy T-duality. Specifically, $(2,0)$ LSTs are T-dual to $(1,1)$ LSTs \cite{Kim:2015gha, Kim:2018gak}. $(2,0)$ LSTs have well-known $ADE$ classification, but one can impose an outer-automorphism twist on a T-dual circle. We speculate that the corresponding T-dual theory is a $(1,1)$ LST with a non-simply-laced gauge group. 
 It would be interesting if one can check the duality by explicitly computing the elliptic genera of 6d twisted little strings. Similarly, we can start with (1,1) LST with ADE group and compactify with an automorphism twist on T-dual circle.    The resulting (2,0) theories seem to lie beyond the ADE classification. It would be interesting whether these theories make any sense.

  Another interesting direction is to explore the blowup equation for the little string theories, in particular, the blowup equation for 6d $(1,1)$ LSTs. The main difference between LSTs and SCFTs is that there is an additional non-shrinkable circle, which may lead to non-local objects in the blowup equations. However, notice that a 6d $(1,1)$ little string theory can be regarded as an elliptic version of the 5d $\mathcal{N}=1^*$ theories, which means the blowup equation for $(1,1)$ little strings is simply an elliptic lift of \eqref{blowup5d3}. We hope to present it in the near future.

 \par 
 
Twisted strings in 6d $(1,0)$ SCFTs are also interesting objects to study. Twisted compactification of 6d SCFTs yields 5d KK theories, which are parents of large class of 5d SCFTs \cite{Bhardwaj:2019fzv}. It has been recently found that the blowup equations are strong methods to obtain the BPS spectra of those 5d theories on \(\mathbb{R}^4\times S^1\) \cite{Kim:2020hhh}. However, non-critical strings in twisted 6d SCFTs have not been studied extensively so far. We expect that the modular property of twisted strings in $(1,0)$ SCFTs can be also obtained from folding procedure similar to (\ref{eq385_folding}), and the modular bootstrap would be a useful tool to explore the physics on those strings.

 \acknowledgments
We are grateful to Amir-Kian Kashani-Poor, Hee-Cheol Kim, Sung-Soo Kim, Albrecht Klemm, Thorsten Schimannek and Kaiwen Sun for useful discussions. We would like to thank Amir-Kian Kashani-Poor and Thorsten Schimannek for their reading of the draft and providing inspiring comments. ZD, JN, KL and XW are  supported by KIAS Individual Grant PG076901, PG076401, PG006904 and QP079201 respectively. KL  is also supported   
in part by the National Research Foundation of Korea Grant NRF-2017R1D1A1B06034369.  
 
\newpage 
\appendix

\section{Modular forms}
\label{appendix_modular_forms}
In this appendix, we briefly summarize necessary results about the theory of modular forms \cite{Zagierbook, MR0344216, EZ}.
\subsection*{Elliptic Modular Forms}
\begin{mydef}
A function $f: \mathbb{H} \rightarrow \mathbb{C}$ is called a modular form of integer weight $k$ for a modular group $\Gamma$ if f is holomorphic on $\mathbb{H} \cup \{\infty\}$ and satisfies the following equation
\begin{equation}
    \label{modular}
f\left(\frac{a \tau + b}{c \tau + d}\right) = (c\tau + d)^{k} f(\tau),\ \mathrm{for\ any} \left(\begin{array}{cc}a&b\\c&d\end{array}\right)\in \Gamma 
\end{equation}
\end{mydef}
For example, let us consider the full modular group $\text{SL}(2, \mathbb{Z})$. We can choose the matrix $T =
\begin{pmatrix}
 1 & 1 \\
 0 & 1 \\
\end{pmatrix}$ to find that $f$ is periodic $f(\tau + 1) = f(\tau)$. It is standard to introduce $q = \exp (2\pi i \tau)$, such that  $f$ can be expanded as a power series in $q$,
\begin{equation}
f(\tau) = \sum\limits_{n \geq 0} a_{n} q^{n}\,.
\end{equation}
$f$ being holomorphic yields $n \geq 0$. If furthermore $a_{0} = 0$, $f$ is called a $\mathit{cusp}$ form.

Modular forms naturally form a ring $M_\ast(\Gamma) $, graded by the weight. The important result is that such space is always finitely generated. For the full modular group $\text{SL}(2, \mathbb{Z})$, $M_\ast(\text{SL}(2, \mathbb{Z})) $ is freely generated by $E_4$ and $E_6$ \cite{Zagierbook, MR0344216}, where $E_k$ is the famous Eisenstein series,

\begin{equation}
    E_{k}(\tau) = \frac{1}{\zeta(k)}\sum_{\substack{(m, n)\in \mathbb{Z}^2\\(m,n) \neq (0,0)}} \frac{1}{(m + n \tau)^k}\,, \quad
        E_k|_{q = 0} = 1 - \frac{2k}{B_k}\sum\limits_{n = 1}^{\infty}\sigma_{k - 1}(n)\, q^n\,,
\end{equation}
with $B_k$ the $k^{\text{th}}$ Bernoulli number and $\sigma_{i}(n)$ the sum of the $i^\text{th}$ powers of the positive divisors of $n$.

Moreover, in this paper it is equally important to consider congruence subgroups $\Gamma_0(N)$ of $\text{SL}(2, \mathbb{Z})$, defined as follows,
\begin{align}
\Gamma_0(N)=\Bigr\{
\begin{pmatrix}
a & b \\ c & d
\end{pmatrix} \in \text{SL}(2,\mathbb{Z}): c \equiv 0\ (\text{mod}\ N)
\Bigr\}\,.
\end{align}
As a special case, $\Gamma_0(1) = \text{SL}(2, \mathbb{Z})$. The ring of modular forms for $\Gamma_0(N)$ will be denoted simply as $M_\ast(N)$. Since the group is smaller, it is possible to have more generators. We introduce
\begin{equation}
    E_{2}^{(N)}(\tau) = -\frac{24}{N-1} \frac{1}{2\pi i}\partial_\tau  \log\left(\frac{\eta(\tau)}{\eta(N\tau)}\right)\,,
\end{equation}
where $\eta(\tau)$ is the Dedekind eta function
$\eta(\tau) = q^\frac{1}{24}\prod\limits_{i = 1}^{\infty}(1 - q^i)$. It is not difficult to show that $E_{2}^{(N)}$ is an element of weight two in $M_\ast(N)$. In the main text, we need to consider $N = 2,3,4$, for which we have the following result,
\begin{equation}
    \begin{aligned}
    M_\ast(2) &= \langle E_{2}^{(2)}(\tau),E_4(\tau)\rangle\,,\\
    M_\ast(3) &= \langle E_{2}^{(3)}(\tau),E_4(\tau),E_6(\tau)\rangle\,,\\
    M_\ast(4) &= \langle E_{2}^{(2)}(\tau),E_{2}^{(4)}(\tau),E_4(\tau),E_6(\tau)\rangle\,.
    \end{aligned}
\end{equation}
However, in actual computations, we find it enough in mostly of the cases to use the modular forms $\widehat{E}_k(\tau) := E_k(N\tau)$. It is not difficult to show that $\widehat{E}_k(\tau)$ indeed belongs to $M_\ast(N)$. 

\subsection*{Jacobi Modular Forms}
\label{appendixJMF}
\begin{mydef}\label{defjmf}
A Jacobi modular form of weight $k$ and index $m$ for a modular group $\Gamma$ is a function $\phi:\mathbb{H}\times \mathbb{C}\rightarrow \mathbb{C}$ that depends on a modular 
	parameter $\tau\in \mathbb{H}$ and an elliptic parameter $z\in \mathbb{C}$. It transforms 
	under the action of $\Gamma$ on $\mathbb{H} \times \mathbb{C}$ as
    \begin{equation}
	\tau \mapsto \tau_\gamma=\frac{a \tau+ b}{c \tau + d}, \quad z \mapsto z_\gamma=\frac{z}{c \tau + d}\quad  {\rm with} \quad  
	\left(\begin{array}{cc}a&b\\ c& d \end{array}\right) \in \Gamma  \ ,
	\label{PSL2} 
    \end{equation} 
	as  
	\begin{align}\label{modtran} 
    \phi_{k,m}\left(\tau_\gamma, z_\gamma\right) &= (c \tau + d)^k e^{\frac{2 \pi i m c z^2}{c \tau + d}} \phi_{k,m}(\tau,z)\,, \\
	\phi_{k,m}(\tau,z +\lambda \tau+ \mu) &= e^{- 2 \pi i m (\lambda^2 \tau+ 2 \lambda z)}\phi_{k,m}(\tau,z) \quad \forall  \lambda, \mu \in \mathbb{Z}\,.\label{elltrans}
	\end{align}
\end{mydef}

Let us first consider $\Gamma$ to be $\text{SL}(2, \mathbb{Z})$. Choosing $T = \begin{pmatrix} 1 & 1\\
0 & 1\\
\end{pmatrix}$ and $\lambda = 0, \mu = 1$ in equations (\ref{modtran}) and (\ref{elltrans}) respectively we learn that the Jacobi form is invariant under both shifts $\tau \rightarrow \tau + 1$ and $z\rightarrow z + 1$, so it's legitimate to write down the expansion,
\begin{equation} 
	\phi(\tau,z)_{k,m}=\sum_{n,r} c(n,r) q^n y^r\,, \quad {\rm with} \ \ q=e^{2 \pi i \tau},\ \  y=e^{2 \pi i z} \, .
\end{equation}
The coefficient $c(n,r)$ actually only depends on $r$ and an $\text{SL}(2, \mathbb{Z})$ invariant combination $4nm - r^2$. We are interested in the so-called weak Jacobi modular forms, which satisfy $c(n,r) = 0$ unless $n \geq 0$. We denote its ring as $J^{w}_{\ast,\ast}(\Gamma)$.

It is shown in \cite{EZ} that the ring of weak Jacobi forms $J^{w}_{\ast,\ast}(\text{SL}(2, \mathbb{Z}))$ with even weight and integral index is freely generated over $M_\ast(\text{SL}(2, \mathbb{Z}))$ by two generators $\varphi_{-2,1} (\tau,z)$ and $\varphi_{0,1} (\tau,z)$. They can be defined in terms of Jacobi theta functions,
\begin{equation}
\begin{aligned}
    A(\tau, z) &\equiv \varphi_{-2,1}(\tau,z) = \frac{ \theta_1(\tau,z)^2} {\eta(\tau)^6} \, , \\ 
	B(\tau, z) &\equiv \varphi_{0,1}(\tau,z) = 4\Big(\frac{ \theta_2(\tau,z)^2} {  \theta_2(\tau, 0)^2 } +\frac{ \theta_3(\tau,z)^2} {\theta_3(\tau, 0)^2 } +\frac{ \theta_4(\tau,z)^2} {  \theta_4(\tau, 0)^2 } \Big). 
\end{aligned}
\end{equation}

By the same token, we also need to consider the Jacobi modular forms for the congruence subgroups $J^{w}_{\ast,\ast}(\Gamma_0(N))$. For us, the most important class of examples is the Jacobi forms $\phi_{k,m}(N\tau,z)$ where $\phi_{k,m}(\tau,z)$ belongs to $J^{w}_{\ast,\ast}(\text{SL}(2, \mathbb{Z}))$. For brevity we just denote it by $\widehat{\phi}_{k,m}$, with $N$ understood from the context. Also note that the actual index of $\phi_{k,m}(\tau,z)$ is reduced to $m/N$, as is shown explicitly in the section $\ref{subsec:4.1}$. 

In the final part, let us show the identities among the generators that are used in our modular ansatz, for $N = 2,3$. For modular forms involving $\tau$ and $2\tau$, we find the following relations,
\begin{equation}\label{App:MFidentites}
\begin{aligned}
\widehat{E}_{4} &= \frac{5}{4}\left(E^{(2)}_{2}\right)^2 -\frac{1}{4}E_4\,,\\
\widehat{E}_{6} &=  \frac{7}{8}\left(E^{(2)}_{2}\right)^3 +\frac{1}{8} E_6\,,\\
A &= \frac{1}{12} \widehat{A} \widehat{B} + \frac{1}{12} E^{(2)}_{2} \widehat{A}^2\,,\\
B &= \frac{1}{12} \widehat{B} \left(E^{(2)}_2 \widehat{A}+\widehat{B}\right)-\frac{1}{16}\widehat{A}^2 \left(\left(E^{(2)}_2\right)^2-E_4\right) \,.\\
\end{aligned}
\end{equation}
In order to prove them, we need to invoke the following known identities involving the Jacobi theta functions,
\begin{equation}
\begin{aligned}
\eta(\tau)^3 &=  \frac{1}{2}(\theta_2\theta_3\theta_4) \,,\\
\theta_3^4 &= \theta_2^4 + \theta_4^4\,,\\
E^{(2)}_{2} &= \frac{1}{2}\theta_3^4 + \frac{1}{2}\theta_4^4\,,\\
E_4 &= \frac{1}{2}(\theta_2^8 + \theta_3^8 + \theta_4^8)\,,\\
E_6 &= \frac{1}{2}(-3\theta_2^8(\theta_3^4+\theta_4^4) + \theta_3^{12} + \theta_4^{12})\,,\\
\widehat{\theta}_2^2 &= \frac{1}{2}\theta_3^2 - \frac{1}{2}\theta_4^2\,,\\
\widehat{\theta}_3^2 &= \frac{1}{2}\theta_3^2 + \frac{1}{2}\theta_4^2\,,\\
\widehat{\theta}_4^2 &= \theta_3\theta_4\,,\\
\widehat{\theta}_1(z)\widehat{\theta}_4(z) &= \frac{1}{2}   \theta_2\theta_1(z)\,, \\
\widehat{\theta}_1(z)^2 + \widehat{\theta}_4(z)^2 &= \theta_3\theta_4(z)\,.
\end{aligned}
\end{equation}

As an example, we present here a proof of the last equality in \eqref{App:MFidentites}.
\begin{equation}
\begin{aligned}
\text{RHS} &= \frac{A}{\widehat{A}}\widehat{B} + \frac{3}{64}\theta_{2}^8 \widehat{A}^2\,,\\
&= \frac{A}{\widehat{A}} \left[ \widehat{A}\left(3\widehat{\theta}_3^2\widehat{\theta}_2^2\frac{\widehat{\theta}_4^2(z)}{\widehat{\theta}_1^2(z)} -  \widehat{\theta}_3^4-\widehat{\theta}_2^4\right)\right] + \frac{3}{64}\theta_{2}^8 \frac{\widehat{\theta}_1^4(z)}{\widehat{\eta}^{12}}\,,\\
&=A \left[\frac{3}{4}\theta_2^{4} \left(\frac{\theta_3 \theta_4(z)}{\widehat{\theta}_1^{2}(z)} - 1\right) -\theta_3^4 + \frac{1}{2}\theta_2^4\right] + \frac{3}{64}\theta_{2}^8 \frac{\widehat{\theta}_1^4(z)}{\widehat{\eta}^{12}}\,,\\
&=A \left[\frac{3}{4}\theta_2^{4} \left(\frac{4\theta_3^2\theta_4^2(z) - 4\theta_3\theta_4(z)\widehat{\theta}_1^2(z)}{\theta_1^2(z)\theta_2^2} -1\right) -\theta_3^4 + \frac{1}{2}\theta_2^4\right] + 12\frac{\widehat{\theta}_1^4(z)}{\theta_3^2\theta_4^2}\,,\\
&= A\left(3\frac{\theta_3^2\theta_3^2\theta_4^2(z)}{\theta_1^2(z)} - \theta_3^4 - \frac{1}{4}\theta_2^4\right) - 12 \frac{\theta_4(z)\widehat{\theta}_1^2(z)}{\theta_3\theta_4^2}+ 12\frac{\widehat{\theta}_1^2(z)(\theta_3\theta_4(z) - \widehat{\theta}_4^2(z))}{\theta_3^2\theta_4^2}\,,\\
&=A\left(3\frac{\theta_3^2\theta_3^2\theta_4^2(z)}{\theta_1^2(z)} - \theta_3^4 - \frac{1}{4}\theta_2^4\right) -3\frac{\theta_2^2\theta_1^2(z)}{\theta_3^2\theta_4^2}\,,\\
&= A\left(3\frac{\theta_3^2\theta_3^2\theta_4^2(z)}{\theta_1^2(z)} - \theta_3^4 - \frac{1}{4}\theta_2^4\right) - \frac{3}{4}A\theta_2^4\,,\\
&= \text{LHS}\,.
\end{aligned}
\end{equation}

For modular forms involving $\tau$ and $3\tau$, the situation becomes more complicated. Experimentally, we find the following identities which we verify up to very high orders in $q$,
\begin{equation}
\begin{aligned}
\widehat{E}_4 &= \frac{1}{9} \left(10 \left(\esan\right)^2-E_4\right)\,,\\
\widehat{E}_6 &= \frac{1}{27} \left(35 \left(\esan\right)^3-7 E_4 \esan-E_6\right)\,,\\
A &= \frac{1}{144}\widehat{A}\left(\left(\esan\right)^2\widehat{A}^2 + 2\esan\widehat{A}\widehat{B} + \widehat{B}^2\right)\,,\\
B &= -\frac{1}{1944}\widehat{A}^3{\left(23 \left(\esan\right)^3-19 E_4 \esan-4 E_6\right) }\,\\
&\ \ \ + \frac{1}{432}\widehat{B}\left[\left(\esan\right)^2\widehat{A}^2 + 2E_4 \widehat{A}^2 + 6\esan\widehat{A}\widehat{B} + 3\widehat{B}^2\right]\,.\\
\end{aligned}
\end{equation}

\clearpage

\section{Elliptic genera and BPS invariants}\label{App:EG}

In this section, we present the numerator of the elliptic genera for the following twisted M-strings: \(B_2\) in table \ref{EG_B2}, \(G_2\) in table \ref{EG_G2}, \(B_3\) in table \ref{EG_B3}, \(( C_3)_{0} \) in table \ref{EG_C3}, and \(F_4\) in table \ref{EG_F4}. The elliptic genera of \((C_1)_{\pi}\) and \((C_2)_{\pi}\) theories is listed in table \ref{tab655_thetapi}. We also list some refined GV invariants of geometries that engineer 5d $\mathcal{N} = 2$ gauge theory of type $G_2$ and $F_4$, obtained from the blowup equations or the elliptic genera, in table \ref{tab:my_label} and \ref{tab:my_label2}.


\begin{table}
\centering
\begin{tabular}{|c||c|}\hline
\(\mathfrak{n}\) & \(\mathcal{N}\) \\ \hline\hline
\dynkin[labels*={1,0}]B2 &
\( {1\over 12}(A_m B_+ - A_+ B_m ) \)  \\ \hline
\dynkin[labels*={0,1}]B2 & 
\({1\over 12}(\widehat{A}_m \widehat{B}_+ - \widehat{A}_+ \widehat{B}_m ) \) \\ \hline
\dynkin[labels*={1,1}]B2 & 
\({1\over 144}( A_m B_- -A_- B_m )(\widehat{A}_m \widehat{B}_+ - \widehat{A}_+ \widehat{B}_m ) \) \\ \hline
\dynkin[labels*={2,1}]B2 & 
\begingroup\renewcommand*{\arraystretch}{1.5}%
\resizebox{13.7cm}{!}{
$\begin{array}{ll} 
&{1\over 2^{17}3^9 }(A_m B_+-A_+ B_m)(\widehat{A}_m \widehat{B}_+-\widehat{A}_+ \widehat{B}_m)
\Bigr(
81 A_-^2 A_+^4 B_- E_4^3 A_m-27 A_-^3 A_+^4 E_4^3 B_m-9 A_+^4 B_-^3 \
E_4^2 A_m-9 A_- A_+^4 B_-^2 E_4^2 B_m-96 A_-^2 A_+^4 B_- E_6^2 A_m
\nonumber \\
&+32 \
A_-^3 A_+^4 E_6^2 B_m-4 A_+^4 B_-^3 E_6 B_m-24 A_- A_+^4 B_-^2 E_4 \
E_6 A_m-54 A_-^3 A_+^3 B_+ E_4^3 A_m-54 A_- A_+^3 B_-^2 B_+ E_4^2 \
A_m+18 A_-^2 A_+^3 B_- B_+ E_4^2 B_m+64 A_-^3 A_+^3 B_+ E_6^2 A_m-6 \
A_+^3 B_-^3 B_+ E_4 B_m-20 A_+^3 B_-^3 B_+ E_6 A_m
\nonumber \\
&-12 A_- A_+^3 B_-^2 \
B_+ E_6 B_m-24 A_-^2 A_+^3 B_- B_+ E_4 E_6 A_m+24 A_-^3 A_+^3 B_+ E_4 \
E_6 B_m+36 A_-^3 A_+^2 B_+^2 E_4^2 B_m-12 A_+^2 B_-^3 B_+^2 E_4 \
A_m-36 A_- A_+^2 B_-^2 B_+^2 E_6 A_m
\nonumber \\
&+36 A_-^2 A_+^2 B_- B_+^2 E_6 \
B_m+24 A_-^3 A_+^2 B_+^2 E_4 E_6 A_m+18 A_-^3 A_+ B_+^3 E_4^2 A_m-6 \
A_- A_+ B_-^2 B_+^3 E_4 A_m+18 A_-^2 A_+ B_- B_+^3 E_4 B_m+12 A_-^2 \
A_+ B_- B_+^3 E_6 A_m+20 A_-^3 A_+ B_+^3 E_6 B_m
\nonumber \\
&+3 A_-^2 B_- B_+^4 \
E_4 A_m+3 A_-^3 B_+^4 E_4 B_m+4 A_-^3 B_+^4 E_6 A_m+2 A_+ B_-^3 B_+^3 \
B_m+B_-^3 B_+^4 A_m-3 A_- B_-^2 B_+^4 B_m
\Bigr)
\end{array}$}
    \endgroup
\\
\hline
\end{tabular}
\caption{Elliptic genera of \(B_2\). Here, \(A_z=A_z(\tau)\), \(\widehat{A}_z=A_z(2\tau)\), and same for \(B_z\).}
\label{EG_B2}
\end{table}

\begin{table}
\centering
\begin{tabular}{|c||c|}\hline
\(\mathfrak{n}\) & \(\mathcal{N}\) \\ \hline\hline
\dynkin[reverse arrows, labels*={0,1}]G2 &
\({1\over 12}(A_m B_+ - A_+ B_m ) \)  \\ \hline
\dynkin[reverse arrows, labels*={1,0}]G2 & 
\({1\over 12}(\widehat{A}_m \widehat{B}_+ - \widehat{A}_+ \widehat{B}_m ) \) \\ \hline
\dynkin[reverse arrows, labels*={1,1}]G2 & 
\({1\over 144}( A_m B_- -A_- B_m )(\widehat{A}_m \widehat{B}_+ - \widehat{A}_+ \widehat{B}_m ) \) \\
\hline
\end{tabular}
\caption{Elliptic genera of \(G_2\). Here, \(A_z=A_z(\tau)\), \(\widehat{A}_z=A_z(3\tau)\), and same for \(B_z\).}
\label{EG_G2}
\end{table}

\begin{table}
\centering
\begin{tabular}{|c||c|}\hline
\(\mathfrak{n}\) & \(\mathcal{N}\) \\ \hline\hline
\dynkin[labels*={1,1,1}]B3 &
\(
{1\over 2^6 3^3}(A_m B_- - A_- B_m)^2(\widehat{A}_+ \widehat{B}_m -\widehat{A}_m \widehat{A}_+)
\)
\\ \hline
\dynkin[labels*={2,1,1}]B3 & 
\begingroup\renewcommand*{\arraystretch}{1.5}%
\resizebox{13.7cm}{!}{
$\begin{array}{ll} 
&{1\over 2^{19} 3^{10}}(A_m B_+ - A_+ B_m)(A_m B_- - A_- B_m)
(\widehat{A}_m \widehat{B}_+ -\widehat{A}_+ \widehat{B}_m)\times \Bigr(
81 A_-^2 A_+^4 B_- E_4^3 A_m-27 A_-^3 A_+^4 E_4^3 B_m-9 A_+^4
   B_-^3 E_4^2 A_m-9 A_- A_+^4 B_-^2 E_4^2 B_m-96 A_-^2 A_+^4
   B_- E_6^2 A_m+32 A_-^3 A_+^4 E_6^2 B_m+4 A_+^4 B_-^3 E_6
   B_m
   \nonumber \\
   &+24 A_- A_+^4 B_-^2 E_4 E_6 A_m-54 A_-^3 A_+^3 B_+
   E_4^3 A_m-54 A_- A_+^3 B_-^2 B_+ E_4^2 A_m+18 A_-^2 A_+^3
   B_- B_+ E_4^2 B_m+64 A_-^3 A_+^3 B_+ E_6^2 A_m-6 A_+^3
   B_-^3 B_+ E_4 B_m+20 A_+^3 B_-^3 B_+ E_6 A_m+12 A_- A_+^3
   B_-^2 B_+ E_6 B_m+24 A_-^2 A_+^3 B_- B_+ E_4 E_6 A_m-24
   A_-^3 A_+^3 B_+ E_4 E_6 B_m
   \nonumber \\
   &+36 A_-^3 A_+^2 B_+^2 E_4^2
   B_m-12 A_+^2 B_-^3 B_+^2 E_4 A_m+36 A_- A_+^2 B_-^2 B_+^2
   E_6 A_m-36 A_-^2 A_+^2 B_- B_+^2 E_6 B_m-24 A_-^3 A_+^2
   B_+^2 E_4 E_6 A_m+18 A_-^3 A_+ B_+^3 E_4^2 A_m-6 A_- A_+
   B_-^2 B_+^3 E_4 A_m+18 A_-^2 A_+ B_- B_+^3 E_4 B_m-12
   A_-^2 A_+ B_- B_+^3 E_6 A_m
   \nonumber \\
   &-20 A_-^3 A_+ B_+^3 E_6 B_m+3
   A_-^2 B_- B_+^4 E_4 A_m+3 A_-^3 B_+^4 E_4 B_m-4 A_-^3
   B_+^4 E_6 A_m+2 A_+ B_-^3 B_+^3 B_m+B_-^3 B_+^4 A_m-3 A_-
   B_-^2 B_+^4 B_m
\Bigr)
\end{array}$}
    \endgroup 
    \\
\hline
\end{tabular}
\caption{Elliptic genera of \(B_3\). Here, \(A_z=A_z(\tau)\), \(\widehat{A}_z=A_z(2\tau)\), and same for \(B_z\).}
\label{EG_B3}
\end{table}

\begin{table}
\centering
\begin{tabular}{|c||c|}\hline
\(\mathfrak{n}\) & \(\mathcal{N}\) \\ \hline\hline
\dynkin[labels*={1,1,1}]C3 & 
\(
{1\over 2^6 3^3}(A_- B_m -A_m B_-)
(\widehat{A}_- \widehat{B}_m - \widehat{A}_m \widehat{B}_-)(\widehat{A}_+ \widehat{B}_m -\widehat{A}_m \widehat{B}_+)
\)
\\ \hline
\dynkin[labels*={1,1,2}]C3 & 
\begingroup\renewcommand*{\arraystretch}{1.5}%
\resizebox{13.7cm}{!}{
$\begin{array}{ll} 
& {1\over 2^{19}3^{10}}(A_m B_+ - A_+ B_m)
(\widehat{A}_m \widehat{B}_+ -\widehat{A}_+ \widehat{B}_m)(\widehat{A}_m \widehat{B}_- -\widehat{A}_- \widehat{B}_m)
\Bigr(81 A_-^2 A_+^4 B_- E_4^3 A_m-27 A_-^3 A_+^4 E_4^3 B_m-9 A_+^4
   B_-^3 E_4^2 A_m-9 A_- A_+^4 B_-^2 E_4^2 B_m-96 A_-^2 A_+^4
   B_- E_6^2 A_m+32 A_-^3 A_+^4 E_6^2 B_m+4 A_+^4 B_-^3 E_6
   B_m+24 A_- A_+^4 B_-^2 E_4 E_6 A_m
   \nonumber \\
   &-54 A_-^3 A_+^3 B_+
   E_4^3 A_m-54 A_- A_+^3 B_-^2 B_+ E_4^2 A_m+18 A_-^2 A_+^3
   B_- B_+ E_4^2 B_m+64 A_-^3 A_+^3 B_+ E_6^2 A_m-6 A_+^3
   B_-^3 B_+ E_4 B_m+20 A_+^3 B_-^3 B_+ E_6 A_m+12 A_- A_+^3
   B_-^2 B_+ E_6 B_m+24 A_-^2 A_+^3 B_- B_+ E_4 E_6 A_m-24
   A_-^3 A_+^3 B_+ E_4 E_6 B_m
   \nonumber \\
   &+36 A_-^3 A_+^2 B_+^2 E_4^2
   B_m-12 A_+^2 B_-^3 B_+^2 E_4 A_m+36 A_- A_+^2 B_-^2 B_+^2
   E_6 A_m-36 A_-^2 A_+^2 B_- B_+^2 E_6 B_m-24 A_-^3 A_+^2
   B_+^2 E_4 E_6 A_m+18 A_-^3 A_+ B_+^3 E_4^2 A_m-6 A_- A_+
   B_-^2 B_+^3 E_4 A_m+18 A_-^2 A_+ B_- B_+^3 E_4 B_m-12
   A_-^2 A_+ B_- B_+^3 E_6 A_m
   \nonumber \\
   &-20 A_-^3 A_+ B_+^3 E_6 B_m+3
   A_-^2 B_- B_+^4 E_4 A_m+3 A_-^3 B_+^4 E_4 B_m-4 A_-^3
   B_+^4 E_6 A_m+2 A_+ B_-^3 B_+^3 B_m+B_-^3 B_+^4 A_m-3 A_-
   B_-^2 B_+^4 B_m
   \Bigr)
\end{array}$}
    \endgroup 
\\ \hline
\end{tabular}
\caption{Elliptic genera of \( (C_3)_{0} \). Here, \(A_z=A_z(\tau)\), \(\widehat{A}_z=A_z(2\tau)\), and same for \(B_z\).}
\label{EG_C3}
\end{table}

\begin{table}
\centering
\begin{tabular}{|c||c|}\hline
\(\mathfrak{n}\) & \(\mathcal{N}\) \\ \hline\hline
\dynkin[labels*={1,1,1,1}]F4 & 
\({1\over 2^8 3^4}
(\widehat{A}_m \widehat{B}_+ - \widehat{A}_+ \widehat{B}_m)
(\widehat{A}_m \widehat{B}_-  - \widehat{A}_- \widehat{B}_m)
(A_m B_- -A_- B_m)^2\)
\\ \hline
\dynkin[labels*={2,1,1,1}]F4 &
\begingroup\renewcommand*{\arraystretch}{1.5}%
\resizebox{13.7cm}{!}{
$\begin{array}{ll} 
&{1\over 2^{21} 3^{11}}
(A_m B_- -A_- B_m)(A_m B_+ -A_+ B_m)
(\widehat{A}_m \widehat{B}_- -\widehat{A}_- \widehat{B}_m)(\widehat{A}_m \widehat{B}_+ -\widehat{A}_+ \widehat{B}_m)
\times \Bigr(
81 A_-^2 A_+^4 B_- E_4^3 A_m-27 A_-^3 A_+^4 E_4^3 B_m-9 A_+^4
   B_-^3 E_4^2 A_m-9 A_- A_+^4 B_-^2 E_4^2 B_m
   \nonumber \\
   &-96 A_-^2 A_+^4
   B_- E_6^2 A_m+32 A_-^3 A_+^4 E_6^2 B_m-4 A_+^4 B_-^3 E_6
   B_m-24 A_- A_+^4 B_-^2 E_4 E_6 A_m-54 A_-^3 A_+^3 B_+
   E_4^3 A_m-54 A_- A_+^3 B_-^2 B_+ E_4^2 A_m+18 A_-^2 A_+^3
   B_- B_+ E_4^2 B_m+64 A_-^3 A_+^3 B_+ E_6^2 A_m
   \nonumber \\
   &-6 A_+^3
   B_-^3 B_+ E_4 B_m-20 A_+^3 B_-^3 B_+ E_6 A_m-12 A_- A_+^3
   B_-^2 B_+ E_6 B_m-24 A_-^2 A_+^3 B_- B_+ E_4 E_6 A_m+24
   A_-^3 A_+^3 B_+ E_4 E_6 B_m+36 A_-^3 A_+^2 B_+^2 E_4^2
   B_m-12 A_+^2 B_-^3 B_+^2 E_4 A_m-36 A_- A_+^2 B_-^2 B_+^2
   E_6 A_m+36 A_-^2 A_+^2 B_- B_+^2 E_6 B_m
   \nonumber \\
  & +24 A_-^3 A_+^2
   B_+^2 E_4 E_6 A_m+18 A_-^3 A_+ B_+^3 E_4^2 A_m-6 A_- A_+
   B_-^2 B_+^3 E_4 A_m+18 A_-^2 A_+ B_- B_+^3 E_4 B_m+12
   A_-^2 A_+ B_- B_+^3 E_6 A_m+20 A_-^3 A_+ B_+^3 E_6 B_m+3
   A_-^2 B_- B_+^4 E_4 A_m+3 A_-^3 B_+^4 E_4 B_m
   \nonumber \\
   &+4 A_-^3
   B_+^4 E_6 A_m+2 A_+ B_-^3 B_+^3 B_m+B_-^3 B_+^4 A_m-3 A_-
   B_-^2 B_+^4 B_m
   \Bigr)
\end{array}$}
    \endgroup 
\\
\hline 
\end{tabular}
\caption{Elliptic genera of \(F_4\). Here, \(A_z=A_z(\tau)\), \(\widehat{A}_z=A_z(2\tau)\), and same for \(B_z\).}
\label{EG_F4}
\end{table}

\begin{table}
\centering
\begin{tabular}{|c||c|}\hline
\(\mathfrak{n}\) & \(\mathcal{N}\) \\ \hline\hline 
\dynkin[labels*={1}]A1 & \(\frac{1}{2^8 3^2} ( \widehat{A}_m \widehat{B}_+ - \widehat{A}_+ \widehat{B}_m)
(\widehat{A}_m \widehat{B}_- - \widehat{A}_- \widehat{B}_m)
(E_2^{(2)}-E_2^{(4)}) \)\\ \hline
\dynkin[labels*={1,0}]C2 & \({1\over 12}(A_m B_+ - A_+ B_m) \)\\ \hline
\dynkin[labels*={0,1}]C2 &  \({\frac{1}{2^8 3^2}}( \widehat{A}_m \widehat{B}_+ - \widehat{A}_+ \widehat{B}_m)
(\widehat{A}_m \widehat{B}_- - \widehat{A}_- \widehat{B}_m)
(E_2^{(2)}-E_2^{(4)}) \) \\ \hline
\dynkin[labels*={1,1}]C2 & \({\frac{1}{2^{10} 3^3}}(A_m B_- - A_- B_m)( \widehat{A}_m \widehat{B}_+ - \widehat{A}_+ \widehat{B}_m)
(\widehat{A}_m \widehat{B}_- - \widehat{A}_- \widehat{B}_m)
(E_2^{(2)}-E_2^{(4)}) \)\\  \hline 
\multicolumn{1}{c}{} & \multicolumn{1}{c}{}\\
\hline  
\(\mathfrak{n}\) & \(\mathcal{I}(\tau,\ep_1,\ep_2)\) \\
\hline
\end{tabular}
\caption{Elliptic genera of \( (C_1)_{\pi}\) and \( (C_2)_{\pi}\). Here, \(A_z=A_z(2\tau)\), \(\widehat{A}_z=A_z(4\tau)\), and same for \(B_z\). And $E_{2,2}=E_{2}^{(2)}(\tau),\, E_{2,4}=E_{2}^{(4)}(\tau)$. (The definition of $\mathcal{I}(\tau,\ep_1,\ep_2)$ can be found in section \ref{sec:thetapi}.)}
\label{tab655_thetapi}
\end{table}

\begin{table}[ht]
    \centering
    \begin{tabular}{|c|cccccc|}\hline
         $\{(1,2),2,2\}$&$2j_+=$0&1&2&3&4&5  \\\hline
         $2j_-=$0&&16&&15&&1\\
         1&6&&18&&8&\\
         2&&3&&6&&2\\
         3&0&&0&&1&\\
         \hline
    \end{tabular}

     \begin{tabular}{|c|ccccc|}\hline
         $\{(1,2),2,3\}$&$2j_+=$0&1&2&3&4  \\\hline
         $2j_-=$0&3&&7&&2\\
         1&&4&&5&\\
         2&0&&1&&1\\
         \hline
    \end{tabular}    

     \begin{center}
\begin{tabular}{|c|cccccccc|}\hline
         $\{(1,2),3,2\}$&$2j_+=$0&1&2&3&4&5&6&7  \\\hline
         $2j_-=$0&&97&&98&&17&&0\\
         1&52&&129&&73&&7&\\
         2&&47&&72&&31&&1\\
         3&3&&15&&25&&8&\\
         4&&0&&3&&6&&2\\
         5&0&&0&&0&&1&\\
         \hline
    \end{tabular}  
    
     \begin{tabular}{|c|ccccccc|}\hline
         $\{(1,2),3,3\}$&$2j_+=$0&1&2&3&4&5&6  \\\hline
         $2j_-=$0&19&&50&&19&&0\\
         1&&39&&44&&9&\\
         2&7&&22&&19&&2\\
         3&&1&&6&&5&\\
         4&0&&0&&1&&1\\
         \hline
    \end{tabular}
    \end{center}
     \begin{tabular}{|c|cccccccccc|}\hline
         $\{(1,2),4,2\}$&$2j_+=$0&1&2&3&4&5&6&7&8&9  \\\hline
         $2j_-=$0&&486&&510&&141&&7&&0\\
         1&206&&719&&467&&81&&1&\\
         2&&343&&490&&245&&28&&0\\
         3&54&&179&&229&&96&&7&\\
         4&&18&&62&&82&&28&&1\\
         5&0&&3&&15&&25&&8&\\
         6&&0&&0&&3&&6&&2\\
         7&0&&0&&0&&0&&1&\\
         \hline
    \end{tabular}
    
     \begin{tabular}{|c|ccccccccc|}\hline
         $\{(1,2),4,3\}$&$2j_+=$0&1&2&3&4&5&6&7&8  \\\hline
         $2j_-=$0&118&&261&&131&&11&&0\\
         1&&243&&283&&86&&3&\\
         2&65&&183&&155&&34&&0\\
         3&&36&&80&&62&&9&\\
         4&1&&9&&25&&19&&2\\
         5&&0&&1&&6&&5&\\
         4&0&&0&&0&&1&&1\\
         \hline
    \end{tabular}
    
    \caption{Refined GV invariants from 5d $\mathcal{N} = 2$ theory of type $G_2$, fiber degree 2, 3 and 4 with mass degree 2 and 3 respectively.}
    \label{tab:my_label}
\end{table}

\begin{table}[h]
    \centering
    \begin{tabular}{|c|ccc|}\hline
         $\{(1,1,1,1),4,5\}$&$2j_+=$0&1&2  \\\hline
         $2j_-=$0&46&&16\\
         1&&35&\\
         2&8&&3\\
         \hline
    \end{tabular}
    
\begin{tabular}{|c|cc|}\hline
         $\{(1,1,1,1),4,6\}$&$2j_+=$0&1  \\\hline
         $2j_-=$0&&7\\
         1&5&\\
         \hline
    \end{tabular}    

     \begin{center}
     \begin{tabular}{|c|cccc|}\hline
         $\{(1,1,1,1),5,5\}$&$2j_+=$0&1&2&3  \\\hline
         $2j_-=$0&236&&130&\\
         1&&256&&32\\
         2&86&&64&\\
         1&&16&&4\\
         \hline
    \end{tabular}
    
     \begin{tabular}{|c|ccc|}\hline
         $\{(1,1,1,1),5,6\}$&$2j_+=$0&1&2  \\\hline
         $2j_-=$0&&46&\\
         1&38&&14\\
         2&&10&\\
         \hline
    \end{tabular}
    \end{center}
     \begin{tabular}{|c|ccccc|}\hline
         $\{(1,1,1,1),6,5\}$&$2j_+=$0&1&2&3&4  \\\hline
         $2j_-=$0&1172&&866&&37\\
         1&&1572&&365&\\
         2&634&&595&&48\\
         3&&219&&93&\\
         4&11&&24&&5\\
         \hline
    \end{tabular}
    
     \begin{tabular}{|c|cccc|}\hline
         $\{(1,1,1,1),6,6\}$&$2j_+=$0&1&2&3  \\\hline
         $2j_-=$0&&310&&34\\
         1&267&&167&\\
         2&&128&&21\\
         3&13&&15&\\
         \hline
    \end{tabular}
    
    \caption{Refined GV invariants from 5d $\mathcal{N} = 2$ gauge theory of type $F_4$, fiber degree 4, 5 and 6 with mass degree 5 and 6 respectively.}
    \label{tab:my_label2}
\end{table}

\clearpage

\section{One-instanton partition functions }\label{app:Z1}
In this appendix, we provide the one-instanton partition function $Z_1^{\text{inst}}=\frac{\left(\sqrt{q_1} Q_m-\sqrt{q_2}\right) \left(\sqrt{q_1}-\sqrt{q_2}Q_m\right)}{\left(q_1-1\right) \left(q_2-1\right) Q_m} \tilde{Z}_1$ for $G_2$, $F_4$, $E_6$, $E_7$, $E_8$ with gauge fugacities turned off. With the notation $v=(q_1 q_2)^{-\frac{1}{2}}=e^{-\ep_+}$, we observe that $\tilde{Z}_{1,G}$ take the form 
\be
\tilde{Z}_{1,G}=\frac{v^{2\hG-2}}{(1-v^2)^{2\hG-2}}\sum_{n=0}^{\hG-1} \chi_{(n)}P_{n}(v)\,,
\ee
with $P_{n}(v)=P_{n}(v^{-1})$. In the massive limit $m\rightarrow \infty$, only $P_{\hG-1}(v)$ survives, and the theory should become the pure gauge theory with gauge group $G$. We verify that our results here indeed are reduced to the Hilbert series for pure gauge theories studied in \cite{Benvenuti:2010pq,Keller:2011ek}. 
{\allowdisplaybreaks
\subsection*{$G_2+1{\text{\bf{Adj}}}$}
\begin{align*}
&\begin{aligned}\tilde{Z}_{1,G2}&=\frac{v^{6}}{\left(1-v^2\right)^{6}}
\Big(\chi_{(0)}(-7v^4+35v^2+196+35 v^{-2}-7v^{-4})+\chi_{(2)}(15 v^2+78+15 v^{-2})\end{aligned}\\
&\begin{aligned}\hspace{2.1cm}+\chi_{(1)}(7v^3-133v-133 v^{-1}+7 v^{-3})+\chi_{(3)}(-{v^3}-{8}{v}-8 v^{-1}-v^{-3})\Big).\end{aligned}\\
\end{align*}
\subsection*{$F_4+1{\text{\bf{Adj}}}$}
\begin{align*}
&\begin{aligned}\tilde{Z}_{1,F4}=\frac{v^{16}}{\left(1-v^2\right)^{16}}
\Big(\chi_{(0)}(-26 v^{12}+390 v^{10}-1651 v^8+51948 v^6\hspace{1.7cm}\\-391820 v^4-5902234 v^2-11784318-5902234 v ^{-2}+\cdots-26 v ^{-16})\end{aligned} \\
&\begin{aligned}\hspace{2.1cm}+\chi_{(1)}(26 v^{11}-416 v^9-8762 v^7-75556 v^5+3376386 v^3+14360034 v+(v\rightarrow v^{-1}))\end{aligned} \\
&\begin{aligned}\hspace{2.1cm}+\chi_{(2)}(-273 v^8+52598 v^6-690716 v^4-7548710 v^2 \hspace{4.7cm}\\-14516294-7548710 v ^{-2}+\cdots-273 v ^{-8})\end{aligned} \\
&\begin{aligned}\hspace{2.1cm}+\chi_{(3)}(v^{11}+283 v^9-7043 v^7+35983 v^5+2016456 v^3+7459088 v+(v\rightarrow v^{-1}))\end{aligned} \\
&\begin{aligned}\hspace{2.1cm}+\chi_{(4)}(-v^{12}-10 v^{10}-3 v^8+4876 v^6-290771 v^4\hspace{4.1cm}\quad\quad\quad\\-2083666 v^2-3747250-2083666 v ^{-2}+\cdots-v ^{-12})\end{aligned} \\
&\begin{aligned}\hspace{2.1cm}+\chi_{(5)}(26 v^9-416 v^7+23842 v^5+328406 v^3+1005966 v+(v\rightarrow v^{-1}))\end{aligned} \\
&\begin{aligned}\hspace{2.1cm}+\chi_{(6)}(-1326 v^6-29731 v^4-145782 v^2-240370-145782 v ^{-2}+\cdots-1326 v ^{-6})\end{aligned} \\
&\begin{aligned}\hspace{2.1cm}+\chi_{(7)}(53 v^7+1531 v^5+10897 v^3+27455 v+(v\rightarrow v^{-1}))\end{aligned} \\
&\begin{aligned}\hspace{2.1cm}+\chi_{(8)}(-v^8-36 v^6-341 v^4-1208 v^2-1820-1208 v ^{-2}+\cdots-v ^{-8})\Big). \end{aligned}
\end{align*}

\subsection*{$E_6+1{\text{\bf{Adj}}}$}
\begin{small}
\begin{align*}
&\begin{aligned}\tilde{Z}_{1,E6}=\frac{v^{22}}{\left(1-v^2\right)^{22}}
\Big(\chi_{(0)}(v^{20}-22 v^{18}+153 v^{16}+98 v^{14}-14993 v^{12}+158136 v^{10}-6926238 v^8-759340 v^6\\+1523232096 v^4+7806498348 v^2+12799549490+7806498348 v ^{-2}+\cdots+v ^{-20})\end{aligned} \\
&\begin{aligned}\hspace{2.1cm}+\chi_{(1)}(-v^{21}+22 v^{19}-231 v^{17}+1618 v^{15}-9681 v^{13}+64658 v^{11}+293625 v^9 \hspace{2cm}\\+38442635 v^7-564702589 v^5-6466371573 v^3-17713660315 v+(v\rightarrow v^{-1}))\end{aligned} \\
&\begin{aligned}\hspace{2.1cm}+\chi_{(2)}(728 v^{14}-15366 v^{12}+273910 v^{10}-11331190 v^8+40992458 v^6+2550510170 v^4\hspace{1cm}\\+12242971676 v^2+19765078892+12242971676 v ^{-2}+\cdots+728 v ^{-14})\end{aligned} \\
&\begin{aligned}\hspace{2.1cm}+\chi_{(3)}(v^{17}-100 v^{15}+1219 v^{13}-41216 v^{11}+890847 v^9+17633516 v^7\hspace{3.4cm}\\-538737817 v^5-4991754986 v^3-13017447838 v+(v\rightarrow v^{-1}))\end{aligned} \\
&\begin{aligned}\hspace{2.1cm}+\chi_{(4)}(-v^{18}+23 v^{16}-175 v^{14}+2980 v^{12}+64205 v^{10}-4291451 v^8+56601413 v^6 \hspace{1.7cm}\\+1259025394 v^4+5370861892 v^2+8434831880+5370861892 v ^{-2}+\cdots-v ^{-18})\end{aligned} \\
&\begin{aligned}\hspace{2.1cm}+\chi_{(5)}(-v^{17}+22 v^{15}-231 v^{13}-13683 v^{11}+331166 v^9-1537592 v^7\hspace{3.5cm}\\-200446915 v^5-1426472661 v^3-3458377541 v+(v\rightarrow v^{-1}))\end{aligned} \\
&\begin{aligned}\hspace{2.1cm}+\chi_{(6)}(v^{14}+628 v^{12}-1771 v^{10}-207239 v^8+20558339 v^6+247969764 v^4\hspace{2.6cm}\\+913982981 v^2+1383354810+913982981 v ^{-2}+\cdots+v ^{-14})\end{aligned} \\
&\begin{aligned}\hspace{2.1cm}+\chi_{(7)}(-v^{15}+23 v^{13}-903 v^{11}+16071 v^9-1430275 v^7\hspace{5.7cm}\\-28606093 v^5-156881483 v^3-348520227 v+(v\rightarrow v^{-1}))\end{aligned} \\
&\begin{aligned}\hspace{2.1cm}+\chi_{(8)}(-v^{14}+22 v^{12}-231 v^{10}+74535 v^8+2217479 v^6\hspace{5.6cm}\\+17433010 v^4+55071963 v^2+79873398+55071963 v ^{-2}+\cdots-v ^{-14})\end{aligned} \\
&\begin{aligned}\hspace{2.1cm}+\chi_{(9)}(-3003 v^9-114959 v^7-1225510 v^5-5303376 v^3-10737012 v+(v\rightarrow v^{-1}))\end{aligned} \\
&\begin{aligned}\hspace{2.1cm}+\chi_{(10)}(79 v^{10}+3695 v^8+50511 v^6+286946 v^4+780340 v^2\hspace{5cm}\\+1082466+780340 v ^{-2}+\cdots+79 v ^{-10})\end{aligned} \\
&\begin{aligned}\hspace{2.1cm}+\chi_{(11)}(-v^{11}-56 v^9-945 v^7-6776 v^5-23815 v^3-43989 v+(v\rightarrow v^{-1}))\Big). \end{aligned}
\end{align*}
\end{small}
\subsection*{$E_7+1{\text{\bf{Adj}}}$}
\begin{tiny}
\begin{align*}
&\begin{aligned}\tilde{Z}_{1,E7}=\frac{v^{34}}{\left(1-v^2\right)^{34}}
\Big(\chi_{(0)}(&-v^{32}+34 v^{30}-561 v^{28}+5984 v^{26}-47839 v^{24}+326535 v^{22}-2369840 v^{20}+192369 v^{18}+319914520 v^{16}\\&-4555578445 v^{14}+45826191560 v^{12}+4297712255489 v^{10}-38846724860953 v^8-1018471434602090 v^6\\&-5997290980617505 v^4-15870630346831724 v^2-21710965763271066-15870630346831724 v ^{-2}+\cdots-v ^{-32})\end{aligned} \\
&\begin{aligned}\hspace{2.1cm}+\chi_{(1)}(&v^{33}-34 v^{31}+561 v^{29}-5984 v^{27}+46243 v^{25}-272271 v^{23}+1220549 v^{21}-84810 v^{19}-13375056 v^{17}-1414730109 v^{15}\\&+48668669215 v^{13}-1754601580986 v^{11}-6682394704228 v^9+476894906994890 v^7+4765294386349491 v^5\\&+18118640735313330 v^3+34067410941177798 v+(v\rightarrow v^{-1}))\end{aligned} \\
&\begin{aligned}\hspace{2.1cm}+\chi_{(2)}(&133 v^{26}-4389 v^{24}+70091 v^{22}-721259 v^{20}-12392579 v^{18}+511484845 v^{16}-10386329537 v^{14}+141736436531 v^{12}\\&+7722900368721 v^{10}-93820139003029 v^8-2070383748708329 v^6-11777765870556048 v^4-30742729289953364 v^2\\&-41891685607140374-30742729289953364 v ^{-2}+\cdots+133 v ^{-26})\end{aligned} \\
&\begin{aligned}\hspace{2.1cm}+\chi_{(3)}(&-133 v^{25}+4522 v^{23}-65968 v^{21}+589722 v^{19}+10331706 v^{17}-602465756 v^{15}+40915328688 v^{13}-1908223340828 v^{11}\\&-24490662408 v^9+561101153401395 v^7+5126808165044564 v^5+18888291604024014 v^3+35082679743154482 v\\&+(v\rightarrow v^{-1}))\end{aligned} \\
&\begin{aligned}\hspace{2.1cm}+\chi_{(4)}(&-v^{28}+34 v^{26}-561 v^{24}-2661 v^{22}+389731 v^{20}-9787317 v^{18}+223445673 v^{16}-8739023767 v^{14}+167083629053 v^{12}\\&+4152767497551 v^{10}-91661902439757 v^8-1558312171083212 v^6-8314456445946167 v^4-21138018078092692 v^2\\&-28586098919651814-21138018078092692 v ^{-2}+\cdots-v ^{-28})\end{aligned} \\
&\begin{aligned}\hspace{2.1cm}+\chi_{(5)}(&v^{29}-34 v^{27}+561 v^{25}-6117 v^{23}+61006 v^{21}-847363 v^{19}+7090105 v^{17}+353568042 v^{15}+9836487068 v^{13}\\&-928207419623 v^{11}+6595777319662 v^9+332856926666326 v^7+2676443644239847 v^5+9388310115593702 v^3\\&+17100982922552817 v+(v\rightarrow v^{-1}))\end{aligned} \\
&\begin{aligned}\hspace{2.1cm}+\chi_{(6)}(&-v^{26}+167 v^{24}-6413 v^{22}+124354 v^{20}-2308041 v^{18}+32158356 v^{16}-3349371897 v^{14}+89887755266 v^{12}\\&+551143731377 v^{10}-49185392946073 v^8-635558671930867 v^6-3112506303222448 v^4-7625247308988574 v^2\\&-10201064010333132-7625247308988574 v ^{-2}+\cdots-v ^{-26})\end{aligned} \\
&\begin{aligned}\hspace{2.1cm}+\chi_{(7)}(&v^{27}-34 v^{25}+428 v^{23}+v^{21}+82818 v^{19}-3359334 v^{17}+278327436 v^{15}-2013396364 v^{13}-184969637203 v^{11}\\&+4802947005758 v^9+111740083684172 v^7+775842462654293 v^5+2556838199829542 v^3+4539811879506726 v\\&+(v\rightarrow v^{-1}))\end{aligned} \\
&\begin{aligned}\hspace{2.1cm}+\chi_{(8)}(&-v^{24}+34 v^{22}-9206 v^{20}+139117 v^{18}-6657806 v^{16}-431411871 v^{14}+19596106410 v^{12}-272765774910 v^{10}\\&-14582938726201 v^8-146126233882236 v^6-648220381372136 v^4-1517814893616390 v^2-2003377943504808\\&-1517814893616390 v ^{-2}+\cdots-v ^{-24})\end{aligned} \\
&\begin{aligned}\hspace{2.1cm}+\chi_{(9)}(&v^{25}-34 v^{23}+561 v^{21}+2661 v^{19}-542244 v^{17}+46406777 v^{15}-978071314 v^{13}+4310719436 v^{11}+1421475102761 v^9\\&+20892071396056 v^7+124675707137247 v^5+382634921027861 v^3+659189242162231 v+(v\rightarrow v^{-1}))\end{aligned} \\
&\begin{aligned}\hspace{2.1cm}+\chi_{(10)}(&-v^{22}+34 v^{20}+41733 v^{18}-1754575 v^{16}+10060655 v^{14}+507682262 v^{12}-105235094029 v^{10}-2280454391446 v^8\\&-18228754753090 v^6-72806364194640 v^4-162029615548868 v^2-210588997820870-162029615548868 v ^{-2}\\&+\cdots+34v^{-20}-v ^{-22})\end{aligned} \\
&\begin{aligned}\hspace{2.1cm}+\chi_{(11)}(&v^{23}-34 v^{21}-978 v^{19}+4048 v^{17}+1238246 v^{15}-35782416 v^{13}+6125971686 v^{11}+191440381209 v^9+2027213791350 v^7\\&+10437327255044 v^5+29690319476282 v^3+49473456685082 v+(v\rightarrow v^{-1}))\end{aligned} \\
&\begin{aligned}\hspace{2.1cm}+\chi_{(12)}(&-v^{20}+1573 v^{18}-52887 v^{16}+869363 v^{14}-293294143 v^{12}-12445611684 v^{10}-171109984354 v^8-1120919963028 v^6\\&-4027894422408 v^4-8493286201536 v^2-10855716786750-8493286201536 v ^{-2}+\cdots-v ^{-20})\end{aligned} \\
&\begin{aligned}\hspace{2.1cm}+\chi_{(13)}(&v^{21}-34 v^{19}+561 v^{17}-5984 v^{15}+11728431 v^{13}+626265991 v^{11}+10870618606 v^9+89085511339 v^7+398918466329 v^5\\&+1050344693949 v^3+1689966022911 v+(v\rightarrow v^{-1}))\end{aligned} \\
&\begin{aligned}\hspace{2.1cm}+\chi_{(14)}(&-374395 v^{14}-23912735 v^{12}-508619265 v^{10}-5112652335 v^8-28093535470 v^6-91075363610 v^4-181772936730 v^2\\&-228337756920-181772936730 v ^{-2}+\cdots-374395 v ^{-14})\end{aligned} \\
&\begin{aligned}\hspace{2.1cm}+\chi_{(15)}(&8778 v^{15}+656754 v^{13}+16692830 v^{11}+201749030 v^9+1338159284 v^7+5263002052 v^5+12840698332 v^3\\&+19939623340 v+(v\rightarrow v^{-1}))\end{aligned} \\
&\begin{aligned}\hspace{2.1cm}+\chi_{(16)}(&-134 v^{16}-11593 v^{14}-345521 v^{12}-4931707 v^{10}-38850151 v^8-182614170 v^6-536726278 v^4-1014596958 v^2\\&-1252490096-1014596958 v ^{-2}+\cdots-134 v ^{-16})\end{aligned} \\
&\begin{aligned}\hspace{2.1cm}+\chi_{(17)}(v^{17}+99 v^{15}+3410 v^{13}+56617 v^{11}+521917 v^9+2889898 v^7+10086066 v^5+22867856 v^3+34289476 v+(v\rightarrow v^{-1}))\Big). \end{aligned}
\end{align*}
\end{tiny}
\subsection*{$E_8+1{\text{\bf{Adj}}}$}
\begin{tiny}
\begin{align*}
&\begin{aligned}\tilde{Z}_{1,E8}=\frac{v^{58}}{\left(1-v^2\right)^{58}}
\Big(\chi_{(0)}(&-v^{56}+58 v^{54}-1653 v^{52}+30856 v^{50}-424022 v^{48}+4567980 v^{46}-40079798 v^{44}+293578994 v^{42}-1824394402 v^{40}\\&+9667731886 v^{38}-42461014998 v^{36}+132909329948 v^{34}+545475286514 v^{32}-18723488678624 v^{30}+199657721586 v^{28}\\&+6070144373847294 v^{26}-158359829821935208 v^{24}+13123293549284565670 v^{22}-304286151278445771080 v^{20}\\&+93807346004579915840 v^{18}+231210520607954818410890 v^{16}+212094374121108650515140 v^{14}\\&-63237430680825110302314085 v^{12}-868401895100025356669075258 v^{10}-5671384696279393198941920061 v^8\\&-22191130664092299841147741812 v^6-56604002226248335411717851348 v^4-98006885388903722205003131524 v^2\\&-117455715138316541428673838364-98006885388903722205003131524 v ^{-2}+\cdots-v ^{-56})\end{aligned} \\
&\begin{aligned}\hspace{2.1cm}+\chi_{(1)}(&v^{57}-58 v^{55}+1653 v^{53}-30856 v^{51}+424270 v^{49}-4582364 v^{47}+40489742 v^{45}-301084032 v^{43}+1923523102 v^{41}\\&-10703585571 v^{39}+51971657828 v^{37}-214401383003 v^{35}+599038220846 v^{33}-2872323233454 v^{31}\\&+168035547337370 v^{29}-4212687410820234 v^{27}+90593403170144332 v^{25}-3239694035198676927 v^{23}\\&-7786725691478936042 v^{21}+2727192357545908810241 v^{19}-71092705066401885095146 v^{17}\\&-1219169195760730033562408 v^{15}+18317174325496896118960535 v^{13}+487044114279693049517489465 v^{11}\\&+4355779358664347762099988593 v^9+21602847681572549810258688754 v^7+67626508813675827650213628830 v^5\\&+141446535661480894042812123181 v^3+203329332624724715036999923952 v+(v\rightarrow v^{-1}))\end{aligned} \\
&\begin{aligned}\hspace{2.1cm}+\chi_{(2)}(&779247 v^{40}-46959451 v^{38}+1390356541 v^{36}-26958891057 v^{34}+1012912518098 v^{32}-19606748317983 v^{30}\\&-64813181649464 v^{28}+8085786419270105 v^{26}-319990736727661081 v^{24}+28351658261313154772 v^{22}\\&-748943196294632770869 v^{20}+1972385457158510035410 v^{18}+537227449737270069701604 v^{16}\\&-173723938370503186697210 v^{14}-155088475288876573865779057 v^{12}-2058214335347861366919588816 v^{10}\\&-13255964412188118810033092960 v^8-51470859163731719012235016574 v^6-130697683902901841473178066633 v^4\\&-225748482048980748563096528628 v^2-270339252171845088398941576788-225748482048980748563096528628 v ^{-2}\\&+\cdots-46959451 v^{-38}+779247 v ^{-40})\end{aligned} \\
&\begin{aligned}\hspace{2.1cm}+\chi_{(3)}(&1763125 v^{39}-100498125 v^{37}+3015389375 v^{35}-46652194655 v^{33}-2473083650543 v^{31}+146019040206949 v^{29}\\&-4267581287915369 v^{27}+140382849105058163 v^{25}-4643148129781819986 v^{23}+17712400328724500496 v^{21}\\&+3404568482471312364678 v^{19}-105871272583204867656060 v^{17}-1480793029776444231785780 v^{15}\\&+29282662577388574143982106 v^{13}+695633112048480861279448560 v^{11}+6037571298947608844865633794 v^9\\&+29495304460718792477325388428 v^7+91525807890191640869111761302 v^5+190469083642574070772294370556 v^3\\&+273160662888213408798843762586 v+(v\rightarrow v^{-1}))\end{aligned} \\
&\begin{aligned}\hspace{2.1cm}+\chi_{(4)}(&-1763125 v^{38}-177215375 v^{36}+9162279865 v^{34}-208077161810 v^{32}+4935311801965 v^{30}-58825224897985 v^{28}\\&-1780691037142485 v^{26}-207802905938617496 v^{24}+23518066935587317824 v^{22}-811213374640556627886 v^{20}\\&+5533969049770967368644 v^{18}+527888804373262377087644 v^{16}-1593968004881123694712236 v^{14}\\&-171243148489213652702326424 v^{12}-2133141108451865760603213212 v^{10}-13373627029479589437198941070 v^8\\&-51151309281941347159443824298 v^6-128735881963573152686069567584 v^4-221297949981119327485353272294 v^2\\&-264607703052426675654332491724-221297949981119327485353272294 v ^{-2}+\cdots-1763125 v ^{-38})\end{aligned} \\
&\begin{aligned}\hspace{2.1cm}+\chi_{(5)}(&-248 v^{45}+14384 v^{43}-379564 v^{41}+5890248 v^{39}+27966805 v^{37}-4187360867 v^{35}+142822460836 v^{33}\\&-3496165515746 v^{31}+73055252132337 v^{29}-2295223662844225 v^{27}+121500006713148412 v^{25}-3826587262546413180 v^{23}\\&+49243850310385394258 v^{21}+2163057027288332541676 v^{19}-93511250346899328197106 v^{17}-884856624270726352654586 v^{15}\\&+29024244939332034225631046 v^{13}+590482437162756854448984348 v^{11}+4889958518686508618285658630 v^9\\&+23315151466889440009721974542 v^7+71320669377037428398402581264 v^5+147198647056877531264934911348 v^3\\&+210297097534691586870217136388 v+(v\rightarrow v^{-1}))\end{aligned} \\
&\begin{aligned}\hspace{2.1cm}+\chi_{(6)}(&-v^{50}+58 v^{48}-1405 v^{46}+16720 v^{44}-62965 v^{42}-703818 v^{40}-4466161 v^{38}+408946517 v^{36}-2713677886 v^{34}\\&-154668482624 v^{32}+4056834767422 v^{30}+81550596556093 v^{28}-7839674180803077 v^{26}-8963349858682126 v^{24}\\&+9880443699715026855 v^{22}-551582734784698348284 v^{20}+7021788132815152972097 v^{18}+318510994245646293322806 v^{16}\\&-2547737397547098912719187 v^{14}-126768143416544898561539490 v^{12}-1451034910174367338520956078 v^{10}\\&-8755863606942300768953853726 v^8-32765797942839101287415742906 v^6-81397403149589411298561278256 v^4\\&-138941704467266571036507101812 v^2-165762816842880396732666145532-138941704467266571036507101812 v ^{-2}\\&+\cdots -1405 v^{-46}+58 v^{-48}-v ^{-50})\end{aligned} \\
&\begin{aligned}\hspace{2.1cm}+\chi_{(7)}(&v^{51}-58 v^{49}+1653 v^{47}-31104 v^{45}+442529 v^{43}-5182555 v^{41}+52546231 v^{39}-435729525 v^{37}+1611219988 v^{35}\\&+36846201591 v^{33}-1078632270370 v^{31}+20445060694419 v^{29}-769965207995803 v^{27}+61133860612275391 v^{25}\\&-2003857524709105545 v^{23}+49590149513671515789 v^{21}+661473714181831599703 v^{19}-55308526729902247488426 v^{17}\\&-209868346865990507881893 v^{15}+20348519237663810642289580 v^{13}+349007598222334144825346438 v^{11}\\&+2721297037268069200651252384 v^9+12561463335559220201605368670 v^7+37688495647788439490516123468 v^5\\&+76912917856872391232824496101 v^3+109307957869017463158752727343 v+(v\rightarrow v^{-1}))\end{aligned} \\
&\begin{aligned}\hspace{2.1cm}+\chi_{(8)}(&-3875 v^{42}+224750 v^{40}-6405375 v^{38}+91392867 v^{36}+271599339 v^{34}-27610467099 v^{32}-1229298897867 v^{30}\\&+111968835142854 v^{28}-5621478945028769 v^{26}+73318727870035658 v^{24}+928195137349802115 v^{22}\\&-237448017788272394473 v^{20}+5123111892756488763486 v^{18}+117755043498443368475248 v^{16}\\&-2191811579728714517043830 v^{14}-67491711160588413066513633 v^{12}-700674240481213891236098524 v^{10}\\&-4032063581415914499690479154 v^8-14674752126088332714846647544 v^6-35848940726399344211324636019 v^4\\&-60636997486772361997887074244 v^2-72132497764650567557687019822-60636997486772361997887074244 v ^{-2}\\&+\cdots+224750 v^{-40}-3875 v ^{-42})\end{aligned} \\
&\begin{aligned}\hspace{2.1cm}+\chi_{(9)}(&30380 v^{39}+1085 v^{37}-50279985 v^{35}+2077984095 v^{33}+60592436475 v^{31}+50736850414 v^{29}-115324935643176 v^{27}\\&+15328472131266604 v^{25}-618025661845255476 v^{23}+27109498314503455289 v^{21}-44072657413395337029 v^{19}\\&-21558720740026716075154 v^{17}+86825411790168279070576 v^{15}+10308894943613770056421591 v^{13}\\&+149234323829450360394565211 v^{11}+1085335460390099834467425870 v^9+4817681814069831288594309950 v^7\\&+14114348349241905710041211705 v^5+28403320942586731746986059885 v^3+40103427088449600479655530690 v\\&+(v\rightarrow v^{-1}))\end{aligned} \\
&\begin{aligned}\hspace{2.1cm}+\chi_{(10)}(&-v^{46}+58 v^{44}-1653 v^{42}+476 v^{40}+1307390 v^{38}-45637109 v^{36}+745768032 v^{34}-7059403107 v^{32}-947925309004 v^{30}\\&+48490585327300 v^{28}-1852983003062332 v^{26}+48036041789475416 v^{24}-1140180289346043561 v^{22}\\&-57535784401162448555 v^{20}+2277608359463371267549 v^{18}+19876286601461769607850 v^{16}\\&-1198250859148174142128764 v^{14}-26192752327971055106031243 v^{12}-245092924622364383540103494 v^{10}\\&-1335669910171192925995840246 v^8-4704118301782910286621534022 v^6-11263165536466981364560476353 v^4\\&-18842784515328885219283938404 v^2-22336374768289540026045428446-18842784515328885219283938404 v ^{-2}\\&+\cdots+58 v^{-44}-v ^{-46})\end{aligned} \\
&\begin{aligned}\hspace{2.1cm}+\chi_{(11)}(&v^{47}-58 v^{45}+1653 v^{43}-31104 v^{41}+469034 v^{39}-6754100 v^{37}+95895546 v^{35}-997979408 v^{33}+75277506284 v^{31}\\&-1850290227551 v^{29}+19246197406010 v^{27}+316139123756781 v^{25}-74934852306761203 v^{23}+8418559870497199835 v^{21}\\&-118755582942355701625 v^{19}-4928377489993175546247 v^{17}+96442029790213667056394 v^{15}\\&+3764219423783047365266815 v^{13}+46417280428974308965869776 v^{11}+313012533488949623853900179 v^9\\&+1329065780351768382842774621 v^7+3787720466043751517803987162 v^5+7498286579668805427696488633 v^3\\&+10505832961466043350900914572 v+(v\rightarrow v^{-1}))\end{aligned} \\
&\begin{aligned}\hspace{2.1cm}+\chi_{(12)}(&-v^{44}+306 v^{42}-15789 v^{40}+426416 v^{38}-5069124 v^{36}-46055392 v^{34}+907248024 v^{32}-170084309061 v^{30}\\&+8821909613529 v^{28}-242492449225431 v^{26}+13156242897840703 v^{24}-588893170810532845 v^{22}\\&-3888606133021353960 v^{20}+592627891512452388075 v^{18}-3505201080890020387158 v^{16}-438581357852438167025291 v^{14}\\&-7366730775368937592398051 v^{12}-62041123378346710028242521 v^{10}-318628930589129018865328736 v^8\\&-1081537394282008756055096665 v^6-2530898934808358428736338491 v^4-4180886987914167423048744943 v^2\\&-4936012096002313577569303188-4180886987914167423048744943 v ^{-2}+\cdots-v ^{-44})\end{aligned} \\
&\begin{aligned}\hspace{2.1cm}+\chi_{(13)}(&v^{45}-58 v^{43}+1405 v^{41}-16472 v^{39}-132924 v^{37}+9013182 v^{35}-152612048 v^{33}+11812706298 v^{31}-462863160411 v^{29}\\&+9008566018950 v^{27}-759566345512723 v^{25}+14103983023370796 v^{23}+1320019076000415382 v^{21}\\&-43411629246390491762 v^{19}-323861258952481637926 v^{17}+40793635843973108953550 v^{15}\\&+978923758458393690336321 v^{13}+10405497224796024740655624 v^{11}+64862930233753203411854027 v^9\\&+262418019157065437267510825 v^7+725260537783755062737546190 v^5+1409508299872588791889658839 v^3\\&+1957739036053515467451479934 v+(v\rightarrow v^{-1}))\end{aligned} \\
&\begin{aligned}\hspace{2.1cm}+\chi_{(14)}(&147250 v^{36}-14201131 v^{34}+363634464 v^{32}-10669131771 v^{30}+405750628884 v^{28}+17338674957834 v^{26}\\&+1290929161309556 v^{24}-121412129268967574 v^{22}+1588014782264264626 v^{20}+68798229659053837236 v^{18}\\&-2951798168594117674774 v^{16}-108874375535843145783584 v^{14}-1477709033531561671041674 v^{12}\\&-11218386723070054658290934 v^{10}-54131001273632232254719384 v^8-176538787668409903172388054 v^6\\&-402829555151057086583058422 v^4-656174683463846496059550902 v^2-771199039104643325765815292\\&-656174683463846496059550902 v ^{-2}+\cdots+147250 v ^{-36})\end{aligned} \\
&\begin{aligned}\hspace{2.1cm}+\chi_{(15)}(&809627 v^{35}-41297735 v^{33}+1213201833 v^{31}-27410717669 v^{29}+361560792904 v^{27}-157266482417736 v^{25}\\&+5880340602342952 v^{23}+40234295931501400 v^{21}-6388337779469638852 v^{19}+158138381108153324756 v^{17}\\&+10143081400532194815924 v^{15}+177878933383196544825692 v^{13}+1649447504688955223691112 v^{11}\\&+9495349309668124610765912 v^9+36506137778468076110328712 v^7+97610737580745902905583160 v^5\\&+185924098244625258119111794 v^3+255784803170473688660432214 v+(v\rightarrow v^{-1}))\end{aligned} \\
&\begin{aligned}\hspace{2.1cm}+\chi_{(16)}(&-v^{40}+58 v^{38}-32033 v^{36}+983269 v^{34}-3684044 v^{32}-396326035 v^{30}-59167467371 v^{28}+8259623614024 v^{26}\\&-86989527158713 v^{24}-9397613052220250 v^{22}+371668609354516474 v^{20}-5535892927080139732 v^{18}\\&-794197228916430096890 v^{16}-18182394168085282136848 v^{14}-206307363525208981322778 v^{12}\\&-1416494858508468159826154 v^{10}-6411527392374038633715810 v^8-20046703815395749452595942 v^6\\&-44523563448993538410617292 v^4-71433359422291534252808998 v^2-83545118232175977334457868\\&-71433359422291534252808998 v ^{-2}+\cdots-v ^{-40})\end{aligned} \\
&\begin{aligned}\hspace{2.1cm}+\chi_{(17)}(&v^{41}-58 v^{39}+1653 v^{37}-476 v^{35}-1337770 v^{33}+45636024 v^{31}+4075068473 v^{29}-197668659009 v^{27}\\&-8586968571167 v^{25}+616688532587580 v^{23}-13270652821774784 v^{21}+63256045950755165 v^{19}\\&+52656758641546779254 v^{17}+1582582366587569041730 v^{15}+21968958783803540167571 v^{13}\\&+179676472310670476847988 v^{11}+955738113117196477428519 v^9+3486305416748212061165423 v^7\\&+9002770192592388934208388 v^5+16784720329113544886566373 v^3+22856526331356062865650122 v\\&+(v\rightarrow v^{-1}))\end{aligned} \\
&\begin{aligned}\hspace{2.1cm}+\chi_{(18)}(&-v^{38}+58 v^{36}-1653 v^{34}+30856 v^{32}-146749540 v^{30}-3681324859 v^{28}+683959984898 v^{26}-21386304062010 v^{24}\\&+186633013567707 v^{22}+5438822594379181 v^{20}-2995173949943424255 v^{18}-117731425214084304706 v^{16}\\&-1993403023612677922737 v^{14}-19370338981413947060273 v^{12}-120778219242461321702703 v^{10}\\&-512615937303211870490020 v^8-1534465349913942084400545 v^6-3312824197210386372533605 v^4\\&-5230497883105625529286955 v^2-6085619763608170162585676-5230497883105625529286955 v ^{-2}\\&+\cdots+58 v^{-36}-v ^{-38})\end{aligned} \\
&\begin{aligned}\hspace{2.1cm}+\chi_{(19)}(&v^{39}-58 v^{37}+1653 v^{35}-30856 v^{33}+277020 v^{31}+456860014 v^{29}-22174219192 v^{27}+276908622382 v^{25}\\&+7195630872251 v^{23}-371217351686290 v^{21}+148787584918035131 v^{19}+7517600770497882736 v^{17}\\&+154222880172226764796 v^{15}+1773059946638390321841 v^{13}+12915107445197815471466 v^{11}\\&+63575991814882647792709 v^9+219883553831912854948949 v^7+547759467553664243167834 v^5\\&+998657902340137353851409 v^3+1345383243896925333540604 v+(v\rightarrow v^{-1}))\end{aligned} \\
&\begin{aligned}\hspace{2.1cm}+\chi_{(20)}(&151125 v^{32}-13499384 v^{30}+217813037 v^{28}+8196562378 v^{26}-464995140626 v^{24}+11558628661480 v^{22}\\&-6564555474138291 v^{20}-413517617757388358 v^{18}-10172196515867592250 v^{16}-137539898061296235842 v^{14}\\&-1165237684087911464249 v^{12}-6627623812928672491028 v^{10}-26389113532247061623200 v^8\\&-75573144625213786098870 v^6-158462387791614962060855 v^4-246054973683874545250400 v^2\\&-284733568122818710251734-246054973683874545250400 v ^{-2}+\cdots+151125 v ^{-32})\end{aligned} \\
&\begin{aligned}\hspace{2.1cm}+\chi_{(21)}(&-3875 v^{33}+73625 v^{31}+7241259 v^{29}-413362897 v^{27}+11087994502 v^{25}-196982088954 v^{23}+258911957331755 v^{21}\\&+19615662926623491 v^{19}+570979585220347732 v^{17}+9011118501559084210 v^{15}+88314231016254076623 v^{13}\\&+577832289860731810369 v^{11}+2638129161370853741932 v^9+8651303430483291392248 v^7\\&+20777892854008046163343 v^5+37021082951919872372783 v^3+49323667904370057599654 v+(v\rightarrow v^{-1}))\end{aligned} \\
&\begin{aligned}\hspace{2.1cm}+\chi_{(22)}(&-v^{34}+3933 v^{32}-226403 v^{30}+6436231 v^{28}-119991270 v^{26}+1648628366 v^{24}-9046167202338 v^{22}\\&-799588990211897 v^{20}-27148940843135026 v^{18}-495746202005214612 v^{16}-5585452053389149006 v^{14}\\&-41828401731486360745 v^{12}-218007820694596785042 v^{10}-815242726868999320264 v^8-2233319018362942312250 v^6\\&-4545943992912889473017 v^4-6939355426450524863256 v^2-7985566609619567261206-6939355426450524863256 v ^{-2}\\&+\cdots+3933 v^{-32}-v ^{-34})\end{aligned} \\
&\begin{aligned}\hspace{2.1cm}+\chi_{(23)}(&v^{35}-58 v^{33}+1653 v^{31}-30856 v^{29}+424270 v^{27}-4582116 v^{25}+274353541462 v^{23}+27736773903360 v^{21}\\&+1083717132719633 v^{19}+22686600167904341 v^{17}+291835895897431767 v^{15}+2487639612268202035 v^{13}\\&+14730121539365731508 v^{11}+62534852080207394086 v^9+194558336911726887864 v^7+450428219780627807278 v^5\\&+784070064521867978771 v^3+1032867960964930233801 v+(v\rightarrow v^{-1}))\end{aligned} \\
&\begin{aligned}\hspace{2.1cm}+\chi_{(24)}(&-7045404534 v^{24}-805014329084 v^{22}-35772400374704 v^{20}-850811428035964 v^{18}-12407651190151244 v^{16}\\&-119680363804925356 v^{14}-801009808071718406 v^{12}-3842573046332617736 v^{10}-13516588835323464576 v^8\\&-35432252561775870696 v^6-70006100353818906870 v^4-105036052816641771420 v^2\\&-120190537433240210820-105036052816641771420 v ^{-2}+\cdots-7045404534 v ^{-24})\end{aligned} \\
&\begin{aligned}\hspace{2.1cm}+\chi_{(25)}(&148775510 v^{25}+19049643406 v^{23}+953419225714 v^{21}+25548811559972 v^{19}+419439941571350 v^{17}\\&+4550734272855678 v^{15}+34244302241654534 v^{13}+184726976160586716 v^{11}+731286035284214648 v^9\\&+2160725895905789460 v^7+4823128264826191932 v^5+8201895969064540290 v^3+10681972147609674990 v+(v\rightarrow v^{-1}))\end{aligned} \\
&\begin{aligned}\hspace{2.1cm}+\chi_{(26)}(&-2480620 v^{26}-353448050 v^{24}-19762775652 v^{22}-592206494114 v^{20}-10872241140216 v^{18}-131897019452008 v^{16}\\&-1109959536372600 v^{14}-6699686456861308 v^{12}-29708346106549856 v^{10}-98485546674356964 v^8\\&-247218326854512912 v^6-474198468063325920 v^4-699307885601386920 v^2-795678443272075320\\&-699307885601386920 v ^{-2}+\cdots-2480620 v ^{-26})\end{aligned} \\
&\begin{aligned}\hspace{2.1cm}+\chi_{(27)}(&30628 v^{27}+4827196 v^{25}+299475624 v^{23}+9968540172 v^{21}+203393692040 v^{19}+2743366921092 v^{17}+25682058074804 v^{15}\\&+172595935369656 v^{13}+853225717878588 v^{11}+3158857871953560 v^9+8875820357073240 v^7+19112754204541620 v^5\\&+31756812829579200 v^3+40891682939217780 v+(v\rightarrow v^{-1}))\end{aligned} \\
&\begin{aligned}\hspace{2.1cm}+\chi_{(28)}(&-249 v^{28}-43186 v^{26}-2955298 v^{24}-108632690 v^{22}-2449463527 v^{20}-36535786736 v^{18}-378554380744 v^{16}\\&-2818846437352 v^{14}-15462313229055 v^{12}-63637439240298 v^{10}-199235067983160 v^8-479385860055630 v^6\\&-893085571375875 v^4-1294715203724820 v^2-1464860072724360-1294715203724820 v ^{-2}+\cdots-249 v ^{-28})\end{aligned} \\
&\begin{aligned}\hspace{2.1cm}+\chi_{(29)}(&v^{29}+190 v^{27}+14269 v^{25}+576213 v^{23}+14284732 v^{21}+234453749 v^{19}+2675683550 v^{17}+21972715186 v^{15}\\&+133126452657 v^{13}+606326972328 v^{11}+2105555153625 v^9+5634990969615 v^7+11714759112330 v^5\\&+19025183027595 v^3+24223919026560 v+(v\rightarrow v^{-1}))\Big). \end{aligned}
\end{align*}
\end{tiny}
}
\clearpage

\bibliographystyle{JHEP}
\bibliography{main}

\providecommand{\href}[2]{#2}\begingroup\raggedright\begin{thebibliography}{10}

\bibitem{Nahm:1977tg}
W.~Nahm, \emph{{Supersymmetries and their Representations}},
  \href{https://doi.org/10.1016/0550-3213(78)90218-3}{\emph{Nucl. Phys.}
  {\bfseries B135} (1978) 149}.

\bibitem{Witten:1995zh}
E.~Witten, \emph{{Some comments on string dynamics}},  in \emph{{Future
  perspectives in string theory. Proceedings, Conference, Strings'95, Los
  Angeles, USA, March 13-18, 1995}}, pp.~501--523, 1995,
  \href{https://arxiv.org/abs/hep-th/9507121}{{\ttfamily hep-th/9507121}}.

\bibitem{Strominger:1995ac}
A.~Strominger, \emph{{Open p-branes}},
  \href{https://doi.org/10.1016/0370-2693(96)00712-5}{\emph{Phys. Lett.}
  {\bfseries B383} (1996) 44}
  [\href{https://arxiv.org/abs/hep-th/9512059}{{\ttfamily hep-th/9512059}}].

\bibitem{Haghighat:2013gba}
B.~Haghighat, A.~Iqbal, C.~Kozçaz, G.~Lockhart and C.~Vafa,
  \emph{{M-Strings}},
  \href{https://doi.org/10.1007/s00220-014-2139-1}{\emph{Commun. Math. Phys.}
  {\bfseries 334} (2015) 779}
  [\href{https://arxiv.org/abs/1305.6322}{{\ttfamily 1305.6322}}].

\bibitem{Kim:2011mv}
H.-C. Kim, S.~Kim, E.~Koh, K.~Lee and S.~Lee, \emph{{On instantons as
  Kaluza-Klein modes of M5-branes}},
  \href{https://doi.org/10.1007/JHEP12(2011)031}{\emph{JHEP} {\bfseries 12}
  (2011) 031} [\href{https://arxiv.org/abs/1110.2175}{{\ttfamily 1110.2175}}].

\bibitem{Tachikawa:2011ch}
Y.~Tachikawa, \emph{{On S-duality of 5d super Yang-Mills on $S^1$}},
  \href{https://doi.org/10.1007/JHEP11(2011)123}{\emph{JHEP} {\bfseries 11}
  (2011) 123} [\href{https://arxiv.org/abs/1110.0531}{{\ttfamily 1110.0531}}].

\bibitem{Jefferson:2018irk}
P.~Jefferson, S.~Katz, H.-C. Kim and C.~Vafa, \emph{{On Geometric
  Classification of 5d SCFTs}},
  \href{https://doi.org/10.1007/JHEP04(2018)103}{\emph{JHEP} {\bfseries 04}
  (2018) 103} [\href{https://arxiv.org/abs/1801.04036}{{\ttfamily
  1801.04036}}].

\bibitem{Bhardwaj:2018yhy}
L.~Bhardwaj and P.~Jefferson, \emph{{Classifying 5d SCFTs via 6d SCFTs: Rank
  one}}, \href{https://doi.org/10.1007/JHEP07(2019)178,
  10.1007/JHEP01(2020)153}{\emph{JHEP} {\bfseries 07} (2019) 178}
  [\href{https://arxiv.org/abs/1809.01650}{{\ttfamily 1809.01650}}].

\bibitem{Bhardwaj:2018vuu}
L.~Bhardwaj and P.~Jefferson, \emph{{Classifying 5d SCFTs via 6d SCFTs:
  Arbitrary rank}}, \href{https://doi.org/10.1007/JHEP10(2019)282}{\emph{JHEP}
  {\bfseries 10} (2019) 282}
  [\href{https://arxiv.org/abs/1811.10616}{{\ttfamily 1811.10616}}].

\bibitem{Bhardwaj:2019fzv}
L.~Bhardwaj, P.~Jefferson, H.-C. Kim, H.-C. Tarazi and C.~Vafa, \emph{{Twisted
  Circle Compactifications of 6d SCFTs}},
  \href{https://doi.org/10.1007/JHEP12(2020)151}{\emph{JHEP} {\bfseries 12}
  (2020) 151} [\href{https://arxiv.org/abs/1909.11666}{{\ttfamily
  1909.11666}}].

\bibitem{Apruzzi:2019vpe}
F.~Apruzzi, C.~Lawrie, L.~Lin, S.~Schäfer-Nameki and Y.-N. Wang, \emph{{5d
  Superconformal Field Theories and Graphs}},
  \href{https://doi.org/10.1016/j.physletb.2019.135077}{\emph{Phys. Lett.}
  {\bfseries B800} (2020) 135077}
  [\href{https://arxiv.org/abs/1906.11820}{{\ttfamily 1906.11820}}].

\bibitem{Apruzzi:2019kgb}
F.~Apruzzi, S.~Schäfer-Nameki and Y.-N. Wang, \emph{{5d SCFTs from Decoupling
  and Gluing}}, \href{https://doi.org/10.1007/JHEP08(2020)153}{\emph{JHEP}
  {\bfseries 08} (2020) 153}
  [\href{https://arxiv.org/abs/1912.04264}{{\ttfamily 1912.04264}}].

\bibitem{Apruzzi:2019opn}
F.~Apruzzi, C.~Lawrie, L.~Lin, S.~Schäfer-Nameki and Y.-N. Wang, \emph{{Fibers
  add Flavor, Part I: Classification of 5d SCFTs, Flavor Symmetries and BPS
  States}}, \href{https://doi.org/10.1007/JHEP11(2019)068}{\emph{JHEP}
  {\bfseries 11} (2019) 068}
  [\href{https://arxiv.org/abs/1907.05404}{{\ttfamily 1907.05404}}].

\bibitem{Apruzzi:2019enx}
F.~Apruzzi, C.~Lawrie, L.~Lin, S.~Schäfer-Nameki and Y.-N. Wang, \emph{{Fibers
  add Flavor, Part II: 5d SCFTs, Gauge Theories, and Dualities}},
  \href{https://doi.org/10.1007/JHEP03(2020)052}{\emph{JHEP} {\bfseries 03}
  (2020) 052} [\href{https://arxiv.org/abs/1909.09128}{{\ttfamily
  1909.09128}}].

\bibitem{Bhardwaj:2020gyu}
L.~Bhardwaj and G.~Zafrir, \emph{{Classification of 5d $ \mathcal{N} $ = 1
  gauge theories}}, \href{https://doi.org/10.1007/JHEP12(2020)099}{\emph{JHEP}
  {\bfseries 12} (2020) 099}
  [\href{https://arxiv.org/abs/2003.04333}{{\ttfamily 2003.04333}}].

\bibitem{Braun:2021lzt}
A.~P. Braun, J.~Chen, B.~Haghighat, M.~Sperling and S.~Yang, \emph{{Fibre-base
  duality of 5d KK theories}},
  \href{https://arxiv.org/abs/2103.06066}{{\ttfamily 2103.06066}}.

\bibitem{Nakajima:2005fg}
H.~Nakajima and K.~Yoshioka, \emph{{Instanton counting on blowup. II.
  K-theoretic partition function}},
  \href{https://arxiv.org/abs/math/0505553}{{\ttfamily math/0505553}}.

\bibitem{Gottsche:2006bm}
L.~Gottsche, H.~Nakajima and K.~Yoshioka, \emph{{K-theoretic Donaldson
  invariants via instanton counting}},
  \href{https://doi.org/10.4310/PAMQ.2009.v5.n3.a5}{\emph{Pure Appl. Math.
  Quart.} {\bfseries 5} (2009) 1029}
  [\href{https://arxiv.org/abs/math/0611945}{{\ttfamily math/0611945}}].

\bibitem{Sun:2016obh}
K.~Sun, X.~Wang and M.-x. Huang, \emph{{Exact Quantization Conditions, Toric
  Calabi-Yau and Nonperturbative Topological String}},
  \href{https://doi.org/10.1007/JHEP01(2017)061}{\emph{JHEP} {\bfseries 01}
  (2017) 061} [\href{https://arxiv.org/abs/1606.07330}{{\ttfamily
  1606.07330}}].

\bibitem{Grassi:2016nnt}
A.~Grassi and J.~Gu, \emph{{BPS relations from spectral problems and blowup
  equations}}, \href{https://doi.org/10.1007/s11005-019-01163-1}{\emph{Lett.
  Math. Phys.} {\bfseries 109} (2019) 1271}
  [\href{https://arxiv.org/abs/1609.05914}{{\ttfamily 1609.05914}}].

\bibitem{Huang:2017mis}
M.-x. Huang, K.~Sun and X.~Wang, \emph{{Blowup Equations for Refined
  Topological Strings}},
  \href{https://doi.org/10.1007/JHEP10(2018)196}{\emph{JHEP} {\bfseries 10}
  (2018) 196} [\href{https://arxiv.org/abs/1711.09884}{{\ttfamily
  1711.09884}}].

\bibitem{Kim:2020hhh}
H.-C. Kim, M.~Kim, S.-S. Kim and K.-H. Lee, \emph{{Bootstrapping BPS spectra of
  5d/6d field theories}},  \href{https://arxiv.org/abs/2101.00023}{{\ttfamily
  2101.00023}}.

\bibitem{Keller:2012da}
C.~A. Keller and J.~Song, \emph{{Counting Exceptional Instantons}},
  \href{https://doi.org/10.1007/JHEP07(2012)085}{\emph{JHEP} {\bfseries 07}
  (2012) 085} [\href{https://arxiv.org/abs/1205.4722}{{\ttfamily 1205.4722}}].

\bibitem{Kim:2019uqw}
J.~Kim, S.-S. Kim, K.-H. Lee, K.~Lee and J.~Song, \emph{{Instantons from
  Blow-up}}, \href{https://doi.org/10.1007/JHEP06(2020)124,
  10.1007/JHEP11(2019)092}{\emph{JHEP} {\bfseries 11} (2019) 092}
  [\href{https://arxiv.org/abs/1908.11276}{{\ttfamily 1908.11276}}].

\bibitem{Gu:2018gmy}
J.~Gu, B.~Haghighat, K.~Sun and X.~Wang, \emph{{Blowup Equations for 6d SCFTs.
  I}}, \href{https://doi.org/10.1007/JHEP03(2019)002}{\emph{JHEP} {\bfseries
  03} (2019) 002} [\href{https://arxiv.org/abs/1811.02577}{{\ttfamily
  1811.02577}}].

\bibitem{Gu:2019dan}
J.~Gu, A.~Klemm, K.~Sun and X.~Wang, \emph{{Elliptic blowup equations for 6d
  SCFTs. Part II. Exceptional cases}},
  \href{https://doi.org/10.1007/JHEP12(2019)039}{\emph{JHEP} {\bfseries 12}
  (2019) 039} [\href{https://arxiv.org/abs/1905.00864}{{\ttfamily
  1905.00864}}].

\bibitem{Gu:2019pqj}
J.~Gu, B.~Haghighat, A.~Klemm, K.~Sun and X.~Wang, \emph{{Elliptic blowup
  equations for 6d SCFTs. Part III. E-strings, M-strings and chains}},
  \href{https://doi.org/10.1007/JHEP07(2020)135}{\emph{JHEP} {\bfseries 07}
  (2020) 135} [\href{https://arxiv.org/abs/1911.11724}{{\ttfamily
  1911.11724}}].

\bibitem{Gu:2020fem}
J.~Gu, B.~Haghighat, A.~Klemm, K.~Sun and X.~Wang, \emph{{Elliptic Blowup
  Equations for 6d SCFTs. IV: Matters}},
  \href{https://arxiv.org/abs/2006.03030}{{\ttfamily 2006.03030}}.

\bibitem{DelZotto:2016pvm}
M.~Del~Zotto and G.~Lockhart, \emph{{On Exceptional Instanton Strings}},
  \href{https://doi.org/10.1007/JHEP09(2017)081}{\emph{JHEP} {\bfseries 09}
  (2017) 081} [\href{https://arxiv.org/abs/1609.00310}{{\ttfamily
  1609.00310}}].

\bibitem{Gu:2017ccq}
J.~Gu, M.-x. Huang, A.-K. Kashani-Poor and A.~Klemm, \emph{{Refined BPS
  invariants of 6d SCFTs from anomalies and modularity}},
  \href{https://doi.org/10.1007/JHEP05(2017)130}{\emph{JHEP} {\bfseries 05}
  (2017) 130} [\href{https://arxiv.org/abs/1701.00764}{{\ttfamily
  1701.00764}}].

\bibitem{DelZotto:2017mee}
M.~Del~Zotto, J.~Gu, M.-X. Huang, A.-K. Kashani-Poor, A.~Klemm and G.~Lockhart,
  \emph{{Topological Strings on Singular Elliptic Calabi-Yau 3-folds and
  Minimal 6d SCFTs}},
  \href{https://doi.org/10.1007/JHEP03(2018)156}{\emph{JHEP} {\bfseries 03}
  (2018) 156} [\href{https://arxiv.org/abs/1712.07017}{{\ttfamily
  1712.07017}}].

\bibitem{Kim:2018gak}
J.~Kim, K.~Lee and J.~Park, \emph{{On elliptic genera of 6d string theories}},
  \href{https://doi.org/10.1007/JHEP10(2018)100}{\emph{JHEP} {\bfseries 10}
  (2018) 100} [\href{https://arxiv.org/abs/1801.01631}{{\ttfamily
  1801.01631}}].

\bibitem{DelZotto:2018tcj}
M.~Del~Zotto and G.~Lockhart, \emph{{Universal Features of BPS Strings in
  Six-dimensional SCFTs}},
  \href{https://doi.org/10.1007/JHEP08(2018)173}{\emph{JHEP} {\bfseries 08}
  (2018) 173} [\href{https://arxiv.org/abs/1804.09694}{{\ttfamily
  1804.09694}}].

\bibitem{Duan:2018sqe}
Z.~Duan, J.~Gu and A.-K. Kashani-Poor, \emph{{Computing the elliptic genus of
  higher rank E-strings from genus 0 GW invariants}},
  \href{https://doi.org/10.1007/JHEP03(2019)078}{\emph{JHEP} {\bfseries 03}
  (2019) 078} [\href{https://arxiv.org/abs/1810.01280}{{\ttfamily
  1810.01280}}].

\bibitem{Duan:2020cta}
Z.~Duan and J.~Nahmgoong, \emph{{Bootstrapping ADE M-strings}},
  \href{https://doi.org/10.1007/JHEP02(2021)057}{\emph{JHEP} {\bfseries 02}
  (2021) 057} [\href{https://arxiv.org/abs/2009.03626}{{\ttfamily
  2009.03626}}].

\bibitem{Duan:2020imo}
Z.~Duan, D.~J. Duque and A.-K. Kashani-Poor, \emph{{Weyl invariant Jacobi forms
  along Higgsing trees}},  \href{https://arxiv.org/abs/2012.10427}{{\ttfamily
  2012.10427}}.

\bibitem{Huang:2015sta}
M.-x. Huang, S.~Katz and A.~Klemm, \emph{{Topological String on elliptic CY
  3-folds and the ring of Jacobi forms}},
  \href{https://doi.org/10.1007/JHEP10(2015)125}{\emph{JHEP} {\bfseries 10}
  (2015) 125} [\href{https://arxiv.org/abs/1501.04891}{{\ttfamily
  1501.04891}}].

\bibitem{Huang:2020dbh}
M.-X. Huang, S.~Katz and A.~Klemm, \emph{{Towards Refining the Topological
  Strings on Compact Calabi-Yau 3-folds}},
  \href{https://arxiv.org/abs/2010.02910}{{\ttfamily 2010.02910}}.

\bibitem{Cota:2019cjx}
C.~F. Cota, A.~Klemm and T.~Schimannek, \emph{{Topological strings on genus one
  fibered Calabi-Yau 3-folds and string dualities}},
  \href{https://doi.org/10.1007/JHEP11(2019)170}{\emph{JHEP} {\bfseries 11}
  (2019) 170} [\href{https://arxiv.org/abs/1910.01988}{{\ttfamily
  1910.01988}}].

\bibitem{Cota:2020zse}
C.~F. Cota, A.~Klemm and T.~Schimannek, \emph{{State counting on fibered CY-3
  folds and the non-Abelian Weak Gravity Conjecture}},
  \href{https://arxiv.org/abs/2012.09836}{{\ttfamily 2012.09836}}.

\bibitem{Haghighat:2015ega}
B.~Haghighat, S.~Murthy, C.~Vafa and S.~Vandoren, \emph{{F-Theory, Spinning
  Black Holes and Multi-string Branches}},
  \href{https://doi.org/10.1007/JHEP01(2016)009}{\emph{JHEP} {\bfseries 01}
  (2016) 009} [\href{https://arxiv.org/abs/1509.00455}{{\ttfamily
  1509.00455}}].

\bibitem{Lee:2018urn}
S.-J. Lee, W.~Lerche and T.~Weigand, \emph{{Tensionless Strings and the Weak
  Gravity Conjecture}},
  \href{https://doi.org/10.1007/JHEP10(2018)164}{\emph{JHEP} {\bfseries 10}
  (2018) 164} [\href{https://arxiv.org/abs/1808.05958}{{\ttfamily
  1808.05958}}].

\bibitem{Lee:2018spm}
S.-J. Lee, W.~Lerche and T.~Weigand, \emph{{A Stringy Test of the Scalar Weak
  Gravity Conjecture}},
  \href{https://doi.org/10.1016/j.nuclphysb.2018.11.001}{\emph{Nucl. Phys.}
  {\bfseries B938} (2019) 321}
  [\href{https://arxiv.org/abs/1810.05169}{{\ttfamily 1810.05169}}].

\bibitem{Lee:2020gvu}
S.-J. Lee, W.~Lerche, G.~Lockhart and T.~Weigand, \emph{{Quasi-Jacobi forms,
  elliptic genera and strings in four dimensions}},
  \href{https://doi.org/10.1007/JHEP01(2021)162}{\emph{JHEP} {\bfseries 01}
  (2021) 162} [\href{https://arxiv.org/abs/2005.10837}{{\ttfamily
  2005.10837}}].

\bibitem{Vafa:1997mh}
C.~Vafa, \emph{{Geometric origin of Montonen-Olive duality}},
  \href{https://doi.org/10.4310/ATMP.1997.v1.n1.a6}{\emph{Adv. Theor. Math.
  Phys.} {\bfseries 1} (1998) 158}
  [\href{https://arxiv.org/abs/hep-th/9707131}{{\ttfamily hep-th/9707131}}].

\bibitem{kac1990infinite}
V.~G. Kac, \emph{Infinite-dimensional Lie algebras}. Cambridge university
  press, 1990.

\bibitem{Dorey:1996hx}
N.~Dorey, C.~Fraser, T.~J. Hollowood and M.~A.~C. Kneipp, \emph{{S duality in
  N=4 supersymmetric gauge theories with arbitrary gauge group}},
  \href{https://doi.org/10.1016/0370-2693(96)00773-3}{\emph{Phys. Lett.}
  {\bfseries B383} (1996) 422}
  [\href{https://arxiv.org/abs/hep-th/9605069}{{\ttfamily hep-th/9605069}}].

\bibitem{Argyres:2006qr}
P.~C. Argyres, A.~Kapustin and N.~Seiberg, \emph{{On S-duality for
  non-simply-laced gauge groups}},
  \href{https://doi.org/10.1088/1126-6708/2006/06/043}{\emph{JHEP} {\bfseries
  06} (2006) 043} [\href{https://arxiv.org/abs/hep-th/0603048}{{\ttfamily
  hep-th/0603048}}].

\bibitem{Kim:2004xx}
S.~Kim, K.-M. Lee, H.-U. Yee and P.~Yi, \emph{{The N = 1* theories on R**(1+2)
  x S1 with twisted boundary conditions}},
  \href{https://doi.org/10.1088/1126-6708/2004/08/040}{\emph{JHEP} {\bfseries
  08} (2004) 040} [\href{https://arxiv.org/abs/hep-th/0403076}{{\ttfamily
  hep-th/0403076}}].

\bibitem{Hori:1998iv}
K.~Hori, \emph{{Consistency condition for five-brane in M theory on R**5 / Z(2)
  orbifold}}, \href{https://doi.org/10.1016/S0550-3213(98)00728-7}{\emph{Nucl.
  Phys.} {\bfseries B539} (1999) 35}
  [\href{https://arxiv.org/abs/hep-th/9805141}{{\ttfamily hep-th/9805141}}].

\bibitem{Gimon:1998be}
E.~G. Gimon, \emph{{On the M theory interpretation of orientifold planes}},
  \href{https://arxiv.org/abs/hep-th/9806226}{{\ttfamily hep-th/9806226}}.

\bibitem{Hanany:2000fq}
A.~Hanany and B.~Kol, \emph{{On orientifolds, discrete torsion, branes and M
  theory}}, \href{https://doi.org/10.1088/1126-6708/2000/06/013}{\emph{JHEP}
  {\bfseries 06} (2000) 013}
  [\href{https://arxiv.org/abs/hep-th/0003025}{{\ttfamily hep-th/0003025}}].

\bibitem{Witten:1979ey}
E.~Witten, \emph{{Dyons of Charge e theta/2 pi}},
  \href{https://doi.org/10.1016/0370-2693(79)90838-4}{\emph{Phys. Lett.}
  {\bfseries 86B} (1979) 283}.

\bibitem{Sen:1994yi}
A.~Sen, \emph{{Dyon - monopole bound states, selfdual harmonic forms on the
  multi - monopole moduli space, and SL(2,Z) invariance in string theory}},
  \href{https://doi.org/10.1016/0370-2693(94)90763-3}{\emph{Phys. Lett.}
  {\bfseries B329} (1994) 217}
  [\href{https://arxiv.org/abs/hep-th/9402032}{{\ttfamily hep-th/9402032}}].

\bibitem{Lee:1996if}
K.-M. Lee, E.~J. Weinberg and P.~Yi, \emph{{Electromagnetic duality and SU(3)
  monopoles}}, \href{https://doi.org/10.1016/0370-2693(96)00286-9}{\emph{Phys.
  Lett.} {\bfseries B376} (1996) 97}
  [\href{https://arxiv.org/abs/hep-th/9601097}{{\ttfamily hep-th/9601097}}].

\bibitem{Gauntlett:1996cw}
J.~P. Gauntlett and D.~A. Lowe, \emph{{Dyons and S duality in N=4
  supersymmetric gauge theory}},
  \href{https://doi.org/10.1016/0550-3213(96)00218-0}{\emph{Nucl. Phys.}
  {\bfseries B472} (1996) 194}
  [\href{https://arxiv.org/abs/hep-th/9601085}{{\ttfamily hep-th/9601085}}].

\bibitem{Nekrasov:2004vw}
N.~Nekrasov and S.~Shadchin, \emph{{ABCD of instantons}},
  \href{https://doi.org/10.1007/s00220-004-1189-1}{\emph{Commun. Math. Phys.}
  {\bfseries 252} (2004) 359}
  [\href{https://arxiv.org/abs/hep-th/0404225}{{\ttfamily hep-th/0404225}}].

\bibitem{Hwang:2014uwa}
C.~Hwang, J.~Kim, S.~Kim and J.~Park, \emph{{General instanton counting and 5d
  SCFT}}, \href{https://doi.org/10.1007/JHEP07(2015)063,
  10.1007/JHEP04(2016)094}{\emph{JHEP} {\bfseries 07} (2015) 063}
  [\href{https://arxiv.org/abs/1406.6793}{{\ttfamily 1406.6793}}].

\bibitem{Hwang:2016gfw}
Y.~Hwang, J.~Kim and S.~Kim, \emph{{M5-branes, orientifolds, and S-duality}},
  \href{https://doi.org/10.1007/JHEP12(2016)148}{\emph{JHEP} {\bfseries 12}
  (2016) 148} [\href{https://arxiv.org/abs/1607.08557}{{\ttfamily
  1607.08557}}].

\bibitem{Kim:2016foj}
H.-C. Kim, S.~Kim and J.~Park, \emph{{6d strings from new chiral gauge
  theories}},  \href{https://arxiv.org/abs/1608.03919}{{\ttfamily 1608.03919}}.

\bibitem{Gopakumar:1998ii}
R.~Gopakumar and C.~Vafa, \emph{{M theory and topological strings. 1.}},
  \href{https://arxiv.org/abs/hep-th/9809187}{{\ttfamily hep-th/9809187}}.

\bibitem{Gopakumar:1998jq}
R.~Gopakumar and C.~Vafa, \emph{{M theory and topological strings. 2.}},
  \href{https://arxiv.org/abs/hep-th/9812127}{{\ttfamily hep-th/9812127}}.

\bibitem{Iqbal:2007ii}
A.~Iqbal, C.~Kozcaz and C.~Vafa, \emph{{The Refined topological vertex}},
  \href{https://doi.org/10.1088/1126-6708/2009/10/069}{\emph{JHEP} {\bfseries
  10} (2009) 069} [\href{https://arxiv.org/abs/hep-th/0701156}{{\ttfamily
  hep-th/0701156}}].

\bibitem{Aganagic:2011mi}
M.~Aganagic, M.~C.~N. Cheng, R.~Dijkgraaf, D.~Krefl and C.~Vafa, \emph{{Quantum
  Geometry of Refined Topological Strings}},
  \href{https://doi.org/10.1007/JHEP11(2012)019}{\emph{JHEP} {\bfseries 11}
  (2012) 019} [\href{https://arxiv.org/abs/1105.0630}{{\ttfamily 1105.0630}}].

\bibitem{Nekrasov:2002qd}
N.~A. Nekrasov, \emph{{Seiberg-Witten prepotential from instanton counting}},
  \href{https://doi.org/10.4310/ATMP.2003.v7.n5.a4}{\emph{Adv. Theor. Math.
  Phys.} {\bfseries 7} (2003) 831}
  [\href{https://arxiv.org/abs/hep-th/0206161}{{\ttfamily hep-th/0206161}}].

\bibitem{Nekrasov:2003rj}
N.~Nekrasov and A.~Okounkov, \emph{{Seiberg-Witten theory and random
  partitions}}, \href{https://doi.org/10.1007/0-8176-4467-9_15}{\emph{Prog.
  Math.} {\bfseries 244} (2006) 525}
  [\href{https://arxiv.org/abs/hep-th/0306238}{{\ttfamily hep-th/0306238}}].

\bibitem{Hatsuda:2013oxa}
Y.~Hatsuda, M.~Marino, S.~Moriyama and K.~Okuyama, \emph{{Non-perturbative
  effects and the refined topological string}},
  \href{https://doi.org/10.1007/JHEP09(2014)168}{\emph{JHEP} {\bfseries 09}
  (2014) 168} [\href{https://arxiv.org/abs/1306.1734}{{\ttfamily 1306.1734}}].

\bibitem{Wang:2015wdy}
X.~Wang, G.~Zhang and M.-x. Huang, \emph{{New Exact Quantization Condition for
  Toric Calabi-Yau Geometries}},
  \href{https://doi.org/10.1103/PhysRevLett.115.121601}{\emph{Phys. Rev. Lett.}
  {\bfseries 115} (2015) 121601}
  [\href{https://arxiv.org/abs/1505.05360}{{\ttfamily 1505.05360}}].

\bibitem{Choi:2012jz}
J.~Choi, S.~Katz and A.~Klemm, \emph{{The refined BPS index from stable pair
  invariants}}, \href{https://doi.org/10.1007/s00220-014-1978-0}{\emph{Commun.
  Math. Phys.} {\bfseries 328} (2014) 903}
  [\href{https://arxiv.org/abs/1210.4403}{{\ttfamily 1210.4403}}].

\bibitem{MR1228584}
K.~Oguiso, \emph{On algebraic fiber space structures on a {C}alabi-{Y}au
  {$3$}-fold}, \href{https://doi.org/10.1142/S0129167X93000248}{\emph{Internat.
  J. Math.} {\bfseries 4} (1993) 439}.

\bibitem{MR1314743}
P.~M.~H. Wilson, \emph{The existence of elliptic fibre space structures on
  {C}alabi-{Y}au threefolds},
  \href{https://doi.org/10.1007/BF01450510}{\emph{Math. Ann.} {\bfseries 300}
  (1994) 693}.

\bibitem{Schimannek:2019ijf}
T.~Schimannek, \emph{{Modularity from Monodromy}},
  \href{https://doi.org/10.1007/JHEP05(2019)024}{\emph{JHEP} {\bfseries 05}
  (2019) 024} [\href{https://arxiv.org/abs/1902.08215}{{\ttfamily
  1902.08215}}].

\bibitem{MR2927365}
A.~Morrison, S.~Mozgovoy, K.~Nagao and B.~Szendr\H{o}i, \emph{Motivic
  {D}onaldson-{T}homas invariants of the conifold and the refined topological
  vertex}, \href{https://doi.org/10.1016/j.aim.2012.03.030}{\emph{Adv. Math.}
  {\bfseries 230} (2012) 2065}.

\bibitem{MR2545686}
R.~Pandharipande and R.~P. Thomas, \emph{Curve counting via stable pairs in the
  derived category},
  \href{https://doi.org/10.1007/s00222-009-0203-9}{\emph{Invent. Math.}
  {\bfseries 178} (2009) 407}.

\bibitem{Cardy:1986ie}
J.~L. Cardy, \emph{{Operator Content of Two-Dimensional Conformally Invariant
  Theories}}, \href{https://doi.org/10.1016/0550-3213(86)90552-3}{\emph{Nucl.
  Phys.} {\bfseries B270} (1986) 186}.

\bibitem{Kim:2019yrz}
J.~Kim, S.~Kim and J.~Song, \emph{{A 4d $ \mathcal{N} $ = 1 Cardy Formula}},
  \href{https://doi.org/10.1007/JHEP01(2021)025}{\emph{JHEP} {\bfseries 01}
  (2021) 025} [\href{https://arxiv.org/abs/1904.03455}{{\ttfamily
  1904.03455}}].

\bibitem{Strominger:1996sh}
A.~Strominger and C.~Vafa, \emph{{Microscopic origin of the Bekenstein-Hawking
  entropy}}, \href{https://doi.org/10.1016/0370-2693(96)00345-0}{\emph{Phys.
  Lett.} {\bfseries B379} (1996) 99}
  [\href{https://arxiv.org/abs/hep-th/9601029}{{\ttfamily hep-th/9601029}}].

\bibitem{Choi:2018hmj}
S.~Choi, J.~Kim, S.~Kim and J.~Nahmgoong, \emph{{Large AdS black holes from
  QFT}},  \href{https://arxiv.org/abs/1810.12067}{{\ttfamily 1810.12067}}.

\bibitem{Lee:2020rns}
K.~Lee and J.~Nahmgoong, \emph{{Cardy Limits of 6d Superconformal Theories}},
  \href{https://arxiv.org/abs/2006.10294}{{\ttfamily 2006.10294}}.

\bibitem{Seiberg:1994rs}
N.~Seiberg and E.~Witten, \emph{{Electric - magnetic duality, monopole
  condensation, and confinement in N=2 supersymmetric Yang-Mills theory}},
  \href{https://doi.org/10.1016/0550-3213(94)90124-4,
  10.1016/0550-3213(94)00449-8}{\emph{Nucl. Phys.} {\bfseries B426} (1994) 19}
  [\href{https://arxiv.org/abs/hep-th/9407087}{{\ttfamily hep-th/9407087}}].

\bibitem{Klebanov:1996un}
I.~R. Klebanov and A.~A. Tseytlin, \emph{{Entropy of near extremal black
  p-branes}}, \href{https://doi.org/10.1016/0550-3213(96)00295-7}{\emph{Nucl.
  Phys.} {\bfseries B475} (1996) 164}
  [\href{https://arxiv.org/abs/hep-th/9604089}{{\ttfamily hep-th/9604089}}].

\bibitem{Aharony:1999ks}
O.~Aharony, \emph{{A Brief review of 'little string theories'}},
  \href{https://doi.org/10.1088/0264-9381/17/5/302}{\emph{Class. Quant. Grav.}
  {\bfseries 17} (2000) 929}
  [\href{https://arxiv.org/abs/hep-th/9911147}{{\ttfamily hep-th/9911147}}].

\bibitem{Kim:2015gha}
J.~Kim, S.~Kim and K.~Lee, \emph{{Little strings and T-duality}},
  \href{https://doi.org/10.1007/JHEP02(2016)170}{\emph{JHEP} {\bfseries 02}
  (2016) 170} [\href{https://arxiv.org/abs/1503.07277}{{\ttfamily
  1503.07277}}].

\bibitem{Zagierbook}
J.~H. Bruinier, G.~van~der Geer, G.~Harder and D.~Zagier, \emph{The 1-2-3 of
  modular forms}, Universitext. Springer-Verlag, Berlin, 2008,
  \href{https://doi.org/10.1007/978-3-540-74119-0}{10.1007/978-3-540-74119-0}.

\bibitem{MR0344216}
J.-P. Serre, \emph{A course in arithmetic}. Springer-Verlag, New
  York-Heidelberg, 1973.

\bibitem{EZ}
M.~Eichler and D.~Zagier, \emph{The theory of {J}acobi forms}, vol.~55 of
  \emph{Progress in Mathematics}. Birkh\"auser Boston, Inc., Boston, MA, 1985,
  \href{https://doi.org/10.1007/978-1-4684-9162-3}{10.1007/978-1-4684-9162-3}.

\bibitem{Benvenuti:2010pq}
S.~Benvenuti, A.~Hanany and N.~Mekareeya, \emph{{The Hilbert Series of the One
  Instanton Moduli Space}},
  \href{https://doi.org/10.1007/JHEP06(2010)100}{\emph{JHEP} {\bfseries 06}
  (2010) 100} [\href{https://arxiv.org/abs/1005.3026}{{\ttfamily 1005.3026}}].

\bibitem{Keller:2011ek}
C.~A. Keller, N.~Mekareeya, J.~Song and Y.~Tachikawa, \emph{{The ABCDEFG of
  Instantons and W-algebras}},
  \href{https://doi.org/10.1007/JHEP03(2012)045}{\emph{JHEP} {\bfseries 03}
  (2012) 045} [\href{https://arxiv.org/abs/1111.5624}{{\ttfamily 1111.5624}}].

\end{thebibliography}\endgroup

\end{document}